\documentclass{ws-ijmpd}
\usepackage[compress]{cite}
\usepackage[colorlinks=true,citecolor=blue,linkcolor=blue]{hyperref}
\usepackage{float}
\usepackage{xcolor}
\usepackage[perpage]{footmisc}
\usepackage{amsmath}

\newcommand{\be}{\begin{equation}}
\newcommand{\ee}{\end{equation}}
\newcommand{\bea}{\begin{eqnarray}}
\newcommand{\eea}{\end{eqnarray}}

\newcommand{\Mpl}{M_{\textrm{Pl}}}

\def\lam{\lambda}

\def\doi{http://doi.org}

\def\d{\mathrm{d}}

\begin{document}

\title{Spontaneous symmetry breaking in the  late Universe  and glimpses of early Universe phase transitions  {\it \`{a} la}
baryogenesis
\footnote{Dedicated to Professor J V Narlikar on his ${83}^{\text{rd}}$ ~birthday.}  }

\author{M. Sami}
\address{Centre for Cosmology and Science Popularization(CCSP), SGT University, Gurugram 12006, India.\\
International Center for Cosmology, Charusat University, Anand 388421, Gujarat, India.\\
Center for Theoretical Physics, Eurasian National University,
Astana 010008, Kazakhstan.\\
samijamia@gmail.com}
\author{Radouane Gannouji}
\address{Instituto de F\'{\i}sica, Pontificia Universidad Cat\'olica de Valpara\'{\i}so,
Av. Brasil 2950, Valpara\'{\i}so, Chile\\
radouane.gannouji@pucv.cl}

\maketitle

\begin{abstract}
Spontaneous symmetry breaking is the foundation of electroweak unification and serves as an integral part of the model building beyond the standard model of particle physics and it also finds interesting applications in the late Universe. We  review development related to obtaining the late cosmic acceleration from spontaneous symmetry breaking in the Universe at large scales. This phenomenon is best understood through Ginzburg-Landau theory of phase transitions which we briefly describe. Hereafter, we present elements of spontaneous symmetry breaking
in relativistic field theory.
We then discuss the "symmetron" scenario-based upon symmetry breaking in the late Universe which is realized by using a specific form of conformal coupling.
However, the model
is faced with "NO GO" for late time acceleration due to local gravity constraints. We argue that the problem can be circumvented by using the massless $\lambda \phi^4$ theory coupled to massive neutrino matter. 
As for the early Universe, spontaneous symmetry breaking finds its interesting applications in the study of  electroweak phase transition.
To this effect, we first discuss in detail, the Ginzburg-Landau theory of first order phase transitions
and then apply it to electroweak phase transition including technical discussions on bubble nucleation and sphaleron transitions. We provide a pedagogical expositions of dynamics of electroweak phase transition and emphasize the need to go beyond the standard model of particle physics for addressing the baryogenesis problem. Review ends with a brief discussion on Affleck-Dine mechanism and spontaneous baryogenesis. Appendixes include technical details on essential ingredients of baryogenesis, sphaleron solution, one loop finite temperature effective potential and dynamics of bubble nucleation.
\end{abstract}




\section{Introduction}
Accelerated expansion is generic to our universe; it is believed that the Universe has gone through inflation at early times \cite{Starobinsky:1980te,Guth:1980zm,Linde:1981mu}, and   again entered into the accelerating phase in the recent past \cite{Riess:1998cb,Perlmutter:1998np}. Despite the grand successes, the hot big bang has to its credit, the model has internal inconsistencies: (i) Inconsistencies related to the early universe include, for instance, the flatness and horizon problems  which are beautifully addressed by the paradigm of inflation$-$ brief early phase of accelerated expansion. (ii) The Age puzzle in the hot big bang model is necessarily related to late time evolution and the only known way to address it in the standard lore is provided by the late time cosmic acceleration. 
The phenomenon is confirmed by direct as well as by indirect observations and constitutes one of the most remarkable discoveries of modern cosmology. As for the theoretical understanding, if one adheres to Einstein's theory of general relativity (GR), it necessarily asks for the presence of an exotic matter, repulsive in nature, dubbed {\it dark energy} \cite{Sahni,Copeland,paddy,BambaDE}.
It would be fair to state that the underlying cause of the late time acceleration remains to be the mystery of our times. It is plausible to look for a distinguished physical process  in the late universe with a characteristic mass scale around that of dark energy. And this draws our attention to  neutrino matter, which is one of the most abundant components of the universe  today \cite{Les}. Interestingly, the order of magnitude of neutrino masses  is close to the mass scale of dark energy such that massive neutrinos turn non-relativistic around the present epoch. And certainly, this is a generic  physical process in the late universe with the characteristic mass scale one is looking for. It is tempting to ask whether this process can trigger a late time phase transition in the universe responsible for turning deceleration into  acceleration.   

Let us note that symmetron was one of the first model 
that attempted to realize the idea of late time symmetry breaking or phase transition in the Universe \cite{Hinterbichler:2010es}, {based partially on earlier work such as \cite{Pietroni:2005pv} and also \cite{Olive:2007aj} where a non-universal coupling to matter was considered}, see also Refs.\cite{Bamba:2012yf,cop,others1,others3,Amend,Ant,novikov,Buch} on the related theme. The model is based upon the $\lambda \phi^4$ theory with direct coupling of the field to  matter (coupling is typically proportional to the trace of energy momentum tensor of matter). The underlying symmetry for 
symmetron is  $Z_2$ which is exact in the high density regime (locally), the symmetry is broken in low density regime at large scales giving rise to the true ground state where the field should ultimately settle to mimic  de-Sitter like solution 
of interest to late time acceleration. Unfortunately, the scenario fails due to local gravity constraints which impose stringent constraints on any model which involves direct coupling to matter. Indeed, proper local screening of the extra degree of freedom in this case requires that mass of the scalar field be $\mathcal{O}(10^{4}H_0$) which is too large (compared to $H_0$) to support slow roll around the true ground state obtained after symmetry breaking.
 
The problem faced by the symmetron model can be circumvented by assuming a coupling of the field to massive neutrino matter\cite{Amendola,Wet,staro,H1,sajad} proportional to its trace, which vanishes as long as the neutrinos are relativistic\footnote{Coupling $\propto T_\nu=\rho^\nu(3\omega_\nu-1)$; $\omega_\nu$ $\&$ $\rho^\nu$ being the equation of state parameter and energy density of massive neutrino matter respectively. At early times, neutrino matter is relativistic or $\omega_\nu=1/3$ such that coupling vanishes.}. In this case, coupling builds up dynamically at late times when massive neutrinos become non-relativistic and mimic cold matter ($\omega_\nu=0)$. Another distinguished features of the scenario include: (i) After symmetry breaking, mass of the field gets
 naturally
 linked to the density of massive neutrino matter. (ii) Local gravity constraints do not apply to the neutrino matter, thereby, no extra constraint on mass of the scalar field. 
 
 We should hereby admit that spontaneous symmetry breaking finds its realistic applications in the early Universe, which was hot and composed of plasma of elementary particles. As Universe cooled, it went through various phase transitions with the rearrangement of its ground state. Perhaps one of the most important epochs in the life of Universe was the Electroweak Phase Transition (EWPT) when temperature was $\mathcal{O}(100)$ GeV\cite{Linde:1981zj,Buch,Rbook}. This is a first order  phase transition  which can be described by  one loop finite temperature effective potential computed in the framework of electroweak theory. The EWPT proceeds through bubble nucleation and provides an arena for addressing the baryogenesis problem. Unfortunately, the EWPT is not strong enough to knock out the sphaleron transitions from equilibrium required to produce the observed baryon asymmetry in the Universe\cite{Sakharov,Rbook,Jim,MT,D,AR,AD,SONI,RAGHU}. And this clearly indicates a way out of the standard model.
 
 The plan of the review is as follows.
For pedagogical considerations, we present glimpses of Ginzburg-Landau theory  of phase transitions (section \ref{SGL})  which inspired the idea of spontaneous symmetry breaking in field theory. We then argue that spontaneous symmetry breaking does not take place  in finite systems; quantum tunnelling removes the classical degeneracy of the ground state (section \ref{SQM}). In section \ref{SSS}, we provide detailed  discussion on symmetry breaking in field theory. In Section \ref{sls}, we bring out the details of the scenario that uses the direct coupling of the  scalar field to  matter and show that coupling to massive neutrino matter can evade "NO GO" faced by the "symmetron" model.
Section \ref{euniverse} is devoted to the applications of spontaneous symmetry breaking to early Universe. This section contains a pedagogical exposition on the dynamics of electroweak phase transition.
Section \ref{CON} summarises the results of the review. Appendixes include technical details on necessary gradients of baryogenesis, sphaleron solutions, one loop finite temperature effective potential and dynamics of bubble nucleation.
This review is pedagogical in nature, it aims at young researchers not acquainted with high energy physics.  People familiar with spontaneous symmetry breaking may directly jump to section \ref{SCONF} after briefly looking through Ginzburg-Landau theory of phase transitions. Readers interested in the early Universe phase transitions, may skip section \ref{sls}. Review should be read with footnotes which include additional explanations and clarifications. 
Last but not least, a comment on the choice of topics is in order. In section \ref{SSS}, we described in detail:  (1) Spontaneous breaking of $Z_2$ symmetry which is essential for  understating the "symmetron" scenario that uses this concept; (2)  Breaking of continuous  $U(1)$ global symmetry which is a prerequisite for "spontaneous baryogenesis"\cite{SB1,SB2,Ar,sam}. (3) Abelian Higgs model that serves as a foundation 
for grasping the selected aspects of the standard model necessary to understand the dynamics of electroweak phase transition. We hope that section \ref{SSS} would be extremely helpful to cosmologists not acquainted with high energy physics.

We use the metric signature, $(-,+,+,+)$ and  the notation for the reduced Planck mass, $\Mpl^{-2} = 8\pi G$ along with the system of units, $\hbar=c=k_B=1$.

\section{Thermodynamic theory of phase transitions  {\it {\`{a} la}}  Ginzburg-Landau }
\label{SGL}

The mechanism of spontaneous symmetry breaking serves as the foundation of electroweak theory. The underlying idea is essentially inspired by the thermodynamic theory of phase transitions known as Ginzburg$-$Landau theory. It is based upon the following assumptions: (i) The thermodynamic potentials (Gibbs Free energy, Gibbs potential and others) apart from the standard thermodynamic variables such as Pressure (P), Entropy (S), Temperature (T) etc also depend upon an additional parameter dubbed {\it order parameter}, (ii) In the small neighbourhood of phase transition, the thermodynamic potentials can be represented through Taylor series in the order parameter retaining first few terms of the latter. In case of para-ferromagnetic transition, the order parameter is represented by, {\it spontaneous magnetization }(non-vanishing magnetization in absence of external magnetic field), for para-to ferroelectric transition, the order parameter is given by spontaneous polarization, in case of  superconducting phase transition, it is the {\it Cooper pair density} and so on\footnote{In field theory, the order parameter is given by the non-vanishing vacuum expectation value of the  field.}. 

Let us emphasize that the thermodynamics description applies to the equilibrium state of a system which is distinguished in a sense that it corresponds to the minimum of thermodynamic potentials: {\it A thermodynamic system not in equilibrium, left uninterrupted, would ultimately enter into the equilibrium state. } Thus, the extremal property of thermodynamic potentials is essentially associated with the criteria of stability. Let us also note that unlike the case of conservative forces, the work done in thermodynamics is process-dependent quantity, thereby, there is no unique potential; often used thermodynamic potentials include$-$ internal energy (U), Gibbs free energy (F), Gibbs potential (G) and enthalpy (W). 

In what follows, we shall use Gibbs potential which, apart from the standard thermodynamic variables P and T,  should also be a function of an additional parameter dubbed {\it order parameter}, 
\begin{equation}
 G=G(P,T,x;\alpha),
\end{equation}
where $x$ is an external field and $\alpha$ denotes the order parameter. In what follows, we shall spell out the functional dependence of G upon the order parameter  which would determine fo us the type  of the phase transition, namely, the "first order" or the "second order". 
\subsection{Second order phase transitions}
\label{2order}
According to Ginzburg$-$Landau assumption, the Gibbs potential can be expanded into Taylor series in $\alpha$ in the neighbourhood of the critical point,
\begin{equation}
\label{GL}
 G(P,T,x;\alpha)=G_0( P,T,x)-\alpha x+\frac{1}{2}a(T,P) \alpha^2+  \frac{1}{4}b(T,P) \alpha^4+..
\end{equation}
where by assumption, $b>0$ which is necessary to ensure the stability of the system (G should be bounded from below). Secondly, it is sufficient to retain the first few terms of the Taylor series.
{Thirdly, in the case under consideration, adhering to reflection symmetry (in the absence of external field), one keeps only even powers of the order parameter in the expression of the Gibbs potential (\ref{GL}).}
    In what follows, we shall be interested in understanding the features
   of the system in absence of the external field. Keeping terms up to fourth order in $\alpha$\footnote{The Gibbs potential (\ref{GL1}) would describe the second order phase transition which is the rationale behind the assumption.}, we have
\begin{equation}
\label{GL1}
G(P,T;\alpha)-G_0(P,T)=\frac{1}{2}a(P,T) \alpha^2+  \frac{1}{4}b(P,T) \alpha^4
\end{equation}
Let us note that the physical quantities are given by the first and second derivatives of  thermodynamic potentials, which implies, in particular, that $G$ must be continuous, otherwise the physical quantities would be ill defined. It is imperative that $a(P,T)$, $b(P,T)$ are continuous functions of $P$ and $T$.
 Since G should be minimum in the state of thermodynamic equilibrium, for fixed values of T and P, we have,
 \begin{equation}
 \frac{\partial G}{\partial \alpha}=0;~~~~\frac{\partial^2 G}{\partial \alpha^2}>0     \end{equation}
 which implies,
 \begin{equation}
 \label{alphaeq}
\alpha[a(T,P)+b(T,P)\alpha^2]=0;~~~
a(T,P)+3b(T,P)\alpha^2>0
 \end{equation}
 that allows us to determine the order parameter along with stability conditions, dictated by the inequality in expression (\ref{alphaeq}),
 \begin{equation}
     \alpha(T,P) =
    \begin{cases}
    0  ,\quad \quad \quad \quad \quad \quad a(T,P)>0\ , \\
    \pm \sqrt{-\frac{a(T,P)}{b(T,P)}}, \quad  \quad a(T,P)<0 
    \end{cases}
 \end{equation}   
 From the continuity of functions, $a$ and $b$,
 Eq.(6) tells us that the order parameter vanishes for $a(T,P)>0$ but picks up the nonzero value for $a(T,P)<0$. In the second case, the state of minimum energy or ground state,  is doubly degenerate. Since $a(T,P)$ is a continuous function of its variables, thereby, while changing sign, it passes through zero for certain values of $T=T_c, P=P_c$ dubbed the critical point,
 \begin{equation}
   a(T_c,P_c)=0  \to \alpha(T_c,P_c)=0,
   \label{acrit1}
 \end{equation}
and this implies that the order parameter is continuous at the critical point which is  the distinguished features of the second order phase transition.
 \begin{figure}[ht]
\centering
\includegraphics[scale=.4]{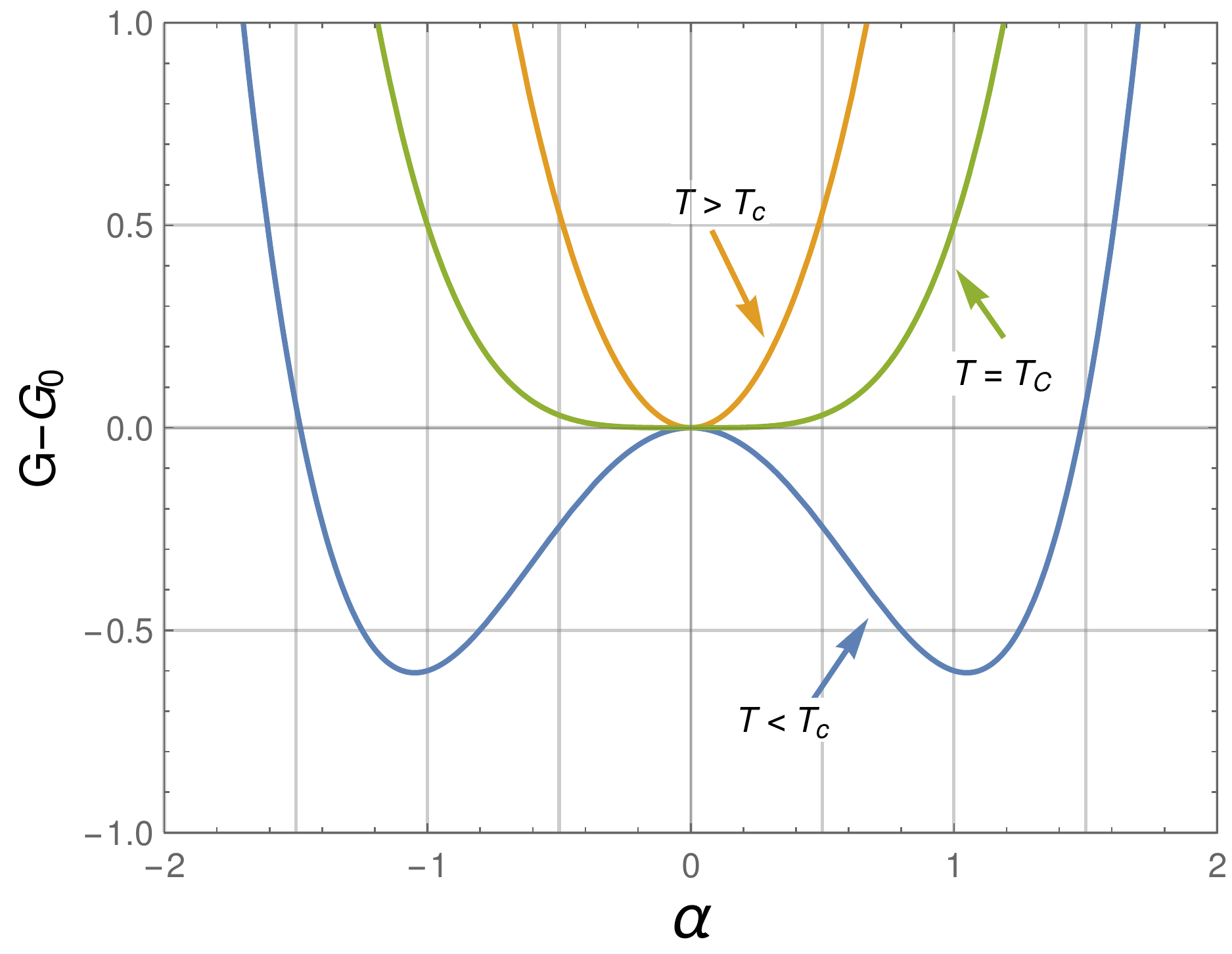}
 \caption{Plot of the Gibbs potential versus the order parameter $\alpha$. For $T\geq T_c$, the minimum of the potential is given by $\alpha=0$. As temperature drops below $T_c$, the doubly degenerate symmetric minima appear for $\alpha=\pm \sqrt{-a/b}$, minima follows the reflection symmetry ($\alpha\to -\alpha) $ of the Ginzburg$-$Landau expression (\ref{GL}) written, for simplicity, in one dimension. } 
\label{GLf}
\end{figure}

\begin{figure}[ht]
\centering
\includegraphics[scale=.62]{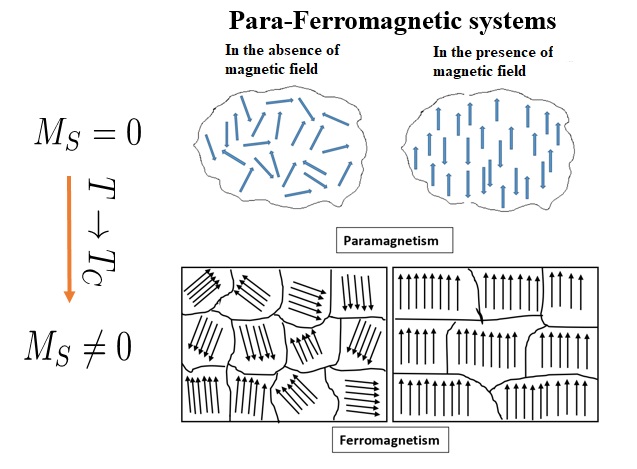}
\caption{A cartoon diagram which shows phase transition from paramagnetic state to ferromagnetic state as the temperature drops below the critical temperature.  After phase transition, the system acquires a domain structure such that the magnetization is non-zero in absence of an external magnetic field dubbed spontaneous magnetization $M_s$ which obviously vanishes in case of paramagnetic state. Ferromagnetic state in three dimensions adheres to rotational symmetry which is lost when a particular direction of magnetization is chosen. In this case, the order parameter is spontaneous magnetization with three components. } 
\label{ferro}
\end{figure}

For phenomenological reasons, we will assign temperatures larger than the critical value, $T\geq T_c$, to the phase of the system with vanishing order parameter. The order parameter picks up the nonzero values below the critical point $T<T_c$. The two phases of the thermodynamic system are clearly distinguished by the behaviour of the order parameter across the critical point or the phase transition.
The situation is well summarized in Fig.\ref{GL}. For a familiar, namely, the magnetic system, this implies that as the system cools below $T_c$, the true ground state is the one with non-vanishing order parameter$-$ spontaneous magnetization, $M_S\neq 0$ signaling phase transition from para ($M_S=0$) to ferromagnetic state ($M_S\neq 0$), see Fig.\ref{ferro}. For simplicity, we considered a one dimensional system. Thus the direction of $M_S$ "Up" or "Down" given by the sign of the order parameter ($M_S=\pm \sqrt{-a/b}$). In order to investigate the system further, we need to choose one of the ground states\footnote{For instance, as a next step, we should understand the effect of the external field on the ferromagnetic properties of the system. }; the moment we do so, the "Up"$-$ "Down" symmetry ($\alpha\to -\alpha$) of (\ref{GLf}) is lost$-$ which corresponds to {\it spontaneous symmetry breaking.} Ferromagnetism is one of the first examples of symmetry breaking.

 To understand the behaviour of the order parameter in the neighbourhood of the critical point ($T<T_c$), one expands $a(T,P)$ series retaining the linear term,
 \begin{equation}
 \label{acrit}
 a(T,P)=\left(\frac{\partial a}{\partial T}  \right)_{c}(T-T_c)  \to \alpha=\sqrt{\frac{\left(\frac{\partial a}{\partial T}  \right)_c}{b_c} }\left(T_c-T\right)^{1/2}\sim \left(T_c-T\right)^{1/2}
 \end{equation}
 The nature of the phase transition is defined by the behaviour of the first and second derivatives of $G$ (physical quantities of interest) at the critical point. To this effect, one defines the jump of a physical quantity across the critical point,
 \begin{equation}
     [X]=\left(X_{\alpha \neq 0}-X_{\alpha= 0}\right)|_{T=T_c}
 \end{equation}
 As mentioned before $G(P,T;\alpha)$ is continuous at the critical point,
 \begin{equation}
  G(P,T;\alpha)=G_0(P,T)-\frac {a^2(T,P)}{4b(T,P)}
 \end{equation}
 The first derivative of $G$ with respect to T gives,
\begin{equation}
S_\alpha-S_0 =\frac{a(2a'b-ab')}{4b^2}= \frac{\alpha(2a'b-ab')}{4b}\to [S]=0~~(\alpha(T=T_c)=0) 
\end{equation}
 where prime denotes derivative with respect to temperature and $S=-\partial G/\partial T$. One can easily verify that the first derivative with respect to $P$ is also continuous across the critical point.
 Similarly one computes the jump of the second derivatives, for instance,
 \begin{equation}
[C_P]=\left[ T\frac{\partial}{\partial T}(S_\alpha-S_0) \right]|_{T=T_c}    =\frac{(a'^2 T)_{c}}{2b_c}\neq 0
 \end{equation}
where we used the fact that, $a'(T=T_c)\neq 0$ and $a(T=T_c)=0$ or $\alpha(T_c)=0$ (see Eqs.(\ref{acrit1}) and (\ref{acrit}))\footnote{Same conclusion applies to the mixed derivatives (isothermal/adiabatic comprehensibility)}. The aforesaid defines the type of phase transition$-$ {\it Phase transition is termed as of first/second order if the first/second derivatives of the thermodynamic potential are discontinuous at the critical point. }Thus, the Ginzburg-Landau theory based upon expression (\ref{GL}) describes the phase transitions of second order.
\subsection{First order phase transitions} 
\label{1order}
\begin{figure}[ht]
\centering
\includegraphics[scale=.5]{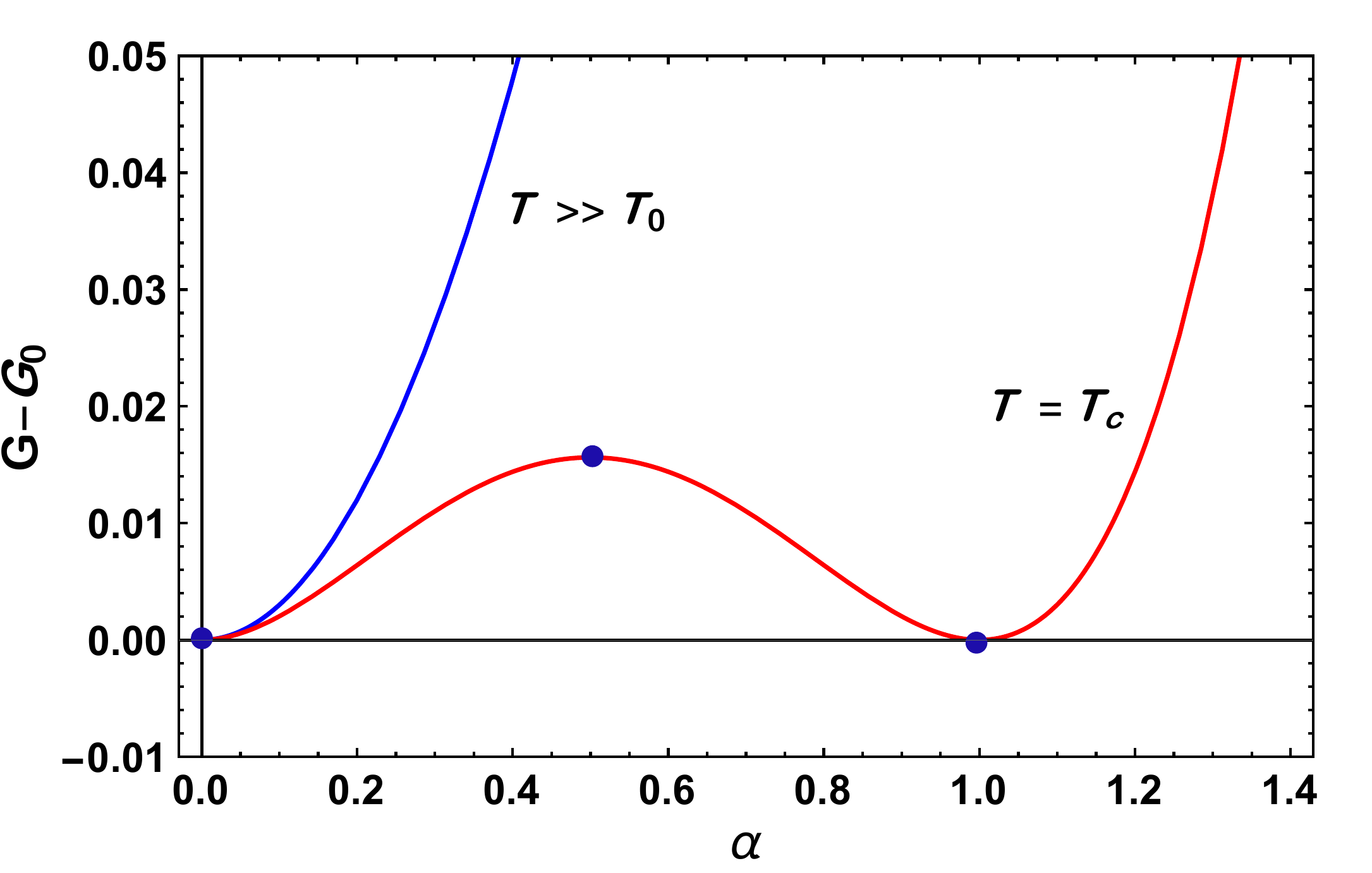}
\caption{Figure displays the qualitative behaviour of the Gibbs potential (\ref{GL1O}) versus $\alpha$ for $T>>T_0, (T: a(T_0)=0$; $a(T)$ being a monotonously increasing function of $T$). The critical curve for $T=T_c$ corresponds to the expression (\ref{CC}) 
which assumes specific values of the coefficient $b$ along the with normalization, $\alpha(T_c)=1$. In the first case, the unique minimum is located at $\alpha=0$. In the second case, the Gibbs potential has two minima, one at $\alpha=0$ and the other at $\alpha=\alpha(T_c)=1$ with a maximum in between at $\alpha=\alpha(T_c)/2=1/2$. } 
\label{1}
\end{figure}

In the Ginzburg-Landau framework, the first-order phase transition can be captured by adding the next higher order term in $\alpha$ $(\mathcal{O}({\alpha^6})$) to the Gibbs potential (\ref{GL}).  If we do not adhere to reflection symmetry, adding a third order term in $\alpha$ (with a suitable coefficient) to expression (\ref{GL}) would suffice for the first order phase transition.
\begin{equation}
\label{GL1O}
 G(P,T;\alpha)-G_0(P,T)=\frac{1}{2}a(P,T) \alpha^2 -\frac{1}{3}c(P,T)\alpha^3+  \frac{1}{4}b(P,T) \alpha^4
\end{equation}
where $b,c>0$ by assumption (c(T)=0 corresponds to second order phase transition). {We do not consider a linear term in the order parameter, $\alpha$, because any such term can be eliminated by a shift, $\alpha\rightarrow\alpha+$constant}. For phenomenological reasons, we also assume $a(T)$ to be a monotonously increasing function of temperature. The temperature $T_0$, at which $a(T)$ vanishes, plays an important role.
Let us sketch  the qualitative behaviour  of the thermodynamic system described by (\ref{GL1O}). To this effect, we imagine the thermodynamic system in hot background. As the system gradually cools, it undergoes
various phase transformations associated with the rearrangement of its ground state. For instance transition from vapour to liquid and liquid to solid serve examples of phase transitions, other example includes the transition  from para to ferromagnetic state below Curie temperature. 
In case, $T\gg T_0$, we have a unique ground state corresponding to $\alpha=0$ (Fig.\ref{1}). Indeed, in this case, the influence of the cubic term in (\ref{GL1O}) is undermined by the quadratic term. Something generically different happens when  temperature approaches a particular value when the ground state becomes degenerate such that we have two minima, one at $\alpha=0$ as before and the other at $\alpha\neq0$; the  two minima are separated by a maximum, see Fig.\ref{1}. At this temperature, there is a perfect balance between the cubic and the quadratic terms. Interestingly, the order parameter vanishes for $T$ larger than this value whereas it assumes non-zero values at it.
Thus the order parameter has a jump at this point which qualifies for {\it critical point} and corresponding temperature is {\it critical temperature}, $T_c$. We immediately see here the difference from the second order phase transition where $\alpha(T_c)=0$,
\begin{equation}
     \alpha(T) =
    \begin{cases}
    0  ,  \quad \quad T<T_c\ , \\
    \frac{2c}{3b} \neq 0 \quad    \quad  T=T_c~~~(c>0, b>0) 
    \end{cases}
 \end{equation}   
\begin{figure}[ht]
\centering
\includegraphics[scale=.55]{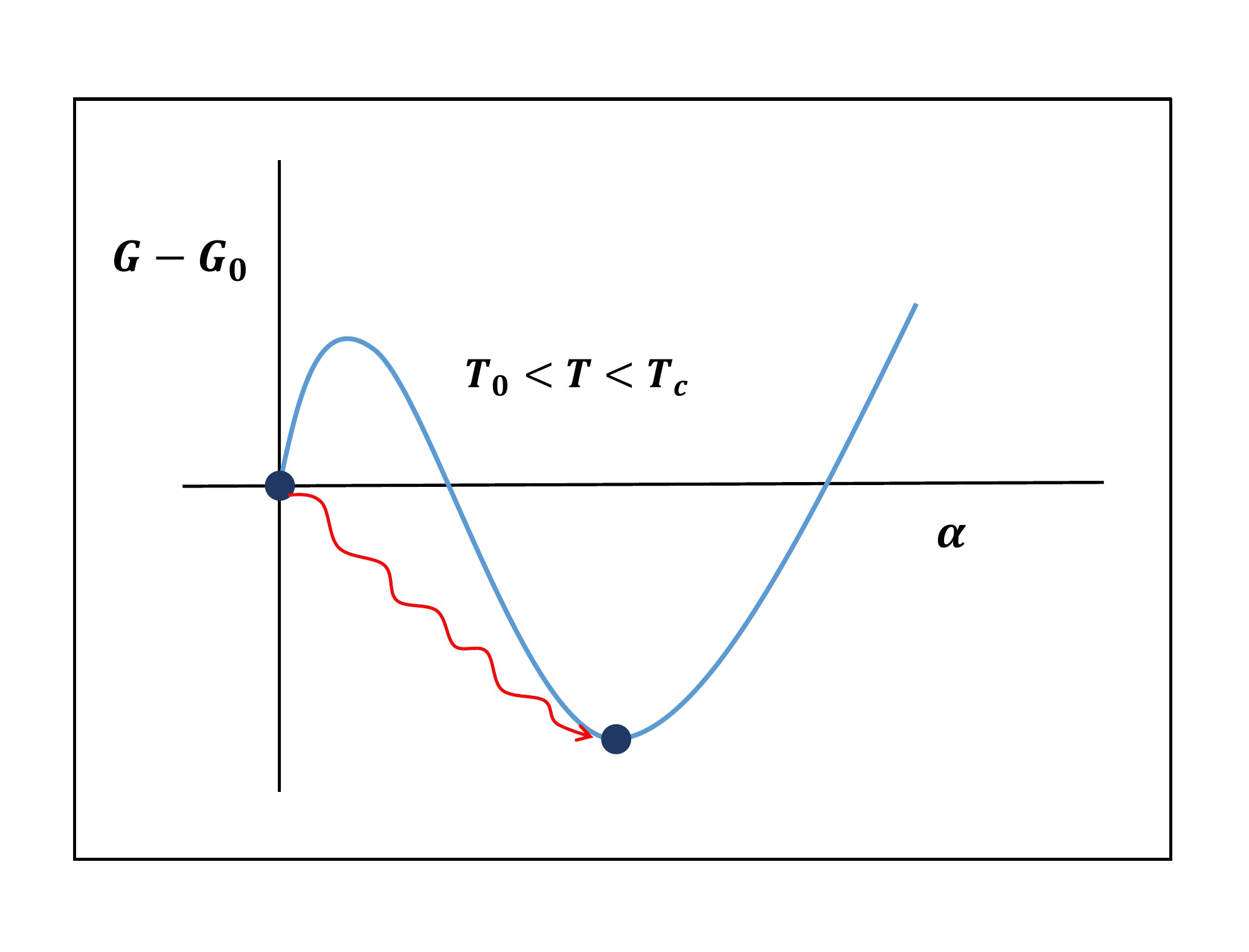}
\caption{Schematic plot of the Gibbs potential versus the order parameter $\alpha$ for a very special case when $T_0<T<T_c$. In this case, we have a global minimum or true ground state at $\alpha \neq 0$ and a meta-stable state at $\alpha=0$ separated from the global minimum by a maximum in between. Classically, no transition is possible between them at zero temperature. Quantum tunnelling from the meta-stable state to the ground state is depicted by a zigzag line. However, at finite temperature, thermal fluctuations could go over the barrier giving rise to major contribution to this process$-$ bubble nucleation. }
\label{2}
\end{figure}

Let us derive a relation  between coefficients $a,b,c$ in the expression of the Gibbs potential (\ref{GL1O}) at the critical point and express the order parameter through them. These general relations would allow us to understand the broad features of dynamic of a first order phase transition described by (\ref{GL1O}).

Let us note the the minimum of the Gibbs potential at $\alpha\neq 0$ in Fig.\ref{2} is such that the Gibbs potential also vanishes there. The corresponding temperature is the {\it critical temperature}, $T_c$ by definition. And using (\ref{GL1O}), this implies that,

\begin{eqnarray}
\label{minimum}
&& a(T_c)\alpha-c(T_c)\alpha^2+b(T_c)\alpha^3=0 \label{GV0}
\\
&& \frac{1}{2}a(T_c) \alpha^2-\frac{1}{3}c(T_c) \alpha^3+\frac{1}{4}b(T_c)\alpha^4=0
\label{GV}
\end{eqnarray}
   giving rising to the following   solution at the critical point ,
   \begin{equation}
   \label{solutioncrit}
 \alpha(T_c)=\frac{2c}{3b};~~a(T_c)=\frac{2c^2}{9b}     
   \end{equation}
These relations at the critical point would play an important role in our discussion on electroweak phase transition in the early Universe.
Let us note that at the critical point, all the coefficients of the  Gibbs potential (\ref{GL1O}), expressed
in terms of $\alpha/\alpha(T_c)$,
are proportional to $b \alpha^4(T_c)$. 
For no loss of generality,
normalizing  the order parameter to one at the critical point ($\alpha(T_c)=1$) and taking $b=1$\footnote{Critical behaviour of the Gibbs potential does not depend upon $b, \alpha(T_c)$; they simply re-scale $G-G_0$ $\&$ $\alpha$: $\alpha\to \alpha/\alpha(T_c)$ $\&$ $G-G_0 \to (G-G_0)/b\alpha^4(T_c)$ }, we have,
\begin{equation}
\label{CC}
(G-G_0)|_{T=T_c}= \frac{1}{4}\alpha^2-\frac{1}{2}\alpha^3+\frac{1}{4}\alpha^4   
\end{equation}
which obviously satisfies Eqs.(\ref{minimum}) $\&$ (\ref{GV});  the critical curve is drawn in Fig.\ref{1}.

{In our discussion, we assume $b>0$ necessary for thermodynamics stability, that is, we do not want an infinite parameter to minimize the thermodynamic potential. We also assume $c>0$ which is not essential. In fact if $c<0$, we can always consider $\alpha\rightarrow -\alpha$. Assuming $c>0$ implies that when $T<T_c$, we have the global minimum for positive order parameter. Apart from the critical temperature, there is another important temperature, namely, $T_0$ where $a(T)$ changes sign such that $a(T)>0$ for  $T>T_0$. Let us confirm that $T_0<T_c$. Indeed, for $a(T)\leq 0$, Eqs.(\ref{minimum},\ref{GV})
have no solution, see Eq.(\ref{solutioncrit}).
 The temperature range  $T_0\leq T \leq T_c$ is of special interest.  Indeed, for $T>T_0$, we have local maximum between the minima  at $\alpha=0$  and $\alpha(T_c)=2c/3b$ (see Fig.\ref{1}) which, however, disappears when  $T=T_0$,
\begin{equation}
\alpha_-(T_0)=\frac{c-\sqrt{c^2-4ab} }{2b}=0 \end{equation}
 At temperatures larger than $T_0$, this local minimum reappears as a local maximum for positive order parameter. In this case, the global minimum remains located at $\alpha_+$(see Fig\ref{2})\footnote{$\alpha_\pm(T_c)$ denote the two non-trivial  roots of Eq.(\ref{minimum}); $\alpha_-(T_c)$ $\&$ $\alpha_+(T_c)$ refer to the smaller and the larger roots respectively.},
\begin{equation}
\alpha_+(T)=\frac{c+\sqrt{c^2-4ab} }{2b} \end{equation}
until the temperature reaches the transition temperature or critical temperature, defined by Eqs.(\ref{GV0},\ref{GV})} and we have the situation shown in Fig.\ref{1}.

Let us note that the phase transition is of second order if $c(T_{c})=0$; on the other hand if $a(T)=0$, the transition from metastable state to the ground state is simply a "smooth crossover". 
{From our analysis, we conclude that $G=G_0$ for $T>T_c$ and  $G(\alpha_+(T))<G_0$ for $T<T_c$. However, the Gibbs function is continuous at $T=T_c$ by definition because we defined $G(\alpha_+(T_c))-G_0=0$ but its first derivative is discontinuous if $c(T_c)\neq 0$. In fact, we find at $T=T_c$ as $T$ approaches $T_c$ from below
\begin{align}
\label{entropy}
    S=S_0-\frac{2c^2}{81b^4}\left(9b^2a'+2c^2b'-4bcc'\right)\Rightarrow [S]_{T_c}=-\frac{\alpha^2}{18b^2}\left(9b^2a'+2c^2b'-4bcc'\right)
\end{align}
where all quantities are evaluated at $T=T_c$. And as $T$ approaches $T_c$ from above, we have $S=S_0$. We see that if $c(T_c)=0$, we would have a second order phase transition while if $c(T_c)\neq 0$ we have a first order transition (assuming that $9b^2a'+2c^2b'-4bcc'\neq 0$
} 
{which is true in general).}
Finally, we notice that $c(T)$ has the dimension of temperature and its functional form depends upon a particular phase transition. For instance, $c(T)\propto T$ in case of electroweak phase transition. In a sense, $c(T_c)/T_c$ defines the strength of a first order phase transition. However, it is better if we use a physical quantity in this context. Indeed,
the order parameter $\alpha(T_c)$ is  a physical quantity that has finite jump across the phase transition. And the amount of entropy produced  during the transition$-$ a measure of non-equilibrium, is proportional to $\alpha^2(T_c)$, see Eq.(\ref{entropy}). Hence
{\it $\alpha(T_c)$ should be traded as a measure of non-equilibrium.}

\begin{figure}[ht]
\centering
\includegraphics[scale=.5]{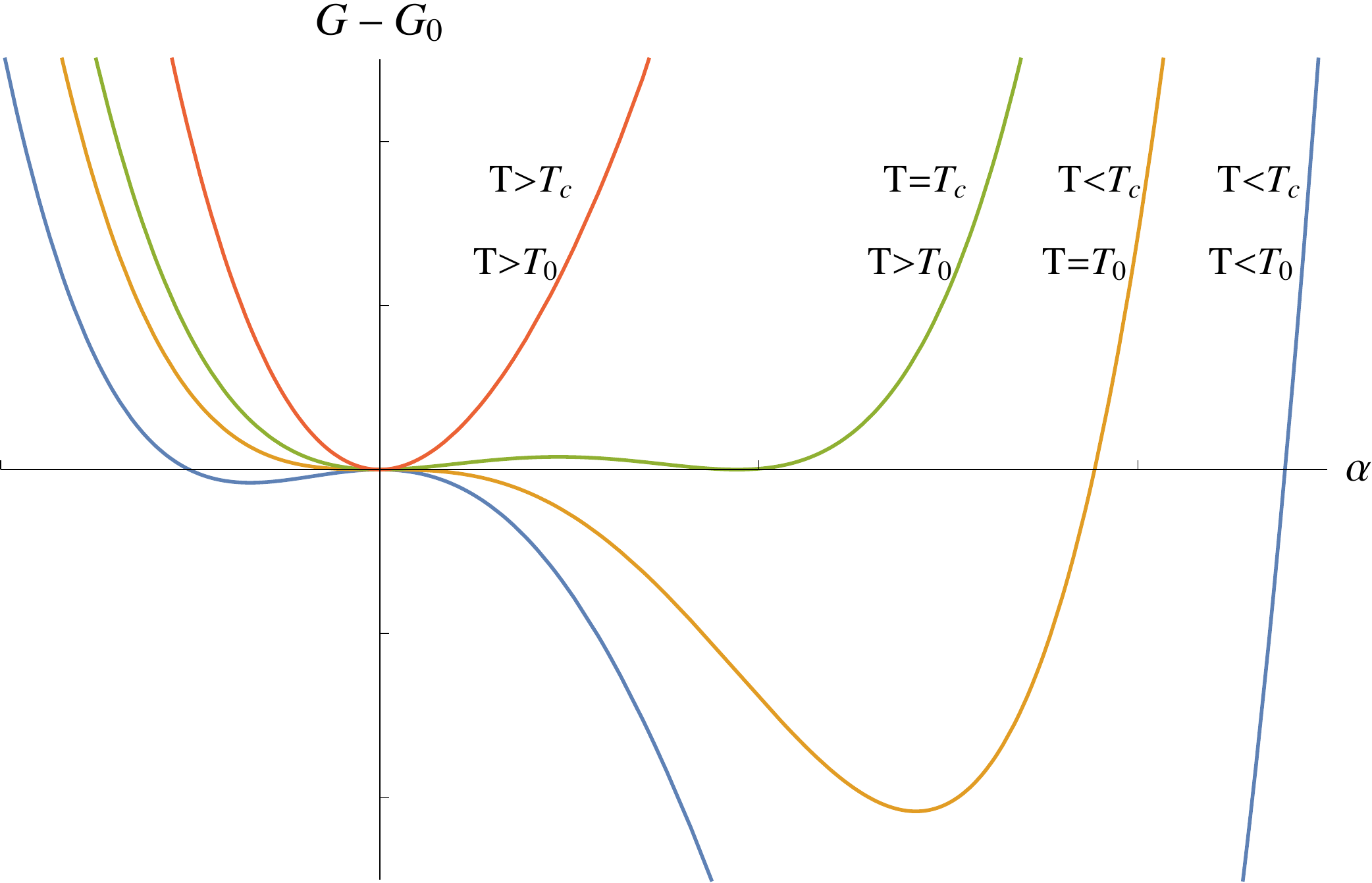}
\caption{ Schematic plot of Gibbs potential (\ref{GL1O}) versus the order parameter $\alpha$. This figure describes the qualitative behaviour of a thermodynamic system described by (\ref{GL1O}) in detail using all different possibilities with regard to  $T_0$ and the critical temperature $T_c$. While drawing the figure, we adapted the numerical values of coefficients in (\ref{GL1O}) from the standard model.}
\label{firstorder}
\end{figure}

We have described the qualitative behaviour  of a thermodynamic system described by (\ref{GL1O}), without fixing the functional form of $a,b$ and $c$, which is summarised in Fig.\ref{firstorder}. The temperature range: $T_0<T<T_c$
is of special interest,   see Fig.\ref{firstorder} or  Fig.\ref{2},  redrawn to highlight the effect.
In this case, the global minimum or the {\it true ground state }
at $\alpha \neq 0$  is separated, from the false vacuum or meta-stable state at $\alpha=0$, by a finite barrier in between. Thereby, if the system initially resides in the false vacuum, it would make a transition to the true ground state with $\alpha \neq 0$ through bubble nucleation, but as we will see, the probability of quantum tunnelling is small. And this characterizes  the {\it first order phase transition}.
If $T=T_0$ ($ a(T_0)=0)$, the bump between the metastable state and the global minimum disappears and we have a {\it smooth crossover}. Hence, barrier is important to realize the first order phase transition. 
The first order phase transitions  proceed through bubble formation of the new phase in the middle of the old. 
Meanwhile the bubbles  expand,  collide and merge and this keeps happening until the old phase disappears completely giving rise to boiling caused by  the latent heat released from bubbles. This  is a  non-equilibrium process in which large entropy is generated depending upon the size of $\alpha(T_c)$. The latter is the distinguished feature of the first order phase transition that makes it wanted in the early Universe.
The process of bubble nucleation is central to first order phase transitions  which we shall discuss in  \ref{bubblenucleation} in detail. 
 \section{Degeneracy of ground state and quantum mechanical consideration}
 \label{SQM}
\begin{figure}[ht]
\centering
\includegraphics[scale=.55]{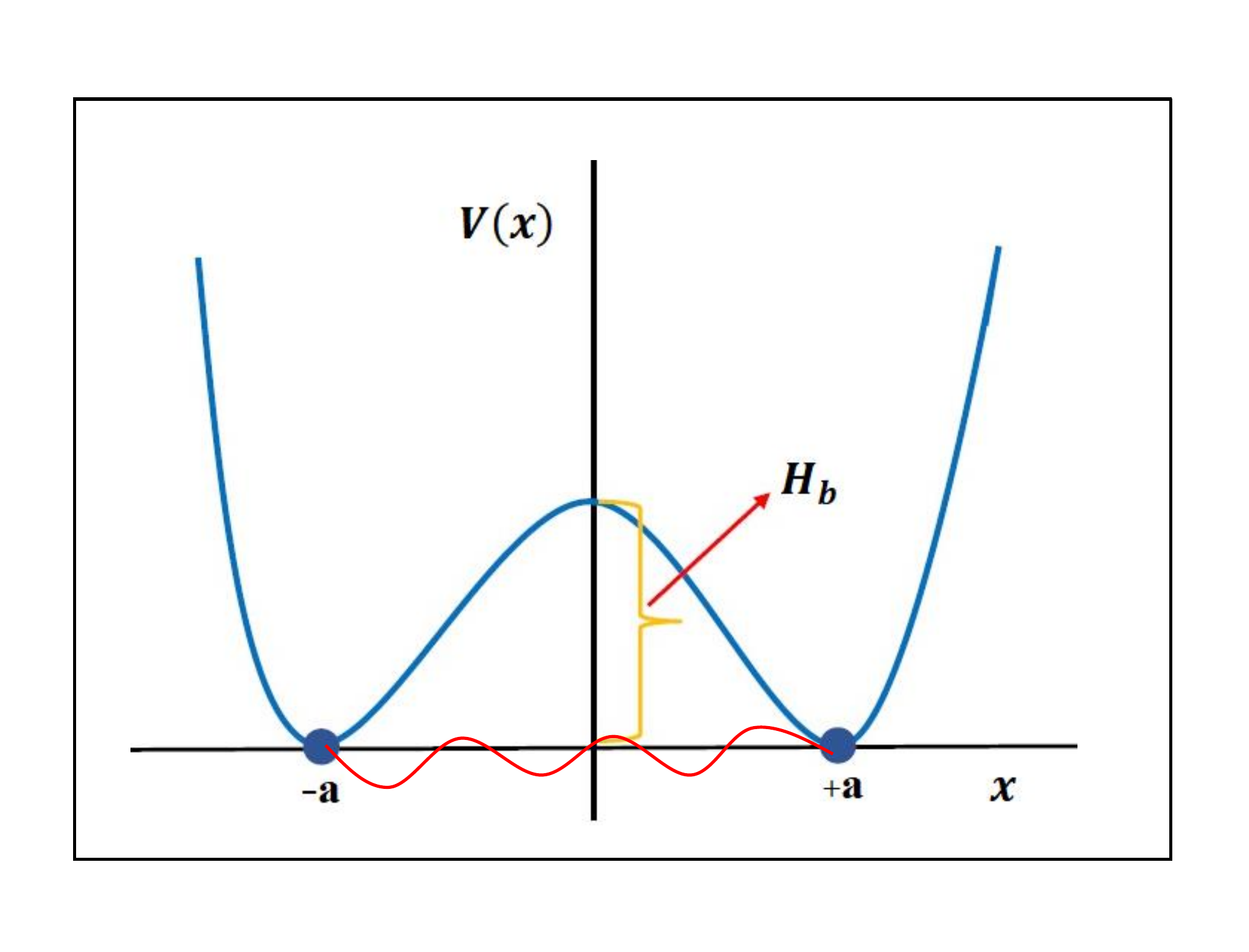}
\caption{ Plot of the double well potential $V(x)$, given by (\ref{potQM}),   versus $x$ for convenient values of constants, $H_b=(\omega a)^2/8$ is the height of the potential barrier between the two vacua at $x=\pm a$. The tunnelling probability can be estimated with help of instanton solutions which connect the vacua. }
\label{QM}
\end{figure}

 As demonstrated in the preceding section, ground state of the thermodynamic system becomes degenerate for $T<T_c$. Question arises, would the quantum tunnelling between degenerate ground states (vacua) lift the degeneracy? To address this question, let us  consider a particle moving in  one dimension in a double well potential,
 \begin{equation}
 V(x)=\frac{\omega^2}{8 a^2} (x-a)^2(x+a)^2  ; H_b=\frac{(\omega a)^2} {8}\,
 \end{equation}
where $H_b$ denotes the height of the potential barrier between minima, see Fig.\ref{QM}. In the limit $a\to \infty$, the problem decomposes into the sum of two independent harmonic oscillators\footnote{In this case, minima are separated by infinite distance with an infinite barrier between them.}. For a large value of $a$, 
two lowest energy wave functions are approximately given by the symmetric and antisymmetric combinations of the harmonic-oscillator wave functions,
\begin{equation}
 \psi_{s,a}\simeq\frac{1}{\sqrt{2} }\left( \psi_0(x-a)\pm  \psi_0(x+a)\right)
 \label{potQM}
\end{equation}
 Due to quantum tunnelling, there is  energy split between the symmetric and antisymmetric states,
 \begin{eqnarray}
 && E_{s,a}=\frac{1}{2}  \hbar \omega \mp \frac{1}{2}\Delta E\\
 && \Delta E=A \hbar \omega e^{- 16H_b/3\hbar\omega}
 \label{QT}
 \end{eqnarray}
 where $A$ is a constant whose exact form is not important for the present discussion. The exponential term in (\ref{QT}) is the tunnelling   probability $P_t\sim e^{- 16H_b/3\hbar\omega}$ between largely separated vacua\footnote{Tunnelling probability can be computed using {\it instantons}-classical solutions that exist for imaginary time (Euclidean space).}. In the case of two/three dimensions, $H_b$ would be replaced by the surface area/volume of the barrier. For an infinite system, one can argue on heuristic grounds that $P_t$ would vanish exponentially. {Indeed, if one is dealing with a field which belongs to  infinite dimensional space, the volume of the barrier is infinite}. We therefore conclude that in  quantum mechanics, the degeneracy is lifted by quantum tunnelling. However, if there is a classical degeneracy in field theory, it can not be lifted by quantum tunnelling and gives rise to spontaneous symmetry breaking\footnote{In scalar field theory with multiple vacua, there exists no instanton which supports the heuristic argument that spontaneous symmetry breaking is generic to infinite systems.}.  The thermodynamic system deals with a large number of particles and practically mimics an infinite system and this answers the question we had posed.

 \section{Spontaneous symmetry breaking in field theory}
 \label{SSS}
 The mechanism of spontaneous symmetry breaking  is at the heart of electroweak unification, see \cite{Lee} and references therein (also \cite{Migdal}, a relatively unknown reference). It was clear from Fermi theory of weak interactions that the mediation in this case should be provided by massive vector bosons and Fermi theory of contact interactions was inconsistent in the high energy regime. However, putting mass by hand destroys the renormalizability of the theory. The situation was rescued by the mechanism of spontaneous symmetry breaking, which  consistently generates the masses of the vector bosons. In what follows, we briefly describe the phenomenon of spontaneous symmetry breaking in field theory.
\subsection{Spontaneous breaking of discrete symmetry}
Let us consider the scalar field Lagrangian,
\begin{equation}
\label{sfl}
  \mathcal{L}=-\frac{1}{2}  \partial_\mu \phi \partial^\mu \phi-V(\phi);~~V(\phi)=-\frac{1}{2}\mu^2\phi^2 +\frac{1}{4} \lambda \phi^4
\end{equation}
 where  the mass term has the wrong sign for $\mu^2>0$. The chosen Lagrangian has reflection symmetry, $\phi\to -\phi$ dubbed $Z_2$ symmetry. The ground state of the system is obtained by minimizing the potential $V(\phi)$. If $\mu^2<0$, the minimum of the potential is given by,
 \begin{equation}
 \phi_0  \equiv \langle \phi\rangle_0 \equiv v=0   ,
\end{equation}
 where $v$ designates the vacuum expectation value of the field. However, we shall be interested
 in the Lagrangian with the wrong mass sign, $\mu^2>0$, when $\phi_0=0$ is no longer the minimum of the potential. Indeed, in this case, tachyonic instability builds up in the system leading it to the true ground state,
 \begin{equation}
 \langle \phi\rangle_0 =\pm v=\pm \sqrt{\frac{\mu^2}{\lambda}}   ,
 \end{equation}
which is doubly degenerate.  In order to develop the perturbation theory, one needs to fix a vacuum state to compute small fluctuations around it.  Thanks to the symmetry of the Lagrangian, we can choose any of the two vacua. However, after we make the choice, the underlying symmetry might be lost$-$ {\it vacuum state breaks the symmetry of the Lagrangian}. Let us choose the ground state, $\phi_0=v$, and compute the mass of the field around it,
 \begin{equation}
m^2=\frac{d^2 \phi}{d\phi^2}|_{\phi=v}=2\mu^2 \Rightarrow~m=\sqrt{2}\mu     \end{equation}
However, the information about our choice of ground state would be reflected in the Lagrangian
 if we rewrite it  through $\sigma$ field (dubbed {\it field shifting}),
 \begin{equation}
  \phi(x)=v+\sigma(x)  \Rightarrow \langle \sigma\rangle_0=0
 \end{equation}
which is equivalent to shifting the minimum at $\sigma=0$ ($\langle \sigma\rangle_0=0$) around which we can study small fluctuations. Lagrangian (\ref{sfl}) expressed in terms of $\sigma$ field acquires the form,
 \begin{equation}
 \label{sigma}
 \mathcal{L}=-\frac{1}{2}  \partial_\mu \sigma \partial^\mu \sigma  -\frac{1}{2}(\sqrt{2}\mu)^2 \sigma^2-\sqrt{\lambda}\mu \sigma^3-\frac{\lambda}{4}\sigma^4+\frac{\mu^4}{4\lambda},
 \end{equation}
 which readily tells us that the Lagrangian in $\sigma$ has the right sign of mass term with mass of the  field given by, $m_\sigma=\sqrt{2} \mu$, which was implicit before the field shifting. One should notice that the mass content of the model is explicit if the Lagrangian is expressed
 through a field with zero vacuum expectation value.
 However, the  Lagrangian (\ref{sigma}) also contains a new cubic interaction which breaks the original symmetry of the Lagrangian. Thus, while choosing one of the ground states and living there, one can not see the original symmetry of the system {\it \`{a} la} {\it secret symmetry} according to {Sidney Coleman} \cite{sidney}. Readers not interested in the early Universe phase transitions may skip the remaining subsections of this section and directly jump to section \ref{sls}.
 \begin{figure}[ht]
\centering
\includegraphics[scale=1.3]{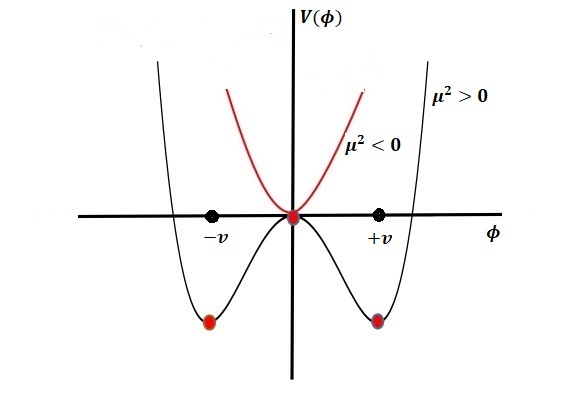}
\caption{Scalar field potential $V(\phi)$ corresponding to  the Lagrangian (\ref{sfl}). In case, $\mu^2<0$, potential has a unique minimum at $\phi=\phi_0=0$. For $\mu^2>0$, due to wrong mass sign, the field potential has two symmetric minima for $\phi=\phi_0= \pm v$. } 
\label{field}
\end{figure}
 \subsection{Breaking of continuous $U(1)$  global symmetry}
 \label{U1global}
 Let us consider a complex scalar field\footnote{We keep a factor $1/2$ in the Lagrangian even if it is usually removed for the complex field if we want to treat $(\phi,\phi^*)$ as independent fields.} with wrong mass sign,
 \begin{eqnarray}
 \label{comphi}
 &&\mathcal{L}= -\frac{1}{2}  \partial_\mu \phi \partial^\mu \phi^* -V(\phi)\\ &&V(\phi)=-\frac{\mu^2}{2}\phi^*\phi+\frac{\lambda}{4}(\phi \phi^*)^2
 \label{Vcomp}
 \end{eqnarray}
 which is invariant under the transformation,
 \begin{equation}
 \label{Ug}
 \phi\to \phi'=e^{i \beta} \phi
 \end{equation}
 dubbed $U(1)$ global with $\beta$ independent of space time coordinates.
 In this case, the minimum of (\ref{Vcomp}) is given by,
 \begin{equation}
 |\phi_0|=v;~~v^2=\frac{\mu^2}{\lambda}    
 \end{equation}
 and vacuum manifold\footnote{Ground state is infinitely degenerate in this case.} is given by a circle of radius
 $v$,
 \begin{equation}
 \label{circle}
\phi_0 \equiv \langle\phi\rangle_0=v e^{i\beta}  \end{equation}
 \begin{figure}[ht]
\centering
\includegraphics[scale=.4]{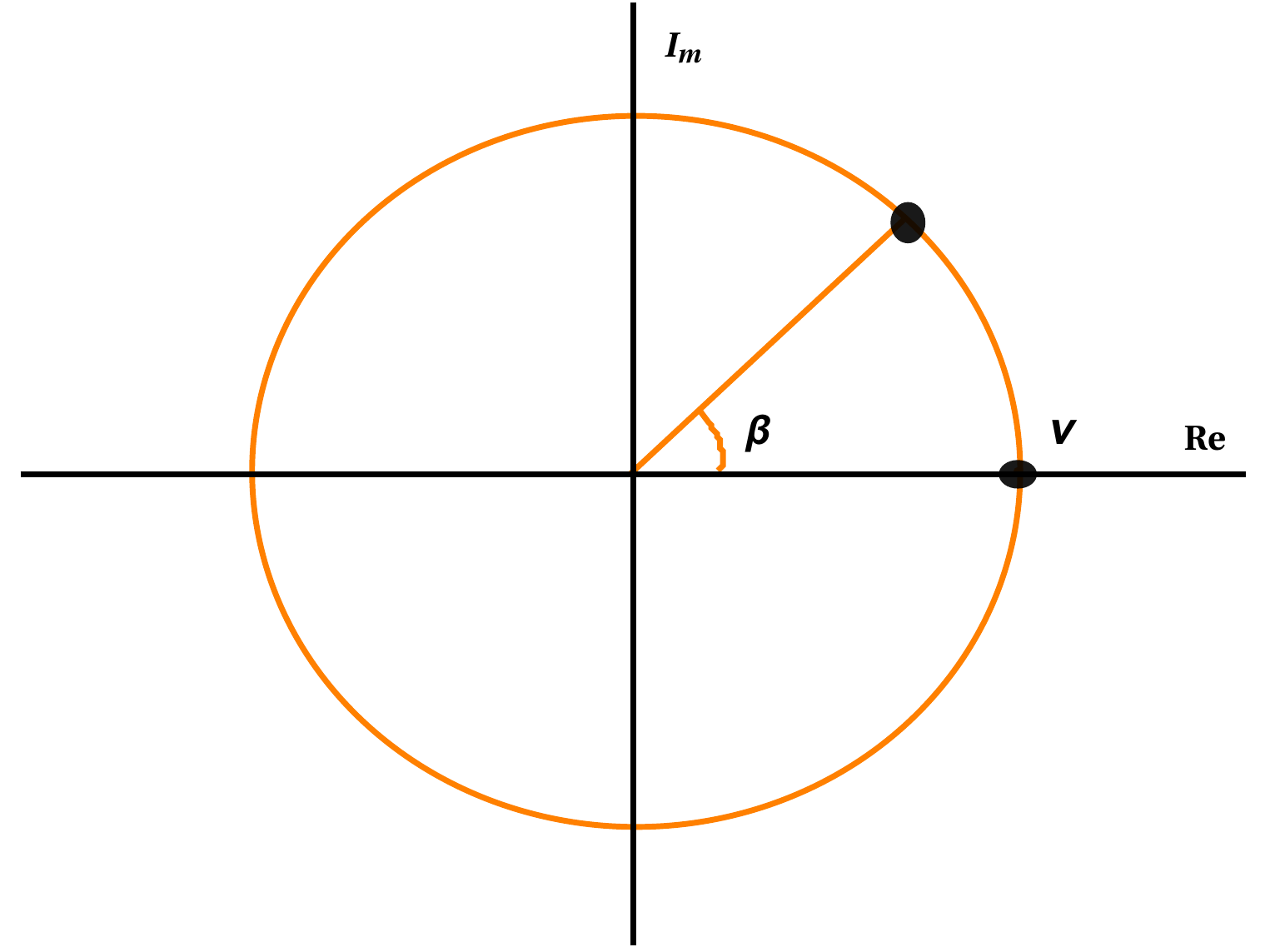}
\caption{Structure of vacuum manifold  (\ref{circle}) associated with the Lagrangian, (\ref{comphi}). All the points on the circle represent legitimate ground states. Choice of a particular vacuum state  implies to fixing $\beta$, the simplest choice is given by $\beta=0$. } 
\label{vacuum}
\end{figure}
In accordance with $U(1)$ symmetry (\ref{Ug}), all the points on this circle are legitimate vacua states which correspond to different values of $\beta$. Choosing a particular vacuum state on this manifold reduces to fixing $\beta$. Let us choose, for convenience $\beta=0$, see Fig.\ref{vacuum}. It would be convenient to write the complex field in Euler form,
 \begin{equation}
 \label{GS}
 \phi=\rho(x) e^{i\theta(x)} \Rightarrow \text{Ground state}:~\langle\rho\rangle_0=v,~~\langle\theta\rangle_0=0   
 \end{equation}
 It may also be instructive to write down the Lagrangian in $(\rho, \theta)$ form,
 \begin{equation}
 \label{comE}
 \mathcal{L}=
 -\frac{1}{2}  \partial_\mu \rho \partial^\mu \rho -\frac{1}{2}\rho^2    \partial_\mu \theta \partial^\mu \theta-V(\rho),    
 \end{equation}
 where the second term represents interaction of $\rho$
and $\theta$.
In the Euler representation, (\ref{Ug}) is equivalent to the following transformation,
\begin{equation}
\theta(x) \to \theta(x)+\beta ,
\end{equation}
under which (\ref{comE}) is invariant. Let us note that the  potential in (\ref{comE}) does not include the $\theta$ field.
We can easily compute the masses of fields in the ground state (\ref{GS}) using (\ref{comE}),
\begin{equation}
  m^2_\rho=\frac {\partial^2 V(\rho)}{\partial \rho^2 }|_{\rho=v}  =2\mu^2;~~
 m^2_{\theta}|_{\rho=v}=0   ,
\end{equation}
which is, however, not explicit\footnote{Specially, the mass of $\rho$ field.}, in the Lagrangian (\ref{comE}). Not surprising, we have not yet convened the Lagrangian that we have selected (\ref{GS}) as our ground state.  Lagrangian would know about our choice  only after we shift the field. In this case, we need to shift only $\rho$ field,
 \begin{equation}
 \phi=(\rho+v)e^{i\theta(x)}
 \end{equation}
 and plugging it in the Lagrangian({\ref{comphi}}).
 we have\footnote{It should be noted that $\theta$ is dimensionless. One could define a dimension-full field, $\varphi=v\theta$ and call the Goldstone  mode such that the kinetic term associated with $\theta$ has standard coefficient when expressed in terms of $\varphi$ .}, 
 \begin{eqnarray}
 \label{Lvrho}
\mathcal{L}&=& -\frac{1}{2}  \partial_\mu \rho \partial^\mu \rho -\frac{1}{2}(v+\rho)^2    \partial_\mu \theta \partial^\mu \theta-V(\rho),\\
 V(\rho)& =& \frac{1}{2}(\sqrt{2}\mu)^2\rho^2+\mu\sqrt{\lambda}\rho^3+\frac{\lambda}{4}\rho^4 -\frac{\mu^4}{4\lambda} 
 \label{vrho}
\end{eqnarray}
Before going ahead, let us clearly spell out the operational definition of "Spontaneous Symmetry Breaking".\\

{\bf Spontaneous  Symmetry Breaking}: { \it Thanks to the underlying symmetry, choosing a preferred ground state from the degenerate vacuum followed by the shifting of the field such that the vacuum expectation value of the shifted field vanishes.}
\\
\\
  We can easily read off the masses of fields from (\ref{Lvrho}) as, $m_\theta=0$ and  $m_\rho=\sqrt{2}\mu$. Thus after symmetry breaking, one of the fields $\rho$ has the standard mass  whereas the $\theta$ field  is massless dubbed Goldstone boson\footnote{It should be noted that the original symmetry of the Lagrangian (\ref{comE}), namely, $\theta(x)\to \theta(x)+\beta$ remains intact after "spontaneous symmetry breaking", see expression, (\ref{Lvrho}).  }.
 This is a general feature of models
 with spontaneous symmetry breaking (in case of global symmetry), namely, the number of Goldstone bosons is related to the number of generators of the underlying symmetry  group which violate the symmetry of the vacuum. For instance, let us consider a Lagrangian invariant under the global $SU(2)$ group (three generators) with a field as complex doublet (four real components). In this case after symmetry breaking, one of the field components acquires the standard mass as before, the remaining three field components are Goldstone bosons. Before proceeding farther, let us clearly specify the meaning of phrases: "Before symmetry breaking"
 and "After symmetry breaking", that we often pronounce\footnote{The chronology related to symmetry breaking here is superficial. It acquires realistic meaning in hot background where chronological ordering is set by the temperature. In that case, one is dealing with field theory at
 finite temperature. The present discussion is restricted to field theory at zero temperature. },
 \begin{eqnarray}
&&\textbf{"Before symmetry breaking" refers to:}~~~~ \mu^2<0\Rightarrow\langle \phi\rangle_0=0 \\
&&\textbf{ "After symmetry breaking"~ refers to:}~~~~ \mu^2>0\Rightarrow \langle \phi\rangle_0=v      
 \end{eqnarray}

 In what follows, we shall examine the spontaneous symmetry breaking  of $U(1)$ local symmetry  giving rise to a miraculous effect known as the Higgs mechanism which allows for consistent generation of gauge boson masses. This mechanism was also used for the generation of a tiny mass of graviton in massive gravity. Unfortunately, massive gravity theories are plagued with formidable problems irrespective of the way mass of the graviton is obtained. We briefly present   the historical account of massive gravity theories before discussing the Higgs mechanism.

 \subsection{Mass of graviton: From Pauli-Fierz theory to date}
 The first theory of massive gravity was proposed by Pauli and Fierz in 1939 \cite{Fierz:1939ix}. It is a linear theory where gravity is represented by a  spin-2 field with a cleverly constructed mass term for graviton such that the theory has 5 degrees of freedom consistent with the group-theoretic framework.  In this theory, the mass term is chosen in a specific way such that the sixth degree of freedom, a ghost, is laid to rest.  Since gravity is weak in the solar system, one might expect that the linear  theory would suffice locally.  However, van Dam, Veltman \cite{vanDam:1970vg} and independently Zakharov \cite{Zakharov:1970cc} discovered in 1970 that Pauli-Firz theory was plagued with a problem dubbed vDVZ discontinuity, namely, the predictions of the theory did not match with that of the General Theory of Relativity in the massless graviton limit.  
In 1971, Vainstein \cite{Vainshtein:1972sx} pointed out  that the assumption of linearity was not valid and a non-linear background should be invoked which would remove the vDVZ discontinuity. Actually, in the non-linear background, the discontinuity disappears but the  sixth degree of freedom becomes alive as the Boulware-Deser ghost \cite{Boulware:1973my} (see \cite{Hinter,samimg} for reviews). 

Let us note that the original motivation for massive gravity was to write down a consistent relativistic equation for spin-2 field  followed by Dirac equation. The contemporary revival of the idea is associated with an attempt to obtain late time acceleration  from a tiny mass of graviton of the order of  $H_0 \sim 10^{-33}eV$ which can be understood using a heuristic argument. Indeed, in case graviton has a mass $m_g$, the gravitational potential of a massive body with mass $M$ at a distance $r$ from the source is given by, $ \Phi(r)=(-GM/r)e^{-m_g r}$ which reduces to Newtonian expression for $r \ll m_g^{-1}$. In case, $r\sim m_g^{-1}$, i.e, at the horizon scale, gravity weakens which is equivalent to putting a positive cosmological constant, to compensate for attractive Newtonian contribution, in the standard lore of FLRW cosmology.  Thus, if a consistent theory of massive gravity is constructed, it could address the puzzle of the millennium$-$ the underlying cause for {\it late time acceleration} or {\it dark energy}.

One might naively think that  Pauli-Fierz theory with a tiny mass of graviton would reconcile locally with the General Theory of Relativity. However, the close scrutiny shows that this assertion is not correct.
It took many years to understand the underlying reason of vDVZ discontinuity and its resolution in the non-linear theory. If the  mass of graviton is non vanishing, irrespective of its smallness, the five degrees of freedom of graviton ($m_g\to 0$) can be visualized using helicity decomposition as : two massless tensor degrees (corresponding to massless graviton)\footnote{This can rigorously be accomplished at the level of Lagrangian.} , two transverse vector degrees  and a longitudinal (massless scalar) degree of freedom, see reviews \cite{Hinter,samimg} for details. In the decoupling limit, relevant to local physics, the vector degrees get decoupled from matter source leaving behind two massless tensor degrees and a massless scalar field  coupled to matter universally. The contribution of the longitudinal degree leads  to nearly doubling of the Newtonian potential giving rise to gross violation of local physics\footnote{The General Theory of Relativity is very accurate in the solar system with an accuracy of one part in $10^{-5}$.}. In the non-linear background, this contribution is taken care of by the ghost, thereby,  resolution of vDVZ introduces another serious problem to be tackled. Where does the ghost come from? There are three degrees of freedom in the pictures, two of them correspond to standard massless graviton and  a scalar degree of freedom. The question boils down to: how many degrees of freedom does a scalar field have? Depending on the structure of the kinetic term, the answer is as many as you like. Indeed, in case, $N$ higher derivative terms are added to the scalar field Lagrangian with a standard kinetic term such that the Lagrangian is non-degenerate, there will be $N$ additional ghosts  dubbed Ostrogradsky ghosts.  Thus in the non-linear background, we have a scalar field with  higher derivative term in addition to standard kinetic term in the decoupling limit. The presence of higher derivative term allows us to locally screen out the effect of longitudinal degree of freedom leaving the local physics intact$-${\it Vainshtein mechanism}\footnote{Modification of gravity due the extra degree of freedom (scalar field)
is locally screened out due to kinetic suppression leaving General Theory of Relativity intact in a large radius (dubbed Vainshtein radius) around a massive body.}. The latter certainly resolves the vDVZ discontinuity \cite{Hinter,samimg}.

Clearly, pushing the massive gravity ideology ahead, requires the treatment of Boulware-Deser ghost. 
The problem was addressed in 2010 by de Rham, Gabadadze and Tolley (dRGT) \cite{cl}  who designed the graviton mass term using the Galileon construction\footnote{The scalar Galileon apart from the standard kinetic term contains higher derivative terms of a specific type such that the equations of motion are still of second order, thereby no Ostrogradsky ghosts. The Galileon system has a well defined structure for a given spacetime dimension. In dRGT, in the decoupling limit, the higher derivative terms associated with the scalar degree of freedom are of Galileon type, thereby no ghosts in the theory. Secondly, in this case, the Vainshtein mechanism screens out the longitudinal degree of freedom locally. } such that the theory is ghost free  {\it \`{a} la} dRGT. Unfortunately, FLRW cosmology is absent in this framework. The reason for the failure was attributed to the assumption that the fiducial metric\footnote{As we pointed out,  gravity is really different from field theory. One might naively treat the space-time metric $g_{\mu\nu}$ as a field, the mas term then requires that we contract the metric $g_{\mu\nu} $ using a fiducial metric, $f_{\mu\nu}$,  namely, $g^{\mu\nu} f_{\mu\nu}$ as $g^{\mu\nu} g_{\mu\nu}$ is constant.} in dRGT model was taken to be a flat spacetime metric. It was then thought to address the problem by replacing the metric by that of a non-trivial background and finally by a dynamical one using a separate action for this fiducial metric$-$  {\it bi-metric gravity} \cite{Hasan}. 
For instance, if flat space-time is replaced by de-Sitter space-time, consistency of quantum theory of spin-2 field on this background asks for a bound on graviton mass, namely, $m_g^2>2 H^2$ dubbed Higuchi bound otherwise theory has a ghost known as Higuchi ghost. In the case of bimetric gravity, this bound is modified such that the effective mass of graviton is larger and larger at earlier times such that it might be difficult to satisfy the bound and reconcile with the  late time acceleration which asks for the effective graviton mass to be $\mathcal{O}( H_0)$, nonetheless, efforts have been made to address the issue\cite{marcos}.  
Theories of massive gravity are also plagued with a number of other problems such as the problem of superluminality and strong coupling.
What is narrated above, clearly tells us that gravity is very different from any other interaction that can be described by a quantum field. In this case, an attempt to address the problem at each stage gives rise to another problem. This, perhaps, tells us that either the mass of graviton is strictly zero or it is challenging to build a consistent theory of massive gravity.

We shall not discuss the Higgs mechanism for graviton mass generation as massive gravity theories are problematic irrespective of the way we introduce the mass of graviton. As mentioned before, spontaneous symmetry breaking finds its important application to EWPT in the early Universe. 
In the discussion to follow, we describe the Higgs mechanism in detail for the Abelian gauge field which would help us to understand the selected aspects of electroweak interaction necessary to discuss the dynamics of EWPT.

  \subsection{The Abelian Higgs Model}
  \label{Abelian}
 We make one step further and consider
 the complex scalar field with the following Lagrangian,
 \begin{equation}
 \mathcal{L}= -\frac{1}{2}  \partial_\mu \phi \partial^\mu \phi^* -V(|\phi|^2);~~V(|\phi|^2)=-\frac{1}{2}\mu^2(\phi^*\phi)+\frac{\lambda}{4}(\phi^*\phi)^2~~~(\mu^2>0)
 \end{equation}
   demanding that the Lagrangian be invariant under the local $U(1)$ phase transformation,
  \begin{equation}
  \label{Ugl}
  \phi\Rightarrow \phi'=e^{i\beta(x)}\phi
  \end{equation}
  where the phase is spacetime dependent.
 The potential term is invariant under the local phase transformation as before but the kinetic term acquires extra terms consisting of derivatives of $\beta(x)$. One then needs to compensate the extra terms by adding a "compensating field" $A_\mu$,
 \begin{eqnarray}
 \label{A}
&& \mathcal{L}=-\frac{1}{4}F_{\mu \nu} F^{\mu \nu} -\frac{1}{2}  (D_\mu \phi) (D^\mu \phi)^* -V(|\phi|^2)\\
&&  D_\mu=\partial_\mu+ieA_\mu \\
&& F_{\mu\nu}=\partial _\mu A_\nu-\partial _\nu A_\mu
\label{fmu}
 \end{eqnarray}
 provided that $A_\mu$ transforms as,
 \begin{equation}
 \label{gt}
 A_\mu \Rightarrow
 A'_\mu  =A_\mu-\frac{1}{e} \partial_\mu \beta(x)
 \end{equation}
 when $\phi$ transforms according to (\ref{Ugl}). We have added the kinetic term for $A_\mu$ (involving $F_{\mu \nu} $) to the Lagrangian (\ref{A})\footnote{Otherwise  $A_\mu$ will be non-dynamical, externally given field. } which is invariant under (\ref{gt}).
 One can easily verify that (\ref{A}) is invariant under local $U(1)$ transformation (\ref{Ugl})  subject to the gradient transformation (\ref{gt})  of $A_\mu$.
 Indeed, let us check how the covariant derivatives $D_\mu\phi$ and $(D_\mu\phi)^*$ transform under
 (\ref{Ugl}) and (\ref{gt}),
 
 \begin{eqnarray}
&&(D_\mu \phi)'\equiv D'_\mu \phi'=(\partial_\mu+ie A'_\mu)e^{i\beta(x)}\phi= \nonumber\\
&&= e^{i\beta(x)}(\partial_\mu+i\partial_\mu \beta(x)+ieA_\mu-i\partial_\mu\beta(x))\phi=e^{i\beta(x)}D_\mu\phi \\
&& (D'_\mu \phi')^*=e^{-i\beta(x)}(D_\mu\phi)^* .
 \end{eqnarray}
 Obviously, the covariant derivative and its complex conjugate transform exactly as  $\phi$ and $\phi^*$ transform\footnote{Which clearly justifies the old nomenclature "compensating field"  for $A_\mu$. } which establishes the invariance of (\ref{A}) under
 transformations (\ref{Ugl},\ref{gt}) referred to as  $U(1)$ gauge transformation ($A_\mu $ is known as $U(1)$ gauge field or Abelian gauge field).
 And we readily recognise $A_\mu$ as an electromagnetic field with gradient invariance (\ref{gt}).
 
  Let us note that the potential $V(|\phi|^2)$ has the wrong mass sign as in (\ref{comphi}). As a result, the true ground state of the system is infinitely degenerate,
\begin{equation}
\label{localg}
\phi_0= v e^{i\beta(x)}  ,
\end{equation}
and represents a circle of radius $v$ in the field space. As said earlier, any point on the circle is a legitimate ground state which can be reached by applying a rotation (transformation (\ref{Ugl})) on $\phi_0(\beta=0)=v$ with an angle $\beta$\footnote{Rotations in the anti-clockwise directions are taken with positive $\beta$.}. Thus the particular choice of the ground state, namely, $\langle\rho\rangle_0=v;~ \langle\theta\rangle_0=0$, can be reached by applying a rotation on $\phi_0$ in (\ref{localg})  with an angle $-\beta$ (see Fig.\ref{vacuum}) which is a symmetry of the Lagrangian (\ref{A}).

It would be instructive to cast the Lagrangian (\ref{A}) in $(\rho,\theta )$ form using the Euler representation for the field: $\phi=\rho(x) e^{i\theta(x)}$,
\begin{equation}
\mathcal{L}=-\frac{1}{4}F_{\mu\nu}F^{\mu\nu}- \frac{1}{2}\partial_\mu \rho  \partial^\mu \rho -\frac{e^2}{2}\left(A_\mu+\frac{1}{e}\partial _\mu\theta(x)\right)^2\rho^2 -V(\rho)  
\label{LE}
\end{equation}
Let us not that (\ref{Ugl}), in the Euler representation, amounts to the following transformation,
\begin{equation}
\theta(x)\to \theta(x)+\beta(x)    \end{equation}
which taking into account (\ref{gt}) readily tells us that Lagrangian  (\ref{LE}) is gauge invariant as it should be.
It is amazing that $\theta(x)$ has disappeared from dynamics altogether, it does not have a kinetic term. It only appears with $A_\mu$ in a specific combination in (\ref{LE}). Defining a new field,
   \begin{equation}
B_\mu\equiv A_\mu +\frac{1}{e}\partial_\mu \theta(x) \,
\label{B}
\end{equation}
we rewrite the Lagrangian (\ref{LE}) through field $B_\mu$\footnote{It should be kept in mind that $F_{\mu\nu}$ does not change by field re-definition (\ref{B})} ,
\begin{eqnarray}
\mathcal{L}&=& -\frac{1}{4}F_{\mu\nu}F^{\mu\nu}-\frac{1}{2}\partial_\mu \rho  \partial^\mu \rho -\frac{e^2}{2}B_\mu B^\mu\rho^2-V(\rho) ,
\label{LBE}
\end{eqnarray}
which has no memory about the $\theta$ field. In variables, $\rho$ and $B_\mu$, gauge invariance does not manifest explicitly$-$ it is hidden  but not lost as (\ref{LBE}) is same as the gauge invariant Lagrangian (\ref{LE}) rewritten in different variables. If
$\mu^2<0$, one has a unique ground state, $\phi_0=0$ or $\langle \rho \rangle_0=0$ and $m_B=0$ and ground state respects the symmetry of the Lagrangian. However, in our case, $\mu^2>0$ and ground state is infinitely degenerate and by virtue of the underlying symmetry, we have chosen one of them, namely, (\ref{GS}), which does not respect the underlying symmetry  {\it \`{a} la} {\it "spontaneous symmetry breaking"}. The field masses in this ground state ($\langle \rho \rangle_0=v $) are given by\footnote{"In the case when ground state respects gauge symmetry, $\langle\rho\rangle_0 = 0$ ($\mu^2<0$) , 
the three independent
components of $B_\mu$ ($B_0$ is non-dynamical, it acts like a Lagrangian multiplier (conjugate momentum corresponding to $B_0$ vanishes)  form a composite representation  made up of two
irreducible
representations: one "massless" spin one field corresponding to two
independent
components of the transverse part and a "massless" scalar field
represented by the
longitudinal component of the vector field $B_\mu$. On the other hand,
when the
ground state violates the symmetry, $\mu^2 >0\to \langle\rho\rangle_0\neq0$  ("spontaneous
breaking
of the symmetry"), as is the case here, these three independent components make one single
irreducible
"massive" spin one representation of the Lorentz group". We thank Romesh Kaul for this comment.}, 
\begin{eqnarray}
\label{Bmass}
 m^2_\rho=\frac {\partial^2 V(\rho)}{\partial \rho^2 }|_{\rho=v}  =2\mu^2;~~
 m^2_{B}|_{\rho=v}&=&e^2v^2
\end{eqnarray}
It is miraculous that after symmetry breaking,  $B_\mu$ becomes massive
with $m_{B}=ev$; mass is given by the vacuum expectation value of $\rho$ field and the electromagnetic coupling. In this case, the field $\phi$ in (\ref{A}) is referred to as the Higgs field. The phenomenon of mass generation of gauge boson(s) through spontaneous symmetry breaking is known as the "Higgs mechanism".
Last but not least, the information about a particular choice of the ground state or "spontaneous symmetry breaking" can be reflected in the Lagrangian by suitably shifting the field $\rho$.

Let us confirm that $\theta(x)$ field in (\ref{LE}) is a pure gauge and can be removed by suitably fixing the gauge transformation. To this effect, an important remark is in order.
As mentioned before, the particular choice of the ground state (\ref{GS}),  becomes possible by applying a gauge transformation to $\phi_0$ in (\ref{localg}),
\begin{equation}
\phi_0\to \phi_0'=\phi_0 e^{-i\beta(x)}=v ,   \end{equation}
with an arbitrary $\beta(x)$.
But the effect of the   choice of the ground state is felt by the Lagrangian only after we shift the field. However, before doing that, let us subject the field to the 
same gauge transformation,
\begin{equation}
\phi\to \phi'=\phi e^{-i\beta(x)}=\rho e^{i\left(\theta(x)-\beta(x)\right)}
\end{equation}
Now, if we make a specific choice, $\beta(x)=\theta(x)$ dubbed {\it unitary gauge}\footnote{Let us note that we do not have this luxury in case of global symmetry; constant $\beta$ can not be identified with $\theta(x)$. }, we have
\begin{equation}
A_\mu \to A'_\mu=A_\mu -\frac{1}{e} \partial_\mu \theta(x) ;~\phi\to\phi'=\rho
\end{equation}
and the field $\theta(x)$ (would be a Goldstone boson in case of global symmetry) completely disappears from the scene in the unitary gauge and we obtain the gauge fixed Lagrangian (see expression, (\ref{LE})),
\begin{equation}
\mathcal{L}=-\frac{1}{4}F_{\mu \nu}F^{\mu\nu}- \frac{1}{2}\partial_\mu \rho  \partial^\mu \rho -\frac{ e^2}{2}\rho^2 A'_\mu A'^\mu  -V(\rho),
\label{LEGF}
\end{equation}
which looks identical to (\ref{LBE}) which is not surprising as it is gauge invariant. Let us recall that gauge invariance of the Lagrangian (\ref{A}) allowed us to eliminate the $\theta$-field as well as to choose the ground state such that $\langle \rho \rangle_0=v$. The masses of the fields in the chosen ground state are given by,
\begin{eqnarray}
\label{mass}
 m^2_\rho&=&\frac {\partial^2 V(\rho)}{\partial \rho^2 }|_{\rho=v}  =2\mu^2\\
 m^2_{A'}|_{\rho=v}&=&e^2v^2,
 \label{mass1}
\end{eqnarray}
as before. An important remark about the number of independent components of $A_\mu$ is in order. When $\mu^2<0$, ground state of the system is unique and $\langle \rho\rangle_0=0$ (symmetry is not broken). In this situation, $A_\mu$ has three independent massless components. In case, of electromagnetic field (two transverse degrees of freedom), the longitudinal components is eliminated using gauge invariance but we have already used that freedom here for knocking out $\theta(x)$. However, in our setting, $\mu^2>0\to\langle \rho\rangle_0=v$   (spontaneous symmetry breaking) and $A_\mu$ is massive.
To summarize, the gauge invariance makes it possible to eliminate one of the components of the complex scalar field which then reincarnates as the  longitudinal component of the massive vector field after symmetry breaking. 
It should also be noted that generation of vector field mass is solely related to symmetry breaking and should not be confused with gauge fixing. This is true that the choice of a particular ground state given by (\ref{GS}) is possible  thanks to the gauge symmetry but the latter is not fixed by this choice. Indeed, the field masses can also be computed in the ground state (\ref{GS}) using the gauge invariant Lagrangian (\ref{LEGF});
gauge fixing for elimination of $\theta$ field can be undertaken thereafter.

Let us note that the mass content of the model is not  explicitly displayed by the Lagrangian (\ref{LEGF}) (or (\ref{LBE})) as the latter knows nothing about our choice of the ground state.
As mentioned before, this information reaches the Lagrangian only after we shift the field. This is, therefore, {\it desirable  though not  mandatory}

to accomplish the field shifting. Due to the specific choice of ground state given in (\ref{GS})  ($\langle\rho\rangle_0=v$), we need to shift only $\rho$ field ($\rho\to \rho+v$ such that vacuum expectation value vanishes for the shifted field \footnote{We used same notation for shifted field, it  should not cause inconvenience}.
The Lagrangian (\ref{LEGF}) then acquires the following form,
\begin{eqnarray}
\label{LUS}
\mathcal{L}&=&-\frac{1}{4}F_{\mu \nu}F^{\mu\nu}- \frac{1}{2}\partial_\mu \rho  \partial^\mu \rho -\frac{ e^2}{2} A'_\mu A'^\mu (v+\rho)^2 -V(v+\rho)\\
V(v+\rho)&=&\frac{1}{2}(\sqrt{2}\mu)^2\rho^2 +\mu\sqrt{\lambda}\rho^3+\frac{\lambda}{4}\rho^4 -\frac{\mu^4}{4\lambda},
\label{VLUS}
\end{eqnarray}
which makes the mass content of the theory explicit.

Let us cast the Lagrangian (\ref{LUS}) in a convenient form,
\begin{equation}
\label{LSF}
\mathcal{L}=-\frac{1}{4}F_{\mu \nu}F^{\mu\nu}- \frac{1}{2}\partial_\mu \rho  \partial^\mu \rho -\frac{ v^2 e^2}{2} A'_\mu A'^\mu  -\frac{1}{2}(\sqrt{2}\mu)^2\rho^2 +\text{Cubic and quartic terms},
\end{equation}
which allows us to immediately read off the
masses of fields given by Eqs.(\ref{mass}) and (\ref{mass1}).
Let us emphasize that the Lagrangian (\ref{LEGF}) is identical to  Lagrangian (\ref{LUS}) (or (\ref{LSF})) in physical content; $\langle \rho\rangle_0=v$ for the former whereas, $\langle \rho_s\rangle_0=0$ after field shifting\footnote{The shifted field $\rho$ is different from $\rho$ though we have denoted it by the same notation. Vanishing of its vacuum expectation should not be confused with the situation when symmetry is exact. Here both the cases correspond to symmetry breaking. }.
As said before, the Lagrangian comes to know about our choice of the ground state after field shifting. Indeed, let us recall that we had chosen the ground state (\ref{GS}) :$ \langle \theta\rangle_0=0;~\langle \rho\rangle_0=v$. From an arbitrary point ($\rho,\beta$) on the circle, one can reach this ground state by applying gauge transformation, $\phi_0=v e^{-i\beta(x)}$, see Fig.\ref{vacuum}. In this case, one should work with the gauge invariant Lagrangian (\ref{LE}). This transformation on the field implies, $\theta(x)\to \theta(x)-\beta(x)$ in (\ref{LE}). In this case, one needs to shift only $\rho$ such that the vacuum expectation value of the shifted  field  vanishes which is explicit in (\ref{VLUS}).
Thus the Lagrangian is well aware about the choice of ground state. Since in the unitary gauge $\theta(x)$ disappears, the ground state before field shifting is specified by $\langle \rho \rangle_0=v$. By virtue of gauge invariance, it is not important from which point on the circle in Fig.\ref{vacuum}, we shift.

Let us conclude the story which begins from the gauge invariant Lagrangian (\ref{A}) and ends with (\ref{LEGF}) or (\ref{LUS})  after symmetry breaking. The gauge invariance allows us to eliminate one of the degrees of freedom associated with a complex scalar field which reappears as a longitudinal component of massive vector field in (\ref{LEGF}) or in (\ref{LSF}). Equivalently, we could use the Lagrangian (\ref{LBE}) where gauge invariance is not explicit. 
{ \it The fact that gauge invariance is hidden is related to either
the choice of variables such as $\rho$ and $B_\mu$ in }(\ref{LBE}) {\it or the  gauge fixing as done in the case of (\ref{LEGF}). And it should not be attributed to "spontaneous symmetry breaking"}.
Actually, we could use the gauge invariant Lagrangian (\ref{LE}), for computing the field masses in the ground state given by (\ref{GS}); the Lagrangian explicitly retains the underlying symmetry after "spontaneous symmetry breaking". However, after  elimination of $\theta(x)$ by choosing a gauge, we shall retrieve the gauge fixed Lagrangian (\ref{LEGF}) where gauge symmetry is not explicit. {\it The ground state does not respect gauge invariance, but Lagrangian does (secretly), thus "spontaneous symmetry breaking" and gauge invariance are two different elements of the theory}\footnote{We thank Romesh Kaul for repeated discussions on related issues.}
The only role, "spontaneous symmetry breaking" plays, lies in the mass generation of the gauge field. Therefore, the nomenclature given to this mechanism might be misleading, as no breaking of gauge symmetry takes place in this case. It could better be called  {\it "secret symmetry"} as suggested by Sidney Coleman.

  What we have witnessed is a general feature of "spontaneous symmetry breaking" in presence of gauge fields.
  In case of SU(2) local symmetry, the Higgs field is  represented by a complex doublet (four real components) interacting with Yang-Mills  field with three massless components. 
 Three of the four components of the complex scalar field (would be Goldstone bosons in case of global symmetry) are gobbled up by three massless gauge bosons after symmetry breaking, making them massive; the fourth component is the Higgs field with standard mass term. After symmetry breaking, the number of components of the Higgs field that were eliminated via gauge invariance get (effectively) attached to massive gauge bosons as their longitudinal components.
  It should be noted that the total number of degrees of freedom before and after "spontaneous symmetry breaking" are the same, they simply redistribute
 
 We mentioned before  that the Fermi theory, to be consistent in a high energy regime, requires three massive gauge bosons as mediators of weak interaction. However, assigning masses to them by hand destroys  renormalizability of the theory. As for the generation of masses through  "spontaneous symmetry breaking", it is really very tricky. Before symmetry breaking, we deal with the massless  $SU(2)$ gauge bosons interacting with Higgs field in the  $\lambda ({\phi}^\dagger \phi)^2$ theory with $\phi$ being a complex doublet; the framework  adheres to gauge symmetry and the theory is renormalizable. Once we assume the scalar field to be with a wrong mass sign, the ground state becomes infinitely degenerate and by virtue of the gauge symmetry, we can choose any of  these with non-vanishing vacuum expectation value $v$ of the Higgs field giving rise to the generation of gauge boson masses. In this process, gauge symmetry is not lost, it remains {\it hidden} or {\it secret} and it is, therefore, not surprising that the Ward identities\footnote{In non-Abelian case, these identities are known as Slavnov-Taylor identities. The anomaly, present in this case, is taken care off by the so called lepton-hadron symmetry. } remain intact after "spontaneous symmetry breaking". On heuristic grounds, it implies that the theory with {\it secret symmetry} allows to consistently generate masses for gauge bosons  {\it \`{a} la}  a renormalizable theory \cite{salam}\footnote{The rigorous proof of renormalizability of gauge theories, with spontaneous symmetry breaking, was provided by G.'t Hooft and M.T. Veltman, see for instance, Ref.\cite{slavnov} for details.}.

\section{Spontaneous symmetry breaking in the late Universe as an underlying cause of dark energy}
\label{sls}
In this section we shall explore the possibility of realizing late time acceleration due to spontaneous symmetry breaking at large scales. Modified theories of gravity provide an arena where the said idea can be accomplished. Conformal  transformation can be used to transform the action from Jordan to Einstein frame where the extra degrees of freedom are directly coupled to matter allowing us to implement the idea of spontaneous symmetry breaking at late times. In the following subsections, we shall discuss these issues in detail.

\subsection{Conformal transformation and non-minimal coupling } 
\label{SCONF}
Conformal transformation plays an important role in model building beyond Einstein gravity\cite{Ant,samrev} (see Ref.\cite{ijmpds} for details on modified theories of gravity). It is believed that modified theories of gravity are equivalent to Einstein's general theory of relativity plus extra degrees of freedom. For instance, in the case of $f(R)$ gravity, we have one extra scalar degree of freedom. Indeed, the $f(R)$ gravity can be transformed to the Einstein frame using a conformal transformation which allows to diagonalize the Lagrangian and clearly read off the degrees of freedom. Let us consider the following action in the Jordan frame (see \cite{Faraoni:1999hp} for a discussion about physical frame),
\begin{align}
    \mathcal{S}=\int {\rm d}^4x \sqrt{-\tilde{g}}\Bigl[\frac{F(\psi)}{2}\tilde{R}-Z(\psi)(\tilde{\nabla}\psi)^2-W(\psi)\Bigr]+\mathcal{S}_m(\tilde{g}_{\mu\nu},\Psi)
\end{align}
where $F(\psi)$ is any coupling to curvature, $Z(\psi)$ generalizes the Brans-Dicke action, $W$ is a potential and $\Psi$ stands for matter fields. Considering the conformal transformation $g_{\mu\nu}=F(\psi)\tilde{g}_{\mu\nu}$, we obtain the action in the Einstein frame (see \cite{Dabrowski:2008kx} for the transformation factors),
\begin{align}
    \mathcal{S}=\int {\rm d}^4x \sqrt{-g}\Bigl[\frac{R}{2}-\frac{1}{2}\Bigl(\frac{2Z(\psi)}{F} +\frac{3F'^2}{2F^2}\Bigr)(\nabla\psi)^2-\frac{W(\psi)}{F^2} \Bigr]+\mathcal{S}_m(g_{\mu\nu}/F,\Psi)
\end{align}
We see that even in the absence of a kinetic term in the Jordan frame, $Z=0$, we obtain a kinetic term in the Einstein frame because of the coupling to curvature. This situation occurs for example in $f(R)$-gravity models. 

Thanks to a redefinition of the scalar field, we can write our action in a canonical form by defining \cite{EspositoFarese:2000ij}
\begin{align}
    \Bigl(\frac{d\phi}{d\psi}\Bigr)^2 &=\frac{2Z(\psi)}{F(\psi)} +\frac{3F'(\psi)^2}{2F(\psi)^2}\\
    V(\phi) &=\frac{W(\psi)}{F(\psi)^2}\\
    A(\phi) &= F^{-1/2}(\psi)
\end{align}
which gives
\begin{equation}
\mathcal{S}=\int{ d^4x\sqrt{-g}\left[ \frac{\Mpl^2}{2}R-\frac{1}{2}(\nabla\phi)^2-V(\phi)  \right]}+ \mathcal{S}_m(A^2(\phi)g_{\mu\nu}, \Psi) \label{E0}
\end{equation}
where we reintroduced a factor $\Mpl^2$. The metric $g_{\mu\nu}$ describes the spacetime in the Einstein frame while quantities with "tilde" refer to the Jordan frame.

In $f(R)$-gravity, we have $Z=0$, which gives 
\begin{align}
    \phi=\sqrt{\frac{3}{2}}\ln F
\end{align}
and therefore a coupling to matter $A(\phi)=e^{-\phi/\sqrt{6}}$.
 
Let us emphasize that conformal transformation gives rise to simplification but it comes with  a price, namely, direct coupling of matter to the field $\phi$ which was not there in the Jordan frame.  In the Jordan frame, $\phi$ is kinetically mixed with the metric which is removed in the Einstein frame.

Obviously,
the energy-momentum tensor of matter in Jordan frame is conserved, 
\begin{equation}
\tilde{\nabla}^\mu \tilde{T}_{\mu\nu}=0.
\end{equation}
Due to the presence of non-minimal coupling in the Einstein frame,  matter and scalar field energy-momentum tensors are not  conserved separately though their sum does. Indeed, varying
action (\ref{E0}) with respect to $g_{\mu\nu}$, we have,
\begin{equation}
 \Mpl^2G_{\mu\nu}=T^{(\phi)}_{\mu\nu} +T_{\mu\nu} \,
\end{equation}
which tells us the sum of the energy momentum tensors is conserved.
The coupling would also manifest in the field equation of motion. 
In order to examine the conservation of individual energy momentum tensors, let us consider the following transformation,
\begin{equation}
\label{tconsef}
 \tilde{T}_{\mu\nu}=-\frac{2}{\sqrt{-\tilde{g}}}\frac{\delta
\mathcal{S}_m}{\delta \tilde{g}^{\mu\nu}}= -\frac{2F(\psi)^2}{\sqrt{-g}}\frac{\delta
\mathcal{S}_m}{\delta g^{\mu\nu}}\frac{\delta g^{\mu\nu}}{\delta \tilde{g}^{\mu\nu}}=F(\psi) T_{\mu\nu}=
A^{-2}(\phi)T_{\mu\nu}.
\end{equation}
Applying the $\tilde{\nabla}$ operator on eq.(\ref{tconsef}) and using conservation of   $ \tilde{T}_{\mu\nu}$, we find,
\begin{equation}
\label{tconsef2}
 \tilde{\nabla}^\mu T_{\mu\nu}-2 \frac{A_{,\phi}}{A^3} T_{\mu\nu}\partial^\mu \phi=0,
\end{equation}
 which after translating $\tilde{\nabla}$ to the Einstein frame gives
\begin{align}
    \tilde{\nabla}^\mu T_{\mu\nu}&= \tilde{g}^{\mu\alpha}\tilde{\nabla}_\alpha T_{\mu\nu}=A^{-2}g^{\mu\alpha}\left(\partial_\alpha T_{\mu\nu}-\tilde{\Gamma}_{\alpha(\mu}^\sigma T_{\nu)\sigma}\right)\\
    &=\frac{1}{A^2}\nabla^\mu T_{\mu\nu}+2 \frac{A_{,\phi}}{A^3}T_{\mu\nu}\partial^\mu \phi-\frac{A_{,\phi}}{A^3}T \partial_\nu \phi
    \label{eq:nablaT}
\end{align} 
where we used
\begin{align}
    \tilde{\Gamma}_{\mu\nu}^\alpha=\Gamma_{\mu\nu}^\alpha+\frac{A_{,\phi}}{A}\left[\delta^\alpha_{\mu} \partial_\nu\phi+\delta^\alpha_{\nu} \partial_\mu\phi-g_{\mu\nu}\partial^\alpha\phi\right]
\end{align}
which reduces (\ref{tconsef2}) using (\ref{eq:nablaT})
to,
\begin{equation}
\label{matcons}
\nabla^\mu T_{\mu\nu}=\frac{A_{,\phi}}{A} T \partial_\nu \phi.
\end{equation}
Since, $T_{\mu\nu}+T^\phi_{\mu\nu}$  should conserve in the Einstein frame, we have \cite{Amend},
\begin{equation}
\label{fieldcons}
\nabla^\mu T^\phi_{\mu\nu}=-
 \frac{A_{,\phi}}{A} T \partial_\nu \phi.
\end{equation}
As mentioned before, coupling also manifests in the field equations. Indeed, varying action (\ref{E0}) with respect to $\phi$, we find,
\begin{eqnarray}
 \label{eqef}
 \Box \phi=- \frac{A_{,\phi}}{A} T+\frac{d V}{d\phi}\ \Rightarrow\
\frac{dV_{\rm eff}}{d\phi}=\frac{dV}{d\phi}-\frac{A_{,\phi}}{A} T
\end{eqnarray}
 
It should further be noted that, in the Einstein frame, the scalar field equation contains an additional term which involves direct coupling of  matter with a field proportional to the trace of matter energy momentum tensor $T$. The latter implies that coupling would vanish for relativistic matter. 

Last but not least, since energy momentum tensor transforms under conformal transformation, particle masses acquire field dependence in the Einstein frame. Keeping in mind the transformation of energy momentum tensor, $\tilde{T}_{\mu\nu}=A^{-2}T_{\mu\nu}$, and $d\tilde{s}^2 =A^2 ds^2$, one can easily identify the particle masses in Einstein frame from
\begin{align}
\tilde{T}^{\mu
\nu}(x) &=\int\frac{\tilde{m}}{\sqrt{-\tilde{g}}}\frac{\d z^\mu}{\d\tilde{ s}}\frac{\d 
z^\nu}{\d\tilde{
s}}\delta\left(x-z(s)\right)\d\tilde{s}\nonumber\\
&=A^{-5}\int\frac{\tilde{m}}{\sqrt{-{g}}}\frac{\d z^\mu}{\d s}\frac{\d z^\nu}{\d 
s}\delta\left(x-z(s)
\right)ds=A^{-6}T^{\mu\nu}(x)
\label{emass}
\end{align}
which gives
\begin{equation}
m=A(\phi) \tilde{m},
\end{equation}
 where particle masses,  $\tilde{m}$ are generic constants in the Jordan frame.
 In fact, the field dependence of mass is an important consequence of conformal transformation which implies explicit microscopic interaction of $\phi$ with matter fields,
\begin{eqnarray}
\mathcal{L}=i\bar{\Psi}\gamma^\mu\partial_\mu\Psi-m(\phi) \bar{\Psi}\Psi
\label{Emass}
\end{eqnarray}
A remark about the Dirac Lagrangian (\ref{Emass}) is in order. For a particular choice of conformal factor, $A=1+\phi^2/M^2$ of interest, the mass term in (\ref{Emass}) would generate an additional, $\phi^2  \bar{\Psi}\Psi$, interaction for field $\phi$.

\subsection{Spontaneous scalarization}

Symmetry breaking has been studied for a long time for compact objects under the name of spontaneous scalarization \cite{Damour:1993hw,Damour:1996ke}. The mechanism triggers a non-zero trivial value of the scalar field near compact objects such as neutron stars which gain some hair. As we have seen (because $T=A^4\tilde{T}$)
\begin{align}
    \Box \phi=-A_{,\phi} A^3 \tilde{T}+V'(\phi)
\end{align}
If $V'(\phi=0)=0$ and $A_{,\phi}=0$ or $A=0$ at $\phi=0$, we have trivially the solution $\phi=0$ to the Klein-Gordon equation, which makes the theory equivalent to general relativity. Spontaneous scalarization occurs when this solution is unstable and evolves to a stable solution for which $\phi \neq 0$. The mechanism can be simply understood by considering $A=e^{\beta\phi^2/2}$ and $V(\phi)=m^2\phi^2/2$
\begin{align}
    \Box \phi=(m^2+\beta e^{2\beta\phi^2}\tilde{\rho})\phi
\end{align}
where we have assumed $\tilde{P}\ll \tilde{\rho}$ and therefore $\tilde{T}=-\tilde{\rho}$. 

Perturbing around the general relativity solution, $\phi=0$, we obtain at the linear order of perturbations
\begin{align}
    \Box \delta \phi=(m^2+\beta \tilde{\rho})\delta\phi
\end{align}
For $\beta$ sufficiently negative $m^2+\beta \tilde{\rho}<0$ which develops a tachyonic instability evolving the scalar field away from zero. At the non-linear level, while the perturbation grows, the term $e^{2\beta\phi^2}$ suppresses ($\beta<0$) the evolution and settles it to a non-zero constant value. 

Notice that in this mechanism, the General theory of Relativity (GR) is usually recovered for low density of matter, while spontaneous scalarization produces a deviation from GR in  high density regime. A similar model has been considered in cosmology, to tackle the dark matter problem in \cite{Chen:2015zmx}, known as "asymmetron". In the next subsections, we shall focus on the  opposite mechanism where GR is recovered in the strong gravity regime.

\subsection{Symmetry breaking in cosmology\protect\footnote{In the rest of this section, we will neglect temperature effects until the second section on early Universe. But it is important to mention that they have been considered for example in \cite{Greenwood:2008qp}. Considering $V(\phi)$ as the zero temperature potential, we can assume that at low temperature, we have an additional thermal mass correction of the following form $a T^2\phi^2$. 
This additional contribution to the potential can produce at some critical temperature a phase transition, from $\phi=0$ to a non-zero value for the scalar field. The different possible future universes are also discussed in the same paper. 
In another paper \cite{Stojkovic:2007dw}, they considered the scalar field as $\phi=\phi^a\lambda_a$ where $\{\lambda_a\}$'s are the generators of SU(3) for which they also study thermal corrections. 
See also \cite{Banihashemi:2018has,Banihashemi:2020wtb} for similar models where the phase transition is triggered by a critical temperature which is similar to a critical redshift 
if the scalar field is in thermal equilibrium with radiation for which we would have $T\propto (1+z)^{-1}$}
}

Compact objects have a rich phenomenology but hereafter, we shall specialize to FLRW spacetimes,
\begin{equation}
 ds^2=-dt^2+a^2(t)\delta_{ij}dx^idx^j   
\end{equation}
In this background, Eqs.(\ref{matcons},\ref{fieldcons}) acquire the following simplified form,
\begin{eqnarray}
\label{frwmcons}
 && \dot\rho+3H(\rho+p)=-\frac{A_{,\phi}}{A}\dot\phi(-\rho+3p) \\
 && \dot{\rho_\phi}+3H(\rho_\phi+p_\phi)=\frac{A_{,\phi}}{A}\dot{\phi}(-\rho+3p)
 \label{frwfcons}
\end{eqnarray}
where, we have assumed matter to be a perfect fluid; scalar field included in action (\ref{E0}) also belongs to this category,
\begin{equation}
T^\mu_\nu=\text{diag}(-\rho,p,p,p); ~~T^{(\phi) \mu} _\nu=\text{diag} (-\rho_\phi,p_\phi,p_\phi,p_\phi) 
\end{equation}
Here $(\rho_i,p_i)$ refer to energy density and pressure.
The scalar field equation simplifies to,
\begin{equation}
 \ddot\phi+3H\dot\phi=-V_{,\phi}+\frac{A_{,\phi}}{A}(-\rho+3p) \, .
 \label{frwfdeq}
\end{equation}
Let us note that (\ref{frwfcons}) can also be obtained from Eq.(\ref{frwfdeq}) multiplying it left right by $\dot{\phi}$.
 Due to coupling, matter energy density $\rho$ does not conserve in Einstein frame, however, $\hat{\rho}=A^{-1}\rho$  does for cold matter $p=0$ and, in this case, it would be convenient to cast the (\ref{frwmcons}) in  $\hat{\rho}$,
 \begin{eqnarray}
\label{frwmconsh}
 && \dot{\hat{\rho}}+3H\hat{\rho}=0 \, \\
&& \ddot\phi+3H\dot\phi=-V_{,\phi}-{A_{,\phi}}{\hat{\rho}} \, .
 \label{frwfdeqh}
\end{eqnarray}
It should be noted that $\hat{\rho}$ is independent of $\phi$ which readily allows us 
 to read off the effective potential from Eq.(\ref{frwfdeqh}) up to an irrelevant constant,
\begin{equation}
 V_{\rm eff}=V(\phi)+A(\phi)\hat{\rho}  \label{effph} 
\end{equation}
In this framework, the effect of coupling is incorporated in the expression of the effective potential. And the only thing left to us now is to make a generic choice for conformal factor $A(\phi)$ in (\ref{effph}). 
\begin{figure}[ht]
\centering
\includegraphics[scale=.56]{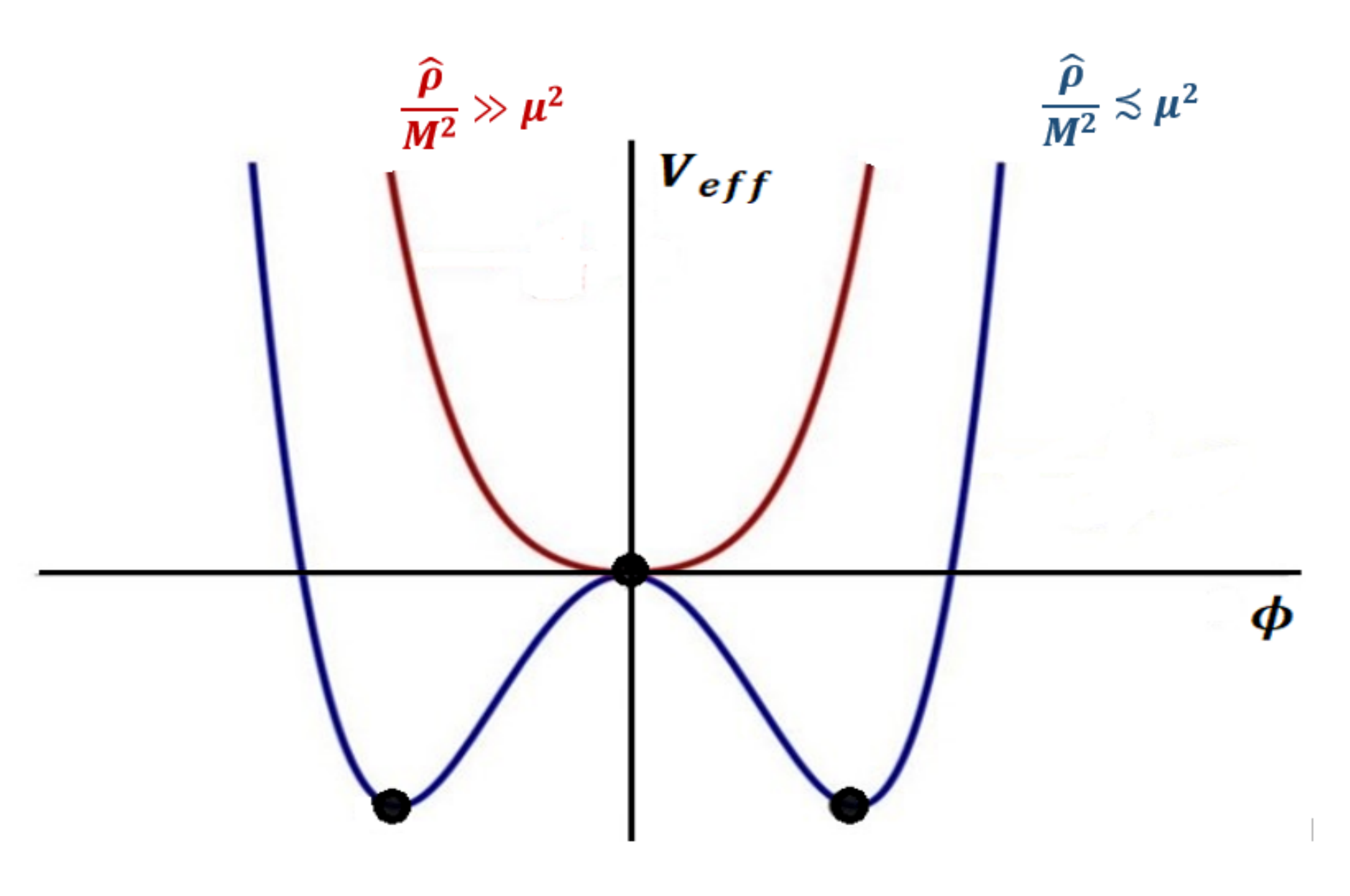}
\caption{Plot of the effective potential (\ref{effpots}) versus the field. For $\hat{\rho}/M^2\gg \mu^2$, the ground state is given by $\phi=0$ However, as density drops below $\mu^2M^2$, a true ground state appears which is doubly degenerate.  } 
\label{symmetron1}
\end{figure}

Notice that we have the same formalism for many other models developed in the literature, such as the chameleon mechanism \cite{Khoury:2003aq,Khoury:2003rn} where
\begin{align}
    A(\phi) &= 1+\alpha\frac{\phi}{M_{Pl}}\\
    V(\phi) &= \frac{M^{4+n}}{\phi^n}
\end{align}
or the Damour-Polyakov mechanism \cite{Damour:1994zq} (dilaton mechanism) which was proposed before the discovery of the accelerated expansion of the Universe and therefore they considered a massless dilaton which decouples from matter by cosmological expansion, known as the "least coupling principle".
\begin{align}
    A(\phi) &= 1+\frac{(\phi-\phi_0)^2}{M}\\
    V(\phi) &= V_0 e^{-\phi/M_{Pl}}
\end{align}

\subsection{Symmetry breaking  in the  Universe at late times$-$ Symmetron}
After a short phase of acceleration, dubbed inflation, the Universe entered the regime of deceleration and continued in that phase before making the transition to accelerated expansion. It is plausible to think that the universe was evolving in a false vacuum with a matter density much larger than the critical energy density. In the "recent past" (around 6 billion years ago), when the former became comparable to the latter, Universe made a transition to the true ground state characterized by a de-Sitter like phase {\it {\`a} la} "Symmetron"\cite{Hinterbichler:2010es}.  The model is based upon $\lambda \phi^4$ theory with wrong mass sign, 
\begin{equation}
V(\phi)=-\frac{1}{2}\mu^2\phi^2 +\frac{1}{4} \lambda \phi^4    
\end{equation}
 directly coupled to matter. The  following   choice for $A(\phi)$ in (\ref{effph}) meets the set goal of the model,
 \begin{equation}
 A(\phi)=1+\frac{\phi^2}{2M^2} \Rightarrow V_{\rm eff}=\frac{1}{2} \left(\frac{\hat{\rho}}{M^2}-\mu^2\right)\phi^2+\frac{\lambda}{4}\phi^4  \label{effpots}
 \end{equation}
 where $M$ is a cut off mass. The model contains two mass scales $ \mu $ and  $ M$ to be estimated from observational constraints. Let us mention that to meet the underlying goal of the model, symmetry breaking should take place around the present epoch. The latter would fix $\mu$; as for $M$, local gravity constraints should constrain it.
 In high density, $\hat{\rho}/M^2\gg \mu^2$, coefficient of quadratic term is positive and minimum of the effective potential (\ref{effpots}) is at $\phi\equiv \phi_0=0$ ($\langle\phi\rangle_0=0$). In this case, $Z_2$ symmetry is exact. However, when $\hat{\rho}\lesssim \mu^2M^2$, the mass term in (\ref{effpots}) acquires wrong sign giving rise to spontaneous symmetry breaking (see Fig.\ref{symmetron1}) \footnote{As seen from Fig.\ref{symmetron1},  $V_{eff}$  at minimum is negative. However, this is not a question of worry as the effective potential is defined up to an irrelevant constant and we can always lift it by adding a suitable constant to (\ref{effpots}).}. Since we wish it to happen around the present epoch, we demand,
 \begin{equation}
\hat{ \rho}\sim \rho_{cr}=\Mpl^2 H^2_0    
 \end{equation}
 and
this fixes the scale $\mu$ in terms of ${M}$,
\begin{equation}
\mu \sim H_0 \frac{\Mpl}{M}   \label{mu} 
\end{equation}
we know that  the field acquires a standard mass, $m=\sqrt{2}\mu$ around the true ground state
 that emerges after symmetry breaking. Acceleration in the ground state puts a restriction on field mass  $m$ or equivalently on cut off, $M$\footnote{One of the slow roll parameters $\eta$ is defined as, $\eta\equiv\Mpl^2 (V_{\phi\phi}/V)^2$ and in slow approximation, $H^2\simeq V/3\Mpl^2$ from where it follows that $m_\phi\equiv m \lesssim H_0$.}
\begin{equation}
\text{Slow role}:~ m \lesssim H_0  \Rightarrow M\gtrsim \Mpl  
\end{equation}
If local gravity constraints fall in this range, the field would roll slowly around the ground state. It is less likely as local physics puts stringent constraints  on  any model with direct coupling to matter. Indeed, a necessary condition to pass local constraints is that our galaxy be screened which translates into (see next section)
\begin{equation}
\text{Local gravity constraints} ~:~~ M\lesssim 10^{-4} \Mpl \Rightarrow m\gtrsim 10^{4}H_0 \,
\end{equation}
which implies that  the field would be rolling very fast ($m\gg H_0)$ around the minimum of the effective potential, all the time overshooting it and the underlying ideology of symmetron gets defeated by the local gravity constraints.

\subsection{Local gravity constraints}

The timelike geodesics for particles are modified in presence of the conformal coupling. We have

\begin{align}
    \ddot x^\mu+\tilde{\Gamma}_{\alpha\beta}^\mu \dot x^\alpha\dot x^\beta=0
\end{align}
where $\tilde{\Gamma}_{\alpha\beta}$ are the Christoffel symbols defined by the metric $\tilde{g}_{\mu\nu}=A^2g_{\mu\nu}$. Using this last relation, we have
\begin{align}
\label{eq:geodesics}
    \ddot x^\mu+\Gamma_{\alpha\beta}^\mu \dot x^\alpha\dot x^\beta+2\dot x^\mu \dot x^\alpha \partial_\alpha \ln A+\partial^\mu\ln A=0
\end{align}
where we have used $\dot x^\alpha \dot x_\alpha=-1$. The second term of eq.(\ref{eq:geodesics}) is the standard gravitational force while the two last terms correspond to the deviation to geodesics of particles which are not conformally coupled to the scalar field. Considering the non-relativistic limit, a test mass, m, will experience an additional force
\begin{align}
    \frac{\vec{F}}{m}=-\vec{\nabla}\ln A=-\frac{A'(\phi)}{A(\phi)} \vec{\nabla}\phi
\end{align}
This force should be screened in the regime where we have tested gravity without founding any fifth force. In order to obtain relevant local constraints on the model, we need to derive $\vec{\nabla}\phi$ in a realistic situation. For simplicity, we study a static spherically symmetric problem, consisting of a central object in the Newtonian regime, characterized by a uniform constant density of matter. The exterior is assumed to be the vacuum. In this case, the fifth force reduces to $-A_{,\phi}\phi'(r)/A$, where $r$ is the radial coordinate. 

The scalar field equation becomes
\begin{align}
    \frac{{\rm d}^2\phi}{{\rm d}r^2}+\frac{2}{r} \frac{{\rm d}\phi}{{\rm d}r}=\frac{{\rm d}V_{\rm eff}}{{\rm d}\phi},\quad V_{\rm eff}=\frac{1}{2}\Bigl(\frac{\hat{\rho}}{M^2}-\mu^2\Bigr)\phi^2+\frac{\lambda}{4}\phi^4
\end{align}

In order to avoid a singularity at $r=0$, we consider the condition $\phi'(r=0)=0$ and the scalar field should recover the cosmological value at large scales $\phi(r=\infty)=\phi_0$ where $\phi_0=\mu/\sqrt{\lambda}$

Inside the spherical object, where the density of matter is assumed much larger than the cosmological matter density, we have $V_{\rm eff}\simeq \frac{\rho}{2M^2}\phi^2$ which reduces the KG equation to
\begin{align}
    \frac{{\rm d}^2\phi}{{\rm d}r^2}+\frac{2}{r} \frac{{\rm d}\phi}{{\rm d}r}-\frac{\rho}{M^2}\phi=0
\end{align}
The general solution is 
\begin{align}
    \phi(r)=\frac{1}{r}\Bigl(\alpha e^{-\sqrt{\rho}r/M}-\beta e^{\sqrt{\rho}r/M}\Bigr)
\end{align}
where $(\alpha,\beta)$ are 2 constants of integration. Imposing the condition $\phi'(r=0)=0$, we obtain $\alpha=\beta$, so we have
\begin{align}
    \phi(r)=\frac{A}{r}\sinh\Bigl({\frac{\sqrt{\rho}}{M}r}\Bigr)
\end{align}
where we have redefined the constant of integration. In the exterior, where we assumed vacuum, the symmetry breaking takes place, so the scalar field is around the value $\phi_0$, the effective potential can be approximated by a harmonic potential, $V_{\rm eff}\simeq m_0^2(\phi-\phi_0)^2/2$, where the $m_0\equiv V''(\phi_0)=2\mu^2$. The solution outside is 
\begin{align}
    \phi(r)=\phi_0+\frac{B}{r}e^{-m_0 r}
\end{align}
where we have suppressed the divergent solution. Assuming continuity of the solution at the radius, $R$, of the object, we find
\begin{align}
      A &= \phi_0\frac{1+m_0 R}{\frac{\sqrt{\rho}}{M}\cosh{\frac{\sqrt{\rho}}{M}R} +m_0 \sinh{\frac{\sqrt{\rho}}{M}R}}\\
        B &= \phi_0 e^{m_0 R}\Bigl(-R+\frac{1+m_0 R}{m_0+\frac{\sqrt{\rho}}{M} \coth{\frac{\sqrt{\rho}}{M}R}}\Bigr)
\end{align}

Analyzing the intermediate regime, where $R\ll r \ll m_0^{-1}$, we have $\phi'(r)\simeq -B/r^2$. Defining the gravitational potential of the source (of mass $M^*$) at its surface as $\Phi= M^* G/R=\rho R^2/6M_{Pl}^2$, we get $\sqrt{\rho}R/M\simeq M_{Pl}\sqrt{\Phi}/M \gg 1$ for large objects, which implies $B\simeq -R\phi_0$. From which we deduce $F/m=R\phi_0^2/(M^2 r^2)$ and finally $F/F_N\simeq \phi_0^2/(M^2\Phi)$ where $F_N=\Phi R/r^2$ is the gravity force.

Assuming that $\phi_0 \simeq M^2/M_{Pl}$ which is necessary to have a fifth force similar to the Newtonian force at cosmological scales \cite{Hinterbichler:2010es} we find $F/F_N\simeq M^2/(M_{Pl}^2\Phi)$, i.e. $M/M_{Pl}\simeq \sqrt{\Phi F/F_N}$. Considering that the fifth force should be screened in the Milky Way, $\Phi\simeq 10^{-6}$, and assuming a screening $F/F_N<10^{-2}$, we get $M<10^{-4} M_{Pl}$.

We see therefore, a limitation of such a form of screening, shared also with the chameleon mechanism. We saw that $F/F_N\simeq \phi_0^2/(M^2\Phi)$ and therefore a bound on the fifth force, implies a bound on the cosmological value of the field fixed by the Newtonian potential of a local object such as the Milky Way. As a consequence, these models have little impact on very large scales as was proved in this no-go theorem \cite{Wang:2012kj}. Because the cosmological field is related to local constraints, the Compton wavelength of such a scalar can be at most Mpc and therefore the deviation from the $\Lambda$CDM background cosmology is negligible at large scales. The accelerated expansion would be due mostly from a varying conformal factor. Nonetheless, this mechanism remains interesting at small scales where deviations from the standard model can be large.

For example, it was shown in Ref.\cite{Winther:2011qb} that the potential governing the dynamics of the matter fields can differ from the lensing potential, and therefore providing a distinctive signature. This effect is stronger in this model as in $f(R)$-gravity. Therefore it is also peculiar to a particular model and not necessarily shared by all modified gravity models. Also the model exhibits interesting stable topological defects \cite{Llinares:2013qbh,Llinares:2014zxa}.

 \subsection{Coupling to neutrino matter }
 Despite the grand success of the hot big bang, namely, the prediction of an expanding  universe, nucleosynthesis and microwave background radiation, the model is plagued with several inconsistencies. One of these related to the age of the Universe should be addressed by late time evolution. The only way to circumvent the problem in the standard lore is by invoking late time acceleration. The latter slows down the  Hubble expansion rate at late stages such that it takes more time to reach the present day value of the Hubble parameter implying a resolution of the age puzzle.
 The phenomenon has been confirmed by direct and indirect observations in recent years. On theoretical grounds, adhering to Einstein's general theory of relativity, cosmic acceleration asks for the presence of an exotic matter repulsive in nature. The mass scale associated with dark energy is $\mathcal{O}(10^{-3})$ eV. Late time cosmic acceleration is an observed reality, but what causes it is a mystery. It is tempting to think whether there is a distinguished physical process in the late Universe with a characteristic mass scale around the mass scale associated with dark energy. And this reminds us about the relic neutrinos. In the Leptonic Era, neutrinos were in thermal equilibrium with the other particles that made up the primordial plasma, namely,
photons, electrons, positrons and nucleons. When the temperature  dropped  to about one MeV, the Universe was about one second old,  neutrinos then
decoupled from the rest species, and since then they are just expanding with expansion with the number density of the order of that of photons in the Universe today. Their masses are typically $\mathcal{O}(10^{-1}-10^{-2}$) eV allowing them to turn non-relativistic around the present epoch. This is what we are looking for, namely, a physical process in the late universe with a characteristic mass scale
of interest to dark energy. In what follows, we shall address the question:  can neutrinos be important to the late time cosmic acceleration?
\begin{figure}[ht]
 \centering
\includegraphics[scale=.4]{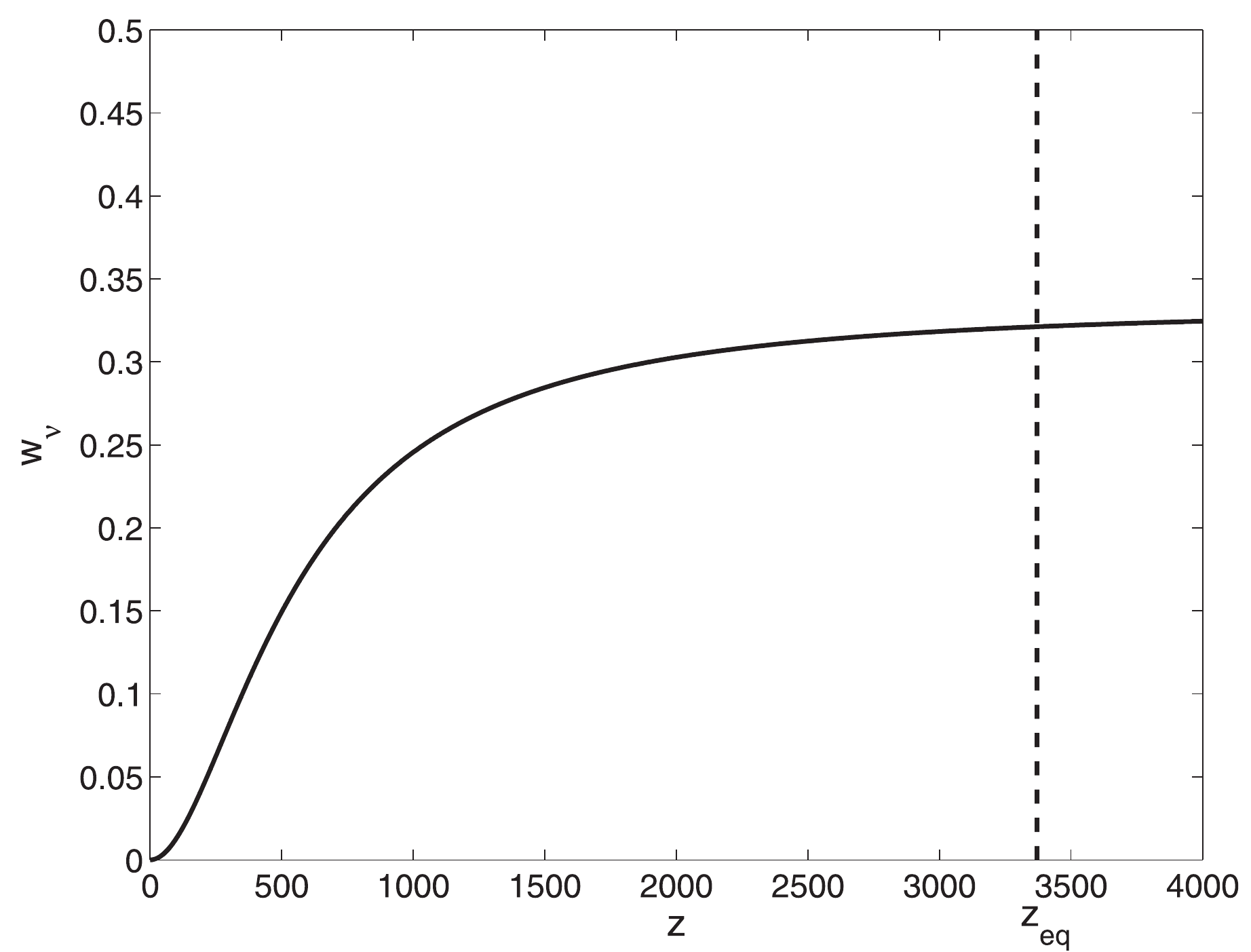}
\caption{Figure shows the equation of state parameter for massive neutrino matter versus the red-shift  in case of $\Lambda$CDM. In matter dominated era,
as seen from the figure, equation of state parameter gradually changes from radiation like to cold matter like.   } 
 \label{wnu}
\end{figure}
 Keeping in mind the  formulated agenda, let us consider the following   action \cite{shn},

\begin{equation}
\mathcal{S}=\int{ d^4x\sqrt{-g}\left[ \frac{\Mpl^2}{2}R-\frac{1}{2}(\nabla\phi)^2-V(\phi)  \right]} +\mathcal{S}_\nu(A^2(\phi)g_{\mu\nu}, \Psi_\nu)    
\label{E}
\end{equation}
where $\mathcal{S}_\nu$ denotes neutrino matter action, $g_{\mu\nu}$ is Einstein frame metric and $\Psi_\nu$ stands for (neutrino) matter field; the standard matter (cold dark matter ) is supposed to be minimally coupled. 

\begin{figure}[ht]
 \centering
\includegraphics[scale=.4]{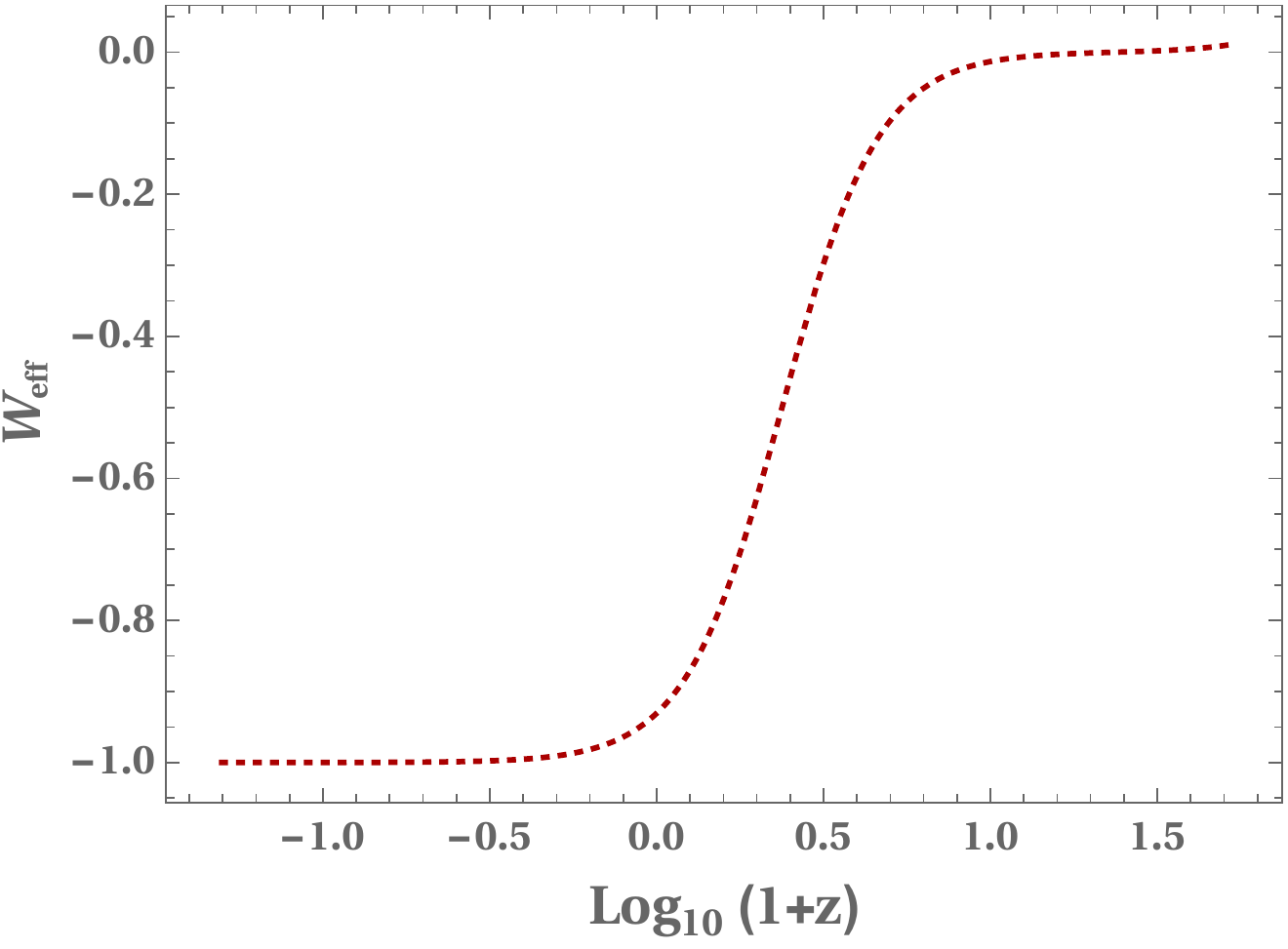}
\caption{Figure displays the numerically estimated effective equation of state $w_{eff}$  versus the redshift ($\lambda =10^{-123}$ and $m_\nu=10^{-2}eV$) in agreement with our analytical claim that field ultimately settles in the minimum after symmetry breaking.}
\label{weff}
\end{figure}

As before, coupling is reflected in the conservation equation for neutrino matter as well as in field equation,
\begin{eqnarray}
\label{frwconsnu}
 && \dot\rho_\nu+3H(\rho_\nu+p_\nu)=\frac{A_{,\phi}}{A}\dot\phi(-\rho_\nu+3p_\nu)\equiv  \frac{A_{,\phi}}{A}\dot\phi \rho_\nu(3\omega_\nu-1)\\
 && \ddot\phi+3H\dot\phi+V_{,\phi}=\frac{A_{,\phi}}{A}(-\rho_\nu+3p_\nu) \equiv \frac{A_{,\phi}}{A}\rho(3\omega_\nu-1).
 \label{frwfdeqnu}
\end{eqnarray}
where $\rho_\nu, p_\nu$ and $\omega_\nu$ represent the energy density, pressure and equation of state parameter for massive neutrino matter.
At early times, neutrino matter is relativistic ($\omega_\nu=1/3$) and RHS of Eqs.(\ref{frwconsnu}) and (\ref{frwfdeqnu}) vanishes. In this case, coupling gradually builds up at late times, see Fig.\ref{weff}. As neutrinos turn non-relativistic around the present epoch, massive neutrino matter mimics cold  matter($\omega_\nu=0$). In this case, it is again convenient as before to work in terms  
of $\hat\rho_\nu=A^{-1} \rho_\nu$,
\begin{equation}
 \dot{\hat\rho}_\nu+3H\hat\rho_\nu=0 \, ,
\end{equation}
and the field evolution  takes the form,
\begin{equation}
 \ddot\phi+3H\dot\phi=-V_{,\phi}-A_{,\phi}\hat\rho_\nu \Rightarrow V_{\rm eff}=V(\phi)+A(\phi)\hat{\rho}_\nu \,
 \label{frwfeq}
\end{equation}
where $\hat{\rho}_\nu$ refers to neutrino matter density today. 
To realize spontaneous symmetry breaking, we shall use massless $\lambda \phi^4$ theory. The choice of the coupling is dictated by the phenomenological considerations keeping in mind the $Z_2 $
symmetry,
\begin{eqnarray}
\label{conf}
 V(\phi)=\frac{\lam}{4}\phi^4;~ 
 A(\phi)=1-\frac{\alpha \phi^2}{2 \Mpl^2}
 \label{eq:pot}
\end{eqnarray}
where $\alpha$ is a constant\footnote{No to be confused with order parameter which is a function of temperature.} to be fixed using observational constraints or some additional requirement ($\alpha$ sets the cutoff scale, $M^2\equiv \alpha^{-1} \Mpl^2$ in Eq.(\ref{conf})).
Using Eqs.(\ref{frwfeq}) $\&$ (\ref{conf}), we then obtain 
effective potential,
\begin{equation}
 V_{\rm eff}=-\frac{\alpha \hat \rho_\nu}{2\Mpl^2}\phi^2 +\frac{\lambda}{4}\phi^4 \,
 \label{eq:pot_eff}
\end{equation}

In absence of coupling (at early stages of evolution), the effective potential given by Eq.(\ref{eq:pot_eff}) has minimum at $\phi=0$ as usual which is no longer the case at late times when coupling builds up dynamically. In this case,  mass term has a wrong  sign and the true minima are now given by,

\begin{equation}
 \phi_{\rm min}=\pm\sqrt{\frac{\alpha}{\lambda}\frac{\hat\rho_{\nu}}{\Mpl^2}} \, .
\end{equation}
We should now repeat the process we carried earlier, namely, choose one of the minima as the ground state, and execute shifting of the field. The latter breaks the original symmetry of the Lagrangian. After that  the mass square of the field is given by ${2} $  times minus the coefficient of the quadratic term in (\ref{eq:pot_eff}),
\begin{equation}
m^2=2\alpha \frac{ \hat\rho_{\nu}}{\Mpl^2};~V_{\rm
min}=-\frac{\alpha^2 \hat\rho^2_{\nu}}{4\lambda \Mpl^4}.
\label{vmin}
\end{equation}
 Since our goal is to have spontaneous symmetry breaking around the present epoch, we would demand that  mass of the field in the true ground state be of the order of $H_0$,
 
\begin{equation}
\label{rhocr0}
m^2=
6 \alpha \Omega_\nu^{(0)} H^2_0\, ,
\end{equation}
Eq.(\ref{rhocr0}) tells us that, not much fine tuning is required for $\alpha$  to achieve slow roll around the true minimum,
\begin{equation}
\alpha \simeq (6 \Omega_\nu^{(0)})^{-1} ,~~~ \Omega_\nu^{(0)}=0.02 (m_\nu/1eV).
\end{equation}
The coupling $\lambda$ in the effective potential can be estimated by asking that,
   $V_{\rm eff}^{\rm min}\simeq \Omega_{DE}^{(0)}\rho_{\rm c0}$,
\begin{equation}
|V_{\rm eff}^{\rm min}|=\frac{\alpha^2 \hat\rho^2_{\nu}}{4\lambda \Mpl^4}\to
\lambda \simeq \frac{\rho_{c0}}{36 \Mpl^4}\left(\frac{1}{4\Omega_{DE}^{(0)}} \right)\simeq 10^{-123}.
\end{equation}
 Such a small numerical value of self coupling might be a problem if we decide to couple the field to any other matter field. We shall come back to this  point later. Secondly, it might look awkward that $V_{\rm min}<0$; it is, however, not problematic
as the
 effective potential (\ref{eq:pot_eff}) is defined up to an irrelevant constant and we can always lift it  appropriately,
\begin{equation}
 V_{\rm eff}=-\frac{\alpha \hat \rho_{\nu}}{2\Mpl^2}\phi^2 +\frac{\lambda}{4}\phi^4+2|V_{\rm eff}^{\rm min}|\, .
 \label{potf}
\end{equation}

We refer back to the symmetron scenario  where the scalar field mass in the true ground state is much larger than $H_0$ ( about $10^4 H_0$) and therefore the field keeps oscillating around the minimum and never settles there.
In our case, the mass being of the order of $H_0$, the field should roll slowly around the minimum.
 Expanding the effective potential  in the small $\delta \phi$  neighbourhood around the
 minimum, we have,
\begin{equation}
\label{ve} V_{\rm eff}\simeq \Omega_{DE}^{(0)}\rho_{\rm
c0}+\frac{\rho_{c0}}{6\Mpl^2} \delta\phi^2
\end{equation}

 The validity of slow roll can be checked by using the following relation,
\begin{equation}
\epsilon\equiv
-\frac{\dot{H}}{H^2}=\frac{\Mpl^2}{2}\left(\frac{V'_{\rm eff}}{V_{\rm
eff}}\right)^2=\frac{3}{2}(1+\omega_\phi),
\end{equation}
Taking into account  the observed numerical value of the equation of state parameter for dark energy to be identified with $\omega_\phi$, we find that, $\delta \phi\lesssim 0.7 \Mpl$ which tells us that the slow is consistent with the approximation used in (\ref{ve})\footnote{Indeed, in a large neighbourhood of the ground state ($\delta \phi\lesssim 0.7 \Mpl$) field rolls slowly.}. Thus, the field rolls slowly around the ground state which is not surprising as the mass of the field is of the order of $H_0$, see Fig.\ref{weff}.

Before closing the review, let us comment on the stability of self coupling $\lambda$, in case, $\phi$ couples to any other field. Strictly speaking, with $\phi$ being a singlet, we can not generate any coupling similar to the standard model. The only microscopic coupling of scalar field with neutrino field gets generated due to field dependence of neutrino masses,
\begin{eqnarray}
\mathcal{L}_\nu=i\bar{\Psi}_\nu\gamma^\mu\partial_\mu\Psi_\nu-m_\nu(\phi) \bar{\Psi}_\nu\Psi_\nu
\label{massnu}
\end{eqnarray}
 where $ m_\nu(\phi)=[1-(\alpha^2/2\Mpl^2)\phi^2]\tilde{m}_\nu$; with $\tilde{m}_\nu$ being the neutrino mass in the Jordan, a generic constant. The latter generates an additional interaction in the Einstein frame,
 \begin{equation}
\mathcal{L}_{\phi\Psi} =\alpha \frac{\tilde{m}_\nu}{\Mpl^2}\phi^2\bar{\Psi}_\nu \Psi_\nu
\label{4v}
 \end{equation}

\begin{figure}[ht]
 \centering
\includegraphics[scale=.4]{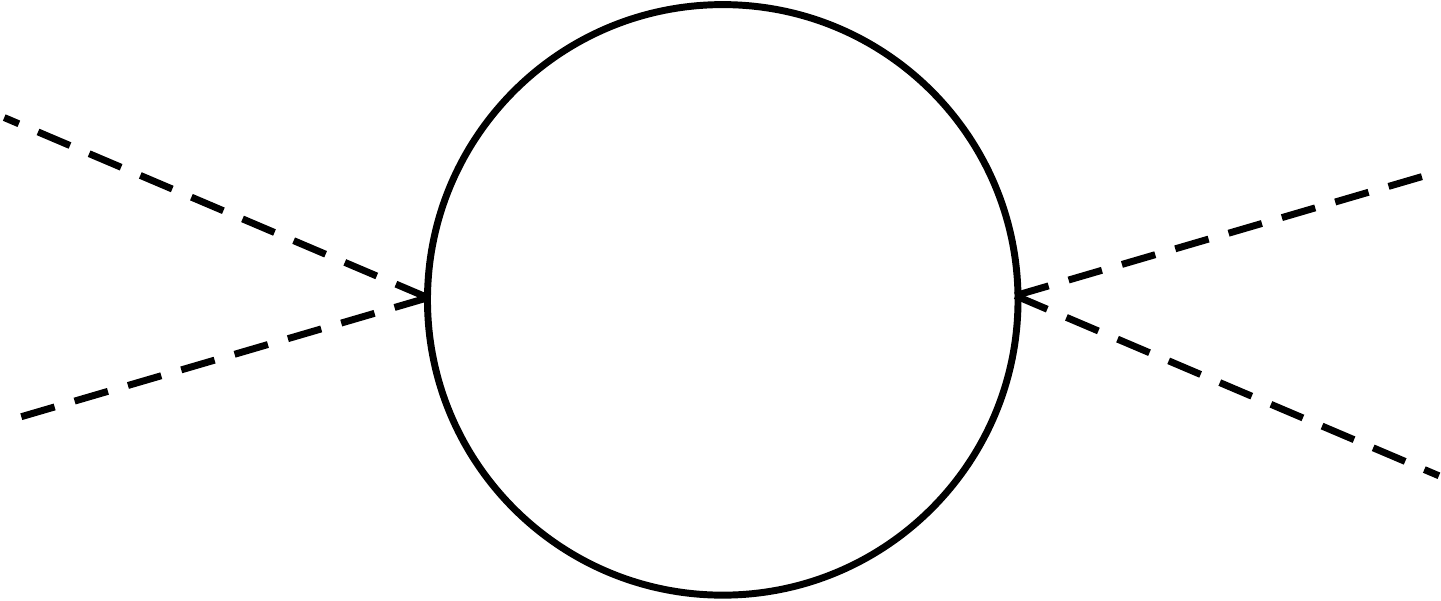}
\caption{One loop diagram due to $\phi\Psi$ interaction  (\ref{4v}). The dashed lines correspond to the $\phi$ field and the closed loop is a fermion loop corresponding to the massive neutrino field. This diagram gives one loop radiative correction to the self coupling $\lambda$.}
\label{circle2}
\end{figure}

The one loop radiative correction to self coupling ${\lambda}$, due to interaction (\ref{4v}), represented by Feynman diagram
in Fig.\ref{circle2},   is given by,
\begin{equation}
\delta \lambda \sim \alpha^2 \frac{\tilde{m}_\nu^2}{\Mpl^2}\frac{\Lambda^2_c  }{\Mpl^2} \ln \Lambda^2_c \,
\end{equation}
where $\Lambda_c$ is UV cut off. Since, for the generic
cut of $\Lambda_c \sim \Mpl$( $\alpha\sim 10^{-2}, \tilde{m}_\nu \sim 10^{-2} eV$) the correction $\delta \lambda \sim 10^{-62}$ is much larger than $\lambda$. Thus the self coupling gets destabilized by the radiative correction. The phenomenon is similar to the quadratic divergence caused by the interaction of the fundamental scalar field with the fermion field. It should, however, be mentioned that such a coupling  is absent in the  framework under consideration\footnote{We do not have $\phi\bar{\Psi}_\nu\Psi_\nu $ interaction here; the only interaction $\phi$ has with fermions is represented by (\ref{4v}) .Field $\phi$ should not be confused with Higgs field; in the scenario under consideration, $\phi$ is a singlet with mass $\mathcal{O}( 10^{-33}) $eV.}.

In the preceding sections we discussed  applications of spontaneous symmetry breaking to late time Universe  and its connection to dark energy. It goes without saying that spontaneous symmetry breaking finds its most realistic application to early Universe, in particular, the electroweak symmetry breaking. In section to follow we shall discuss the dynamics of electroweak phase transitions and related issues.
\newpage
\section{ Phase Transitions in the early Universe: The electroweak symmetry breaking}
\label{euniverse}
Spontaneous symmetry breaking  plays an important role in high energy physics. The  fundamental interactions in nature, with an exception of gravity,  can be described by renormalizable gauge theories, with spontaneous symmetry breaking as an inbuilt mechanism
for mass generation. 
While discussing symmetry breaking, we often pronounced a weird  statement
before and after symmetry breaking which finds a chronological meaning in cosmology.  The early universe is hot composed of plasma of elementary particles. As Universe expands and cools, symmetries break down at critical temperatures resulting into the rearrangement of ground states. For instance, in case of the grand unified theories (GUT),  the underlying symmetries break down to the electroweak symmetry as the Universe cools to temperatures  $\mathcal{O}(10^{16})$ GeV.
 When temperature in the Universe further  drops to the critical value  $\mathcal{O}(100)$ GeV, the electroweak symmetry breaks down to $U(1)$ electromagnetic. In this process, Higgs field, responsible for spontaneous symmetry breaking, acquires non-vanishing expectation value.
 Thinking in the reverse order, the broken symmetries are restored at  higher temperatures in the early Universe which results in a cosmological phase transitions as the Universe cools through the critical temperatures. Thus spontaneous symmetry breaking in the early Universe manifests itself as a phase transition. 
The third example is provided by the QCD phase transition associated with chiral symmetry breaking around 200 MeV.
In what follows, we briefly describe the selected aspects of the standard model required to study the dynamic of electroweak phase transition. In this case, the description is unambiguous as the standard model masses and couplings are known at present to great accuracy which is not the case of GUT. We shall, therefore, not venture into GUT phase transitions. As for the chiral symmetry breaking, numerical studies show that it represents a smooth crossover, nonetheless, it is baryon number conserving and is not of interest in the present context.
\\

\subsection{Selected aspects of standard model}
In section \ref{Abelian}, we discussed in detail the Abelian Higgs model which needs to be generalized here to  $SU(2)_L \times U(1)_Y$ gauge group in case of 
 electroweak theory. This structure is dictated by parity violating nature of weak interactions.
 In this case, the Higgs field is represented by a left handed complex doublet $\Phi$\footnote{The description in this subsection refers to a framework at zero temperature; the temperature corrections would be incorporated in the subsection to follow. },
 \begin{equation}
\mathcal{L}= -\frac{1}{2}  \partial_\mu \Phi \partial^\mu \Phi^* -V(\Phi);~~  V(|\Phi|^2)=-\frac{1}{2}\mu^2(\Phi^*\Phi)+\frac{\lambda}{4}(\Phi^*\Phi)^2 
  \label{HiggsNA}
\end{equation}
Along the lines of section \ref{Abelian},  one replaces the partial derivative, $\partial_\mu $ , in (\ref{HiggsNA}) by the covariant derivative $D_\mu $,
\begin{equation}
\label{LDmu}
\mathcal{L}= -\frac{1}{2} ( D_\mu \Phi) (D^\mu \Phi)^* -V(\Phi)
\end{equation}
defined as,
\begin{equation}
 D_\mu=\partial_\mu+ig\frac{1}2\tau_a W_\mu^a+ig'\frac{1}{2}Y B_\mu  , 
\end{equation}
where $\tau_a, a=1,3$ are three generators of $SU(2)$ and correspondingly, $W_\mu^a, a=1,3$ are three components of non-Abelian gauge field; $Y$ is the weak hypercharge$-$  a quantum number that relates the electric charge and the third component of weak isospin.
Since, we do not have simple group here
but rather a direct product of two groups, we have correspondingly, two gauge couplings, $g$ and $g'$.

The doublet structure of the Higgs field is dictated by the underlying symmetry, $SU(2)_L\times U(1)_Y$,
\begin{equation}
\Phi=\begin{pmatrix} \Phi^+\\ \Phi^0\end{pmatrix} =\frac{1}{\sqrt{2}} \begin{pmatrix} \Phi_1+i\Phi_2\\ \Phi_3+i\Phi_4 \end{pmatrix}
\end{equation}
where the isospin $1/2$ is assigned to the doublet, the electric charge of the upper and the lower components is chosen to ensure that $Y=1$. In that case $\Phi^0$ is charge neutral. This choice would fix $SU(2)_L\times U(1)_Y$ leaving $U(1)_{em}$ untouched.

All the gauge fields are massless at this level. Assuming  all the matter fields to be massless too, one arranges the left handed leptons and quarks in  doublets whereas the right handed fields are considered as singlets.  Then the  Yukawa Lagrangian is introduces  which is necessary for giving masses to leptons and quarks,
\begin{equation}
\label{Yukawa}
\mathcal{L}_Y=-\sum_{leptons}f\left(\bar{\Psi}_L \Phi^\dag  \Psi_R    +\bar{\Psi}_R \Phi  \Psi_L \right); ~~~~ \Psi_R^e\equiv e,~~\Psi^e_L={\begin{pmatrix} \nu_e \\
e        \end{pmatrix}}_L
\end{equation}
where $ e$ and $\nu_e$ refer to electron and electron neutrino wave functions and $f$ is Yukawa coupling. Similar terms are added for quarks.
Analogous to the Abelian Higgs model, we then add  gauge invariant free parts  for $W_\mu^a$ and $B_\mu$ to the system and check that the system is invariant under $SU(2)_L\times U(1)_Y$ gauge group provided that $W_\mu^a$ and $B_\mu$ transform appropriately\footnote{The doublets, $\Psi_L$ and singlets, $\Psi_R$ transform under $SU(2)_L\times U(1)_Y$ local as, $\Psi_L\to \Psi'_L= e^{i\left(\frac{\tau_a}{2} \theta^a(x)+\frac{Y}{2}\beta(x)\right)}\Psi_L;~\Psi_R \to \Psi'_R=\Psi_Re^{i\frac{Y}{2}\beta(x)}$, {where} $\theta^a, a=1,3$ are parameters of $SU(2)$.
$B_\mu \to B'_\mu=B_\mu-\partial_\mu \beta(x);~~W^{'a}_\mu=W^a_\mu-g\epsilon_{abc}\theta^b W_\mu^c-\partial_\mu\theta^a$.}

Before going for symmetry breaking, as a preparatory exercise,    one constructs four convenient combinations using four components of the gauge fields, namely,

\begin{eqnarray}
 && W^{\pm}=\frac{1}{\sqrt{2}} \left( W^1_\mu\mp i W_\mu ^2 \right);~~ Z_\mu=\cos\theta_W  W_\mu^3-\sin \theta_W  B_\mu ,\\
 && A_\mu=\sin \theta_W W_\mu^3+\cos\theta_W B_\mu,
\end{eqnarray}
where $\theta_W$ dubbed Weinberg angle is a free parameter supposed to be determined from observation; $W^\pm$ are two charged gauge fields whereas $Z_\mu$ and $A_\mu$ (would be electromagnetic field after symmetry breaking) are charge neutral. At this level, all the fields are massless. The four transverse gauge fields $W^\pm_\mu, Z_\mu$ and $A_\mu$
have eight degrees of freedom whereas the Higgs field (complex doublet) has four making a total of twelve. 

We then assume that $\mu^2>0$ and go through the process of spontaneous symmetry breaking similar to the Abelian Higgs model. 
 We recall from section \ref{Abelian} that after symmetry breaking, the $U(1)$ gauge field acquires a mass proportional to vacuum expectation value of the Higgs field.
Similarly, in case of $SU(2)_L\times U(1)_Y$ under consideration, 
the Higgs field acquires a non-zero vacuum expectation value $\langle\Phi\rangle_0=v/\sqrt{2}$, giving masses to  other fields such that neutrinos
and $A_\mu$ field are left massless after symmetry breaking.
Let us demonstrate it in case of fermions using the Yukawa Lagrangian. We recall that Higgs field $\Phi$ is a complex doublet with four real components and one can knock out three of them ($\Phi_1,\Phi_2, \Phi_4$) using the unitary gauge and after field shifting we have,
\begin{equation}
\label{hdoublet}
 \Phi=\frac{1}{\sqrt{2}}\begin{pmatrix} 0 \\ {v+h} \end{pmatrix}  
\end{equation} which allows us to rewrite Yukawa Lagrangian (\ref{Yukawa}) for electron in the following form,
\begin{eqnarray}
 \mathcal{L}^e_Y&=&-f_e\frac{1}{\sqrt{2} }\left[ (\bar{\nu}_e, \bar{e})_L {\begin{pmatrix} 0 \\v+h \end{pmatrix}} e_R+\bar{e}_R(0,v+h) {\begin{pmatrix} \nu_e\\ e \end{pmatrix}}_L 
    \right] \nonumber \\
    &=& -\frac{f_e(v+h)}{\sqrt{2}}(\bar{e}_L e_R+\bar{e}_R e_L)
    = -\frac{f_e(v+h)}{\sqrt{2}}\bar{\Psi}_e\Psi_e \nonumber    \\
    &=&-\frac{f_e v}{\sqrt{2}}\bar{\Psi}_e\Psi_e-\frac{f_e }{\sqrt{2}}h\bar{\Psi}_e\Psi_e 
    \label{Ysb}
\end{eqnarray}
where $h$ is the real Higgs field left over after gauge fixing. 
From (\ref{Ysb}), we read off the electron mass as, $m_e=f_e v/\sqrt{2}$ and observe that Higgs coupling to electron is proportional to electron mass.
 Plugging (\ref{hdoublet}) in (\ref{HiggsNA}), one easily obtains the Lagrangian for  $h$ field such that it has a standard mass, namely, $m_H=\sqrt{2}\mu$.
Similarly, we can work out the masses for $Z$ and $W$ bosons after symmetry breaking\footnote{One should notice an extra factor of $1/\sqrt{2}$ in case of $Z(W)$ masses which is related to their interaction with Higgs field which is different from Yukawa coupling.}, 

\begin{eqnarray}
M_W&=& \frac{g}{\sqrt{2}}
\langle \Phi \rangle_0 =\frac{g}{2}v ;~~M_Z=\frac{\sqrt{g^2+g'^2}}{\sqrt{2}} \langle \Phi \rangle_0 = \frac{\sqrt{g^2+g'^2}}{2}v; \nonumber\\ m_F&=&{f}\langle \Phi \rangle_0=\frac{fv}{\sqrt{2}},
\label{masses}
\end{eqnarray}

where $f$ is the Yukawa coupling. We shall use these relations on and off in our forthcoming discussion adding to our list the mass of Higgs boson. All masses in the standard model are given by Higgs field through its expectation value including its own mass! The $A_\mu$ field is left massless and is identified with electromagnetic field.

At the end of the day, let us count the degrees of freedom to :After symmetry breaking,  $Z_\mu, W_\mu^\pm$ become massive (with three degrees of freedom each) leaving
behind a massless $A_\mu$ field (two transverse degrees) and a real scalar (Higgs) field (one degree of freedom). The three components of the Higgs complex doublet got attached to $Z_\mu$ and $W_\mu^\pm$ as their longitudinal components turning them massive. In this process, the total number of degrees of freedom remain same, they simply redistribute. We did not mention here the matter fields as no redistribution of their degrees of freedom takes place in this framework.

Standard model has been put to taste in past years and its predictions have been verified to a high accuracy. There are valid reasons to believe that it might as well provide answer to one of the outstanding problems related to baryon asymmetry in the Universe. 
 Sakharov formulated three necessary conditions for baryogenesis\cite{Sakharov}(see \ref{SakharovE} for details  )\footnote{See Ref.\cite{rathin} on the related theme} : (1) Baryon number  violation, (2) Violation of C and CP  and (3) Departure from
 thermal equilibrium. 
 The first condition is self explanatory as a priory given asymmetry does not sound very attractive and even if it were it would  have been erased by inflation. Obviously, one requires a baryon number violating process.
  C violation is needed to protect the excess production of baryons against the processes that produce the excess of antibaryons. However, $C$ violation alone is not enough. In case $CP$ is exact, the excess of left handed baryons over left handed antibaryons would be balanced by the excess of  right handed antibaryons over the right handed baryons. Clearly, no net asymmetry can be produced in such a situation thereby one needs $CP$ violation also.
   The third condition is extremely important
as the thermal average of Baryon number operator vanishes identically in the equilibrium state provided that $CPT$ is respected. 
And this reminds us about the first order phase electroweak transition which is followed by a  non-equilibrium situation necessary to meet the Sakharov's third condition.
Therefore, it is imperative to study the dynamics of  EWPT and check if it can address the problem. If it does not fully resolve the issue, it might give us important clues as how to go beyond the standard model for addressing the problem.

\subsection{Dynamics of the electroweak phase transition }
To be able to see that 
  spontaneous symmetry breaking manifests itself as phase transition in the early Universe, we need the expression of the effective potential from finite temperature field theory. To this effect, we shall make use of  one loop effective potential computed at temperature $T$ in the framework of standard model, see \ref{effpot} for details ({see also Ref.\cite{Linde:1981zj} and references therein}), 
\begin{equation}
\label{VT}
V_T(\Phi)=\frac{1}{2}a(T)|\Phi|^2-\frac{1}{3}c(T)|\Phi|^3+\frac{b}{4}|\Phi|^4;~~~a(T)=\left(\frac{\beta T^2}{2}-\frac{\mu^2}{2}\right);~~c(T)=\gamma' T;~~b=\lambda~,
\end{equation}
which is specific to first order phase transition, see expression (\ref{GL1O}).
In the standard model, $\beta$, is related to positive linear combination of squares of gauge couplings and Yukawa couplings\footnote{In Abelian Higgs model, after spontaneous symmetry breaking, mass of the $A_\mu$ field is given by, $M_A\sim e v$. Similar result holds here  provided that we carefully identify the gauge couplings for $W_\mu^a$ and $Z_\mu$.}, (see \ref{effpot}), 
\begin{eqnarray}
\label{beta}
 \beta&=&\frac{1}{6}\sum_{Bosons}
 {g_B \frac{M^2}{v^2} }+\sum_{Fermions}{g_F \frac{m^2}{v^2} }= \nonumber\\
 &=&\frac{1}{6}\left(\frac{m^2_H+ 6M^2_W+3 M^2_Z+6m^2_t+ ..}{v^2}\right)
 \end{eqnarray}
 where $g_B$ and $g_F$ denote the degrees of freedom associated with a boson or a fermion respectively and summation runs over all the bosons and fermions in the standard model; clearly, heaviest mass particles contribute most in (\ref{beta}). Coefficients before mass squares in (\ref{beta}) are attributed to helicities  of the particles ($g_B$ or $g_F$)(massive $W^\pm, Z_\mu$ have three degrees of freedom  each whereas color degrees are taken into account for t-quark). It should also be noted that heaviest particles contribute most in (\ref{beta}). 
As for $\gamma$, it is related to positive linear combination of cubes of gauge couplings,
 \begin{eqnarray} 
 \label{gamma}
\gamma=3\left(\frac{1}{12 \pi}\right)\sum_{Bosons}{g_B\frac{M^3 }{v^3}}=
\frac{1}{4\pi}\left(\frac{m^3_H+6M^3_W+3 M^3_z}{v^3}\right)
\end{eqnarray}
Let us note that $\gamma$ can be enhanced, if needed, by going beyond the standard model by including extra scalar fields of masses $\mathcal{O}(100)$ GeV interacting with the Higgs field. This remark will be important in our forthcoming discussion on baryogenesis.

It would be convenient to write down the potential in the following notations,
\begin{eqnarray}
\label{VTFC1}
&& V_T(\Phi)=\frac{\beta}{2}\left(T^2-T^2_0\right)\Phi^2-\frac{1}{3}\gamma T \Phi^3+\frac{\lambda}{4}\Phi^4 \\
&& T_0=v \times \left( \frac{6m^2_H}{m^2_H+6M^2_W+3M^2_Z+6m^2_t} \right)^{1/2}\simeq 147.77 ~\text{GeV}
\label{T0}
\end{eqnarray}
where in (\ref{T0}), we have taken into account the standard model normalization of Higgs vacuum expectation value\footnote{We need to rescale the field, $\Phi\to \Phi/\sqrt{2}$ to match the standard model normalization of vacuum expectation value}.
We used the observed values of masses, $M_W\simeq 80.4, M_Z\simeq 91.2, m_t\simeq 175, v\simeq 247, m_H\simeq 125.3$, we find that $T_0\simeq 147.77$ GeV.

Since the effective potential in (\ref{VT}) has been cast to resemble the Gibbs potential (\ref{GL1O}), it is instructive to write the relations  (\ref{solutioncrit}) in field theory,
\begin{eqnarray}
\label{critpF1}  
 && \Phi(T_c)\equiv \langle \Phi \rangle_{T_c}\equiv \frac{v_c}{\sqrt{2}}=\frac{2c(T_c)}{3 b}=\frac{2 \gamma T_c}{3\lambda}\\
&& a(T_c)=\frac{2c^2(T_c)}{9b}=\frac{2 \gamma^2T_c^2}{9\lambda},
\label{critpF2}    
   \end{eqnarray}
Using  (\ref{VTFC1})
and (\ref{critpF2}), we find the critical temperature\footnote{where we have used $\lambda=m_H^2/2v^2$. We have neglected higher order corrections which in fact can change the final temperature by $+20$GeV. But importantly, the difference between $T_c$ and $T_0$ is always small.},
\begin{equation}
\frac{T^2_c -T^2_0}{T^2_c}=\frac{2\gamma^2 }{9\beta \lambda}   \Rightarrow T_c\simeq 148.04 \text{GeV},
\end{equation}
which is pretty close to $T_0$. From the numerical value of $T_c$, one can infer the  vacuum expectation value at critical temperature, $v_c$, using (\ref{critpF1}),  $v_c\simeq 30 $ GeV which is lower by one order of magnitude compared to Higgs expectation value at zero temperature.

Using the relation   (\ref{critpF1}) and (\ref{critpF2}) and expressing $\Phi$ in dimensionless form, we check that all the coefficients in (\ref{VTFC1}) at the critical point are of the same order, namely, $\lambda v_c^4$. Hence, the effective potential at the critical temperature takes a simple form,
\begin{equation}
\label{VTF}
\frac{V_{T_c}(\Phi)}{\lambda v^2_c}=\frac{1}{4} \Phi^2-\frac{1}{2}\Phi^3+\frac{1}{4}\Phi^4;~~\Phi\equiv\frac{|\Phi|} {v_c} 
\end{equation}
which along with its first derivative vanishes at the critical point, i.e., $\Phi=1$
which is one of the minima of the effective potential. The latter is identical to  (\ref{CC}), see the corresponding Figure \ref{1} which also represents (\ref{VTF}).
Let us recall that
in standard model, the coefficient before the quadratic term, $a(T)$,  is small in the neighbourhood of the critical point. Therefore,  the passage from false vacuum to the true ground state looks like a smooth cross over, see Fig.(\ref{firstorder}). It should be noted that $G-G_0$ in (\ref{GL1O}) is identical to the effective potential (\ref{VT}) or (\ref{VTFC1}) if the coefficients $a,b,c$ are adapted from the standard model. Figure (\ref{firstorder} ) drawn accordingly,  summarizes the behaviour of the effective potential
in detail. The critical behaviour is described  by  the Gibbs potential(\ref{CC}) or by the effective potential (\ref{VTF}), see Fig.\ref{1} and Fig.\ref{firstorder}.
The standard model normalization is met by $\Phi \to \Phi/\sqrt{2}$. 
Obviously, the electroweak phase transition described by the effective potential (\ref{VT}) or (\ref{VTFC1}) is a first order phase transition during which the order parameter jumps from zero to $\Phi(T_c)=2\gamma T_c/3\lambda\neq 0$ giving rise to  jump of entropy at the critical point, see Eq.(\ref{entropy}).
\subsection{Electroweak Baryogenesis: Need of a strongly first order phase transition} 
\begin{figure}[ht]
\centering
\includegraphics[scale=.5]{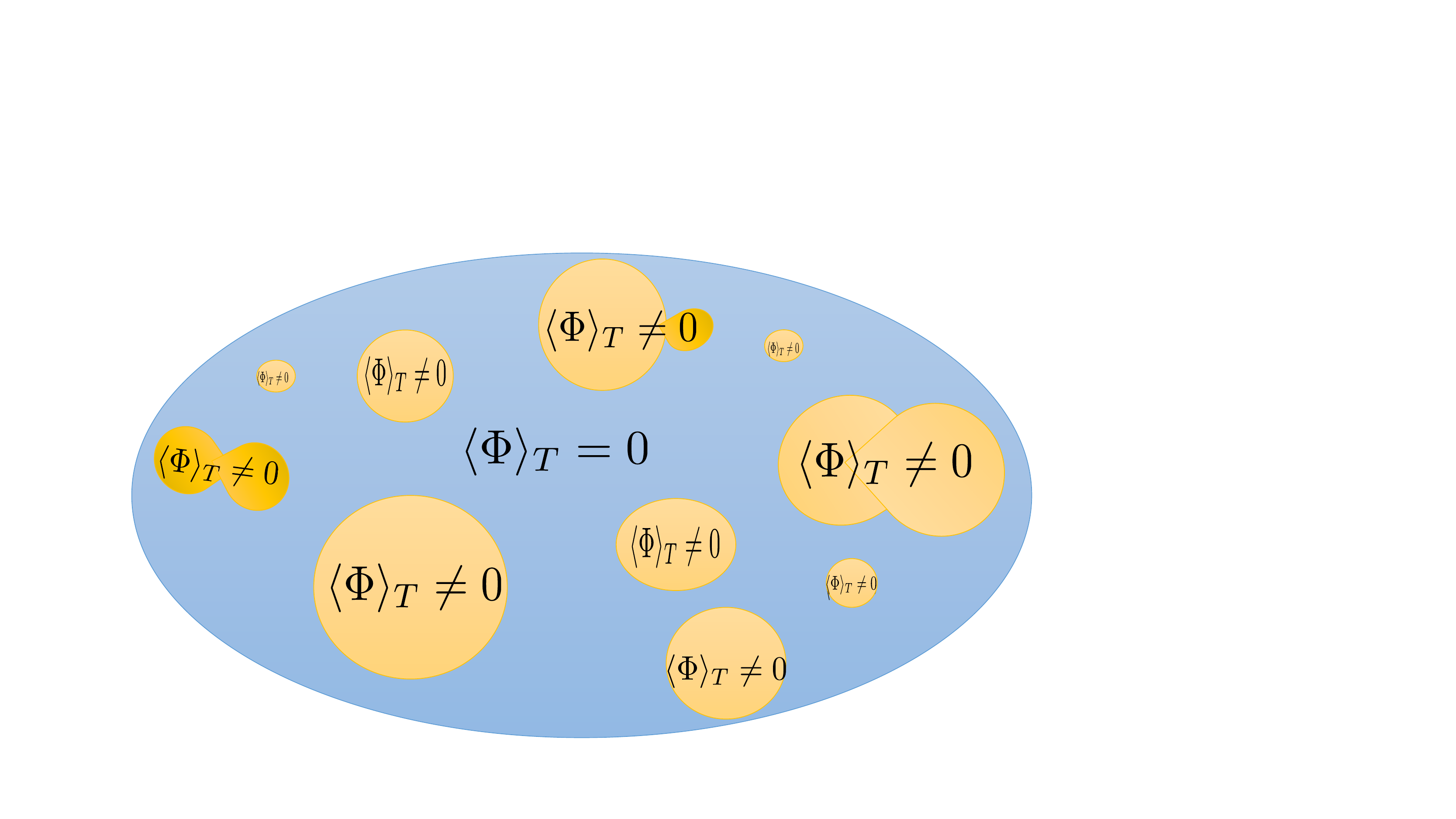}
\caption{Figure shows  the cartoon diagram of vacua bubbles of non-vanishing Higgs expectation value in the  symmetric phase with $\langle\Phi \rangle_{T}=0$ around the critical point. Bubbles expand, collide and merge filling up the whole Universe.} 
\label{vbuubles}
\end{figure}

In previous sections, we discussed the dynamics of electroweak phase transition. In view of Sakharov's criteria of baryogenesis, a notable departure is needed from equilibrium (see Ref.\cite{Sakharov} and \ref{SakharovE} for details). Let us check for the status of equilibrium around the electroweak era when temperature was $\mathcal{O}(100)$ GeV before the phase transition. The relevant interactions around this epoch were weak interactions, their reaction rate is given by,
\begin{equation}
\label{Gammaint}
 \Gamma_{int} =n \sigma \sim \frac{\alpha^2_w T^5}{M^4_{Z,W}}  ;~~H\sim \frac{T^2}{\Mpl} ;~~\alpha_w\equiv \frac{g^2}{4\pi}=1/30
\end{equation}
where $\sigma\sim T^2$ is the cross section of weak processes and $n\sim T^3$ is the number density of relativistic particles   participating in weak reactions.
From  Eq.(\ref{Gammaint}) we read off the estimate,
\begin{eqnarray}
&& \frac{\Gamma_{int}}{H}\simeq \frac{\alpha^2_{w} T^3  \Mpl}{M^4_{Z,W}}\simeq 10^8\times \left(\frac{T}{GeV}\right)^3 ;\nonumber \\
&& \frac{\Gamma_{int}}{H}\gtrsim 1 \Rightarrow\, T\gtrsim  1 MeV
\end{eqnarray}
which tell us that Universe was in perfect equilibrium around the electroweak epoch. However, one of the important necessary requirements of  generation of baryon asymmetry 
is  departure from equilibrium  as the thermal average of baryon number operator vanishes in the equilibrium state if $CPT$ is respected which is the case for electroweak interactions and most of the field theories. And this reminds us about the first order phase transition  when the temperature in the Universe drops  to its critical value, $T_c \simeq 150$ GeV
and electroweak symmetry is broken. This phase transition proceeds through bubble nucleation: 
Bubbles of the new phase, $\langle \Phi\rangle_{T} \neq 0$ form due to thermal fluctuations in the sea of the unbroken phase $\langle \Phi\rangle_{T}=0$. The bubbles  grow if their radius is larger than a critical value\footnote{A bubble  is a classical field configuration of finite energy (soliton) that interpolates between the symmetric and  the broken phases such that below a certain distance (dubbed radius of bubble), $\Phi(T)\neq 0$ and zero otherwise. Such a configuration can be found analytically at the critical temperature, see \ref{bubblenucleation} for details.}. Meanwhile, the different growing bubbles collide and merge and this process continues till the whole Universe is in the broken phase
 and the distinction between the old and new phases disappears, see Fig.\ref{vbuubles}. This process is followed by boiling due the latent heat released by bubbles\footnote{Entropy  released in this process is given by (\ref{entropy}).} which
 is a highly non-equilibrium process that can be helpful to us in the context of baryogenesis, we discuss hereafter.

\begin{figure}[H]
\centering
\includegraphics[scale=.5]{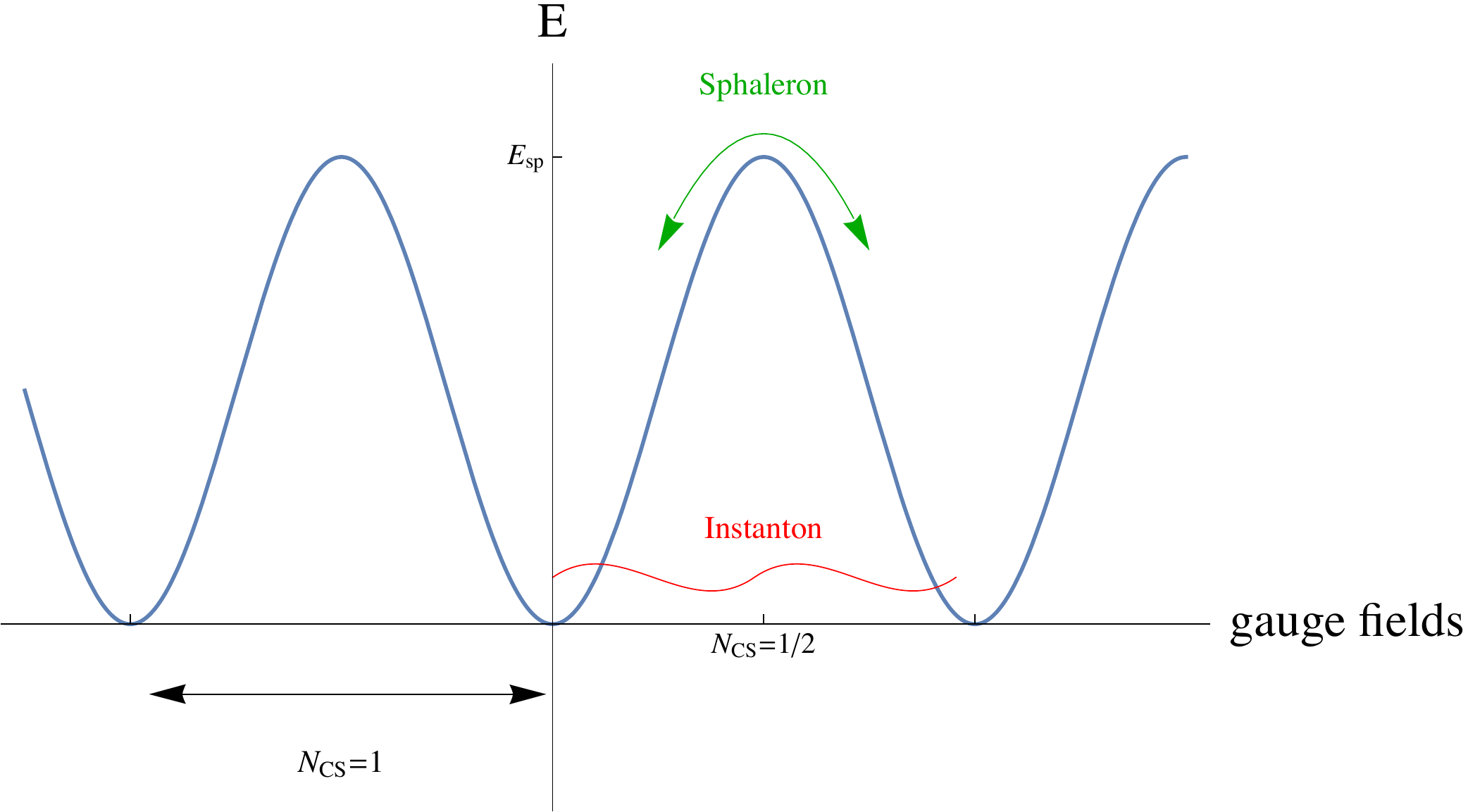}
\caption{Schematic representation of the energy dependence of the gauge configuration with the instanton tunneling through the barrier and the sphaleron path over the barrier. The minima are topologically distinct vacua labelled by a distinct topological number, while the maxima are the unstable sphaleron configuration.} 
\label{sphaleron}
\end{figure}

In electroweak theory, there exist a Baryon number (B)  or Lepton number (L) violating process due to the presence of anomaly such that $B -L$ still conserves. 
The electroweak phase transition, therefore, provides an arena for addressing the baryogenesis problem. Let us note that there exists no chance to cook baryon asymmetry below electroweak scale; the QCD transition associated with chiral symmetry breaking around $200$ MeV turns out to be a smooth cross over and it is also baryon number conserving.
 In gauge theories, vacuum has topological structure, it is infinitely degenerate labelled by Chern-Simons number $N_{CS}$. The transition between the vacua, separated by a barrier, is associated with the change of baryon(lepton) number,
\begin{equation}
 \Delta B   =\Delta L =N_F \Delta N_{CS}
\end{equation}
where $N_F$ is the number of fermion generations which is three in the standard model and $\Delta N_{CS}$ denote the number of units by which the Chern-Simons number changes during a vacuum to vacuum transition. For nearest neighbour transitions, $\Delta N_{CS}= \pm 1$.
The probability of transition between the vacua can be estimated using instantons which connect them in the Euclidean space.
Unfortunately, the  probability of this process is negligibly small at zero temperature\footnote{Indeed, the transition probability is given by: $P(\Delta N_{CS})\sim exp\left(- {4\pi \Delta N_{CS}}{\alpha_W} \right)\sim 10^{-162}; \Delta N_{CS}=1$}. However, an interesting situation arises at finite temperature, relevant to early Universe.
In hot background, it is possible to cross over the barrier (see Fig.\ref{sphaleron}) thanks to the thermal fluctuations {\it {\`{a} la}} {\it sphaleron}, see \ref{Appsphaleron} and \ref{sphaleronsu2} for details. In case of $T>T_c$, the rate of sphaleron transition  is given by,
\begin{equation}
\Gamma_{sp}=\kappa \alpha^5_W T^4 ~~~(\kappa\simeq 25),  \end{equation}
which is established numerically. Then the condition for the sphaleron transition to be in equilibrium, is given by,
\begin{equation}
 \frac{\Gamma_{sp}}{T^3}\gtrsim H\simeq \frac{g^{1/2}_*}{\Mpl} T^2 \Rightarrow T \lesssim 10^{12} GeV
\end{equation}
which implies that sphaleron transitions are in equilibrium for $10^{12} GeV \lesssim T \lesssim 10^2 GeV$\footnote{The lower bound is given by the critical temperature, here, we have indicated only order of magnitudes.}. Thus ultimately, no net asymmetry can be generated in this case as the forward and backward processes take place with the same probability due to $CPT$. Technically, we can estimate the rate at which baryon number is washed out. To this effect, one considers the ratio between transition rates with $\Delta N_{CS}=1 $ and that of with $\Delta N_{CS}=-1$ giving rise to\cite{shapo,AR},
\begin{equation}
\label{expdep}
\frac{ d n_{B} }{dt}= -\frac{13}{2}N_F  \frac{\Gamma_{sp}}{T^3}n_{B}\Rightarrow 
n_{B}(t)=n^0_{B} e^{-\left(\frac{13}{2}N_F  \frac{\Gamma_{sp}}{T^3}\right)t},
\end{equation}
and same argument applies to $L$ and $B+L$.
Eq.(\ref{expdep}) tells us that any a priory existing asymmetry in $B$, $L$ or $B+L$ is washed out by sphaleron at high temperatures and the cleaning process continues till the temperature drops to its critical value. 
For temperatures, below $T_c$, the sphaleron rate is Boltzmann suppressed as the thermal energy is below the energy of sphaleron energy,\cite{Rbook}
\begin{eqnarray}
\label{spratelow}
&& \Gamma_{sp} \sim e^{\frac{-E_{sp}}{T}} ;~~~E_{sp} =B \frac{2 M_W(T)} {\alpha_W};~~B\subset{[1.56-2.72]},\\
&& M_W(T)=\frac{g}{2}\langle \Phi\rangle_T, \nonumber
 \end{eqnarray}
 where the pre-factor of the exponential is proportional $T^4$ on dimensional grounds.
 sphaleron transitions  would continue in equilibrium further below the critical temperature if nothing extra ordinary happens there. 
 However, at $T=T_c$, Universe undergoes a first order electroweak phase transition where the vacuum expectation value of the Higgs field jumps from zero to a non-vanishing value,
 \begin{equation}
 \label{phicrit}
    \frac{ \langle\Phi\rangle_T}{T} =
    \begin{cases}
    0  ,~~\quad   \quad \quad \quad \quad \quad T>T_c\ , \\
    \frac{2 c}{3bT_c}= \frac{2\gamma }{3\lambda}\neq 0,    \quad  T=T_c
    \end{cases}
 \end{equation}   
It should be noted
that sphaleron wash out of a preexisting symmetry stops when the temperature drops to its critical value $\mathcal{O}(100)$ GeV  as $\Gamma_{sp}/T^4\ll 1$ in this case, thereby, $n_{B(L)}(t)\simeq n^0_{B(L)}$, see Eq.(\ref{expdep}).

Recall that the entropy of the system has a jump at $T=T_c$ and amount of its production depends quadratically upon the order parameter, see Eq.(\ref{entropy}).
Hence larger is the amount of entropy produced after phase transition, larger would be the value of the order parameter or Higgs expectation value at the critical point. And this implies that ${\langle \Phi \rangle_{T_c}}/{T_c}$, a measure of non-equilibrium, should be large enough  for keeping the sphaleron transitions out  of equilibrium,
\begin{equation}
 \frac{\langle \Phi \rangle_{T_c}}{T_c}  \gtrsim \frac{\alpha_W^{1/2}}{\sqrt{2\pi }B}\ln\left(\frac{\Mpl}{T_c}\right) \Rightarrow 
\frac{\langle \Phi \rangle_{T_c}}{T_c}  \gtrsim 1
\label{bbound}
\end{equation}
where we considered a convenient value of $B$, namely, $B=1.9$. For the derivation of bound (\ref{bbound}), we used the condition, $\Gamma_{sp}(T_c)\lesssim H(T_c)$ for sphaleron. Larger is the value of
$\langle \Phi \rangle_{T_c}$,
smaller would be the rate of sphaleron transitions at $T=T_c$,
 and the former should be adequately large such that the sphaleron transition process goes out of equilibrium and the bound  (\ref{bbound}) is satisfied.
It is important here to emphasize that despite the Boltzmann suppression  around the critical temperature, sphaleron transitions would continue to be  in equilibrium if  the bound  (\ref{bbound}) is violated. However, they would no longer be in position to wash out a preexisting symmetry and in that sense they shut off around the critical temperature.

Since the strength of the first order phase transition is related to the coefficient of the cubic term, $\gamma$, in (\ref{VTF}) and the order parameter at the critical temperature is linearly dependent upon $\gamma$, the bound (\ref{bbound}) {\it defines a strongly first order phase transition} which is a necessary condition for generation of required baryon asymmetry.
However, in the standard model, using (\ref{phicrit}), we have,
\begin{equation}
\frac{\langle \Phi\rangle_{T_c} }{T_c}=\frac{2\gamma}{3\lambda}=\frac{4\gamma v^2}{3 m^2_H} \simeq  \frac{1}{7},
\end{equation}
that puts an upper bound on Higgs mass, $m_H\lesssim 50$ GeV which is untenable as the observed value of $m_H \simeq 125$ GeV. Hence the standard model misses the bound (\ref{bbound}) by factor of seven approximately.
Let us recall that sphaleron transitions were in equilibrium above the critical temperature and no asymmetry was generated there. Below the critical temperature, the EWPT was supposed to do the job but it is not strong enough to knock out the sphaleron transitions from equilibrium and hence it can not address the baryogenesis problem. But it certainly can give us  clues for circumventing the problem, see remark below Eq.(\ref{gamma}).
Indeed, in theories beyond standard model, enhanced values of $\gamma$, could be obtained allowing us to satisfy the baryogenesis bound (\ref{bbound})\footnote{Enhanced value of $\gamma$ would enhance $\langle \Phi\rangle_{T_c}/T_c$, see Eq.(\ref{phicrit}).}. In particular, one could strengthen $\gamma$  by involving extra scalar fields of masses  $\mathcal{O}(100)$ GeV  interacting with the Higgs field. 
For instance, by adding  a color triplet scalar field,  $\chi$, interacting with the standard model Higgs, we have,
\begin{eqnarray} 
\label{gammaext}
 \gamma&=&\frac{1}{2\pi}\left(\frac{2M^3_W+M^3_z+{3M^3_\chi}}{v^3}\right)=\gamma^{SM}+\frac{3}{2\pi}\left(\frac{M_\chi^3}{v^3}\right)\\
 \frac{\langle \Phi\rangle_{T_c}}{T_c}&=&\left(\langle \Phi\rangle_{T_c}/T_c \right)_{SM}+\frac{1}{\pi\lambda}\left(\frac{M_\chi^3}{v^3}\right)
\end{eqnarray}
where $\gamma^{SM}$ and $\left(\langle \Phi\rangle_{T_c}/T_c\right)_{SM}$ refer to  the standard model values of the respective quantities. Using Eq.(\ref{gammaext}), We find that $M_\chi\simeq 170$ GeV, can match the bound (\ref{bbound}).
Super-symmetric models can also enhance the value of $\gamma$ to the comfortable level. Let us once again note that
 the cubic term with coefficient $\gamma$ in the effective potential (\ref{VTFC1}) is essential for the first order phase transition and the influence of this term is  more, larger is the value of $\gamma$ or equivalently larger is the  expectation value
of the Higgs field at the critical point. 
Thus,
phase transition under consideration is {\it strongly first order} if, $\gamma \gtrsim 7$. 

\subsection{ Electroweak Baryogenesis in a nutshell}

In the standard model, there appears an accidental global  U(1) symmetry which leads to the conservation of the baryon number (B) and lepton number (L). If this conservation is respected at the perturbative level, non-perturbative processes can violate baryon and lepton numbers. Instanton is a non-perturbative process \cite{tHooft:1976snw} that violates B and L  keeping $B-L$ intact and therefore violating $B+L$ number. It is a quantum tunneling process between vacua with different $B+L$ number. Unfortunately this process is highly suppressed. The probability for the process to occur is of the order of $10^{-162}$ (see  \ref{Appsphaleron}). Fortunately, a thermal effect, dubbed sphaleron, can manage the same process, i.e. violation of $B+L$ (see  \ref{Appsphaleron} for technical details on sphaleron). 

It can be shown that the variation of baryon and lepton numbers are proportional to a topological number associated to SU(2) and not U(1). Because, in the standard model we have a group $SU(2)_L$ (the subscript L indicates that we only act on left-chiral fermions), therefore the process  $B+L$ violation produces only left-handed particles. This transition between different vacua involves nine left-handed quarks (three of each generation) and three left-handed leptons (one from each generation). For the viability of this transition, we need to have a temperature $T>E_{sph}$ where $E_{sph}$ is the barrier height which is of the order 10 TeV (see  \ref{Appsphaleron} for more details).

At very high temperatures, the rate of sphaleron interactions can not be calculated analytically, simulations indicate \cite{Bodeker:1999gx} that
\begin{align}
    \frac{\Gamma}{V}\simeq 10^{-6} T^4
\end{align}
where $V$ is a volume, which can be taken of the order of the thermal volume $T^{-3}$, giving $\Gamma \simeq 10^{-6} T$. Considering the Hubble rate during radiation era to be, $3 H^2=\pi^2 g_* T^4/(30 M_{pl}^2)$, where $g_*$ is the number of relativistic degrees of freedom, we find, $H\simeq T^2/M_{pl}$. Therefore, at very high temperatures, we have $H\gg \Gamma$. Sphalerons transitions are dominant and therefore in thermal equilibrium when $H\lesssim \Gamma$ i.e. $ T\lesssim  10^{-6} M_{pl}\simeq 10^{12}$GeV. This analysis is correct before the electroweak phase transition commences at temperature $\mathcal{O}( 100) $GeV. Hence, any preexisting asymmetry above the critical temperature would be washed out by the sphaleon transitions that  transform particles into antiparticles and vice-versa.

As Universe cools to a temperature  $\mathcal{O}(100)$GeV, the Higgs field acquires a non-zero value in some regions of the Universe. Bubbles of broken phase with $\langle \phi \rangle_T \neq 0$ form into a space of unbroken phase with $\langle \phi \rangle_T = 0$ (Fig. \ref{vbuubles}). In the situation under consideration, the three Sakharov conditions are satisfied (see \ref{SakharovE}). The violation of the baryon number is achieved via the sphaleron process, the second condition is naturally achieved in the standard model while the third condition should be met due to the first-order electroweak phase transition associated with  the nucleation and growth of bubbles.

It is interesting to notice that outside the bubbles, the particles have no mass (the Higgs field expectation value is zero), while inside they are massive. The wall of the bubble therefore acts as a potential barrier over which the particles diffuse. A left-handed quark will be reflected by the barrier into a right-handed one. But because of CP-violating processes, we will have different reflection/transmission coefficients for quarks and anti-quarks but also for left-handed and right-handed leptons. Violation of CP is therefore at the origin of a non-zero baryon number around the wall of the bubble. The excess of reflected left-handed quarks in the unbroken phase will be used by the sphaleron process to generate other quarks and leptons. The expanding  bubbles might absorb this excess of baryons before it is washed out by the sphaleron process. Since inside the bubble, this process is highly suppressed\footnote{Inside a bubble, the condition under which the sphaleron process could not wash out totally any excess of baryons, is $\Gamma<H$, which is known as the baryon number conservation condition and occurs for a strongly first order phase transition, see  \ref{Appsphaleron}}, the net excess of baryons remains frozen inside the bubbles. These bubbles grow and merge with other bubbles until they cover all the Universe, ultimately producing a net excess of baryons in the Universe \cite{Cohen:1993nk} which, however, is not sufficient to reconcile with the observed value of the asymmetry. The latter requires a strongly first order phase transition which EWPT is not. And this clearly suggests that the the standard model of particle physics is incomplete and
one needs to look beyond.


The next subsections are devoted to alternative scenarios of baryogenesis
which are inspired by grand unified theories (GUT).
We briefly discuss  Affleck-Dine mechanism and spontaneous baryogenesis  which find their justification in grand unified theories. 
\begin{figure}[ht]
\centering
\includegraphics[scale=0.5]{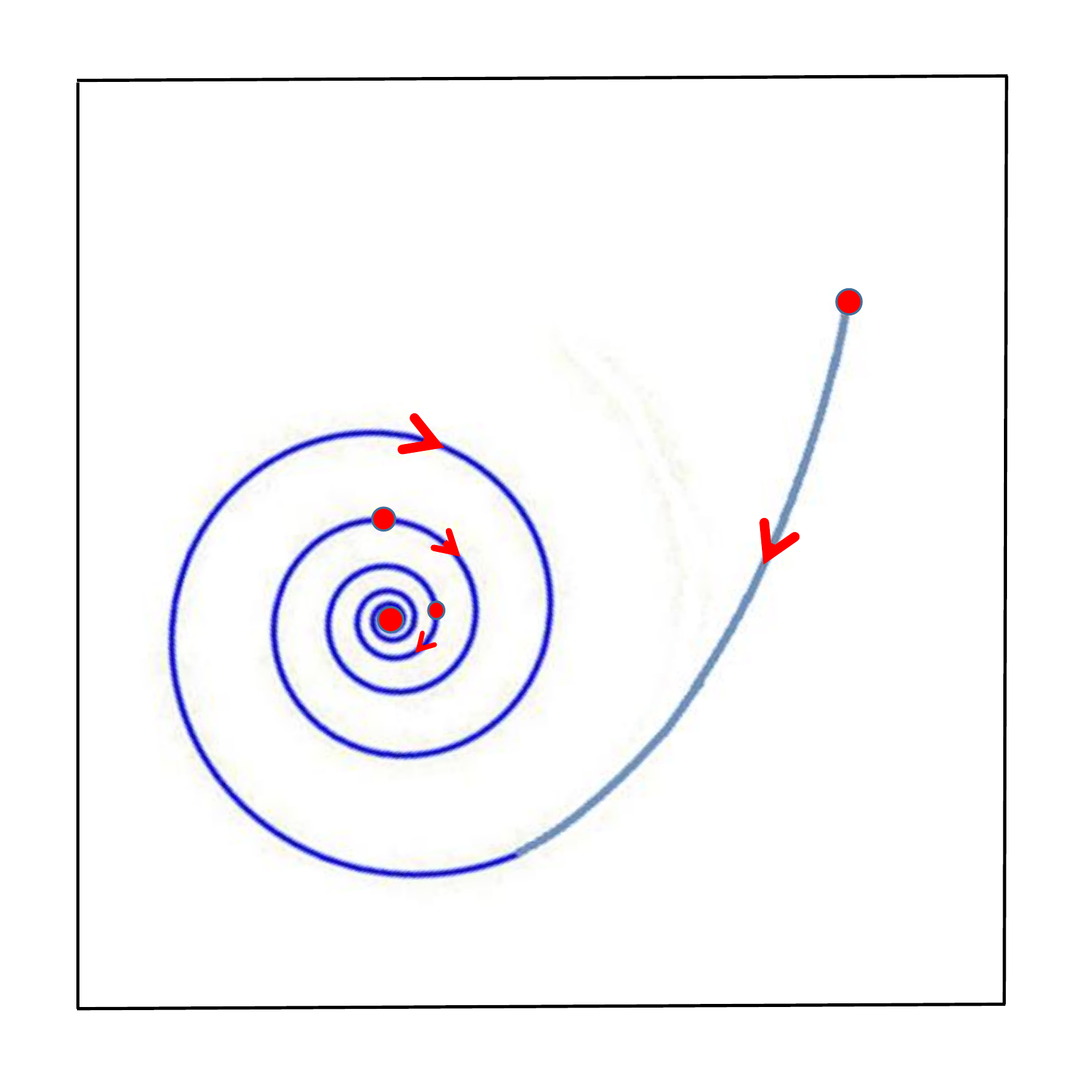}
\caption{Figure depicts the motion of complex Affleck-Dine field $\phi$ which is frozen due to Hubble damping at earlier times. As Hubble parameter becomes comparable to mass of $\phi$, its motion commences such that it ends up circulating around the origin.} 
\label{spiral}
\end{figure}
\subsection{ The Affleck-Dine mechanism for baryogenesis}
The Affleck-Dine (AD)
scenario is inspired by supersymmetric grand unified theories (GUTS). During the process of symmetry breaking, some combinations of squarks and sleptons (scalar fields which carry baryon/lepton number) acquire mass with couplings that violate baryon/lepton number. These features can be incorporated in models of scalar field which is assigned a baryon number. However, the assumptions made in these models can only be justified in the framework of supersymmetry. 

In super-symmetric theories, squarks and sleptons are scalar counterparts of quarks and leptons with  dynamics defined by a potential composed by 2 contributions: F-terms and D-terms
\begin{align}
    V=|F_i|^2+\frac{1}{2}|D^a|^2
    \label{potentialAD}
\end{align}
where the F-term is defined from a superpotential $W(\phi)$ such as $F_i=\partial W/\partial \phi^i$. For a renormalizable theory, the superpotential takes the form
\begin{align}
    W(\phi_i)=m_{ij}\phi_i \phi_j +\lambda_{ijk}\phi_i\phi_j\phi_k
\end{align}
The D-term is defined for a gauge group,
\begin{align}
    D^a=g^a\sum_i \phi_i^* T^a\phi_i,
\end{align}
where $g^a$ denote the coupling constants and $T^a$ are the generators of the group under consideration.  

The potential (\ref{potentialAD}) might have some directions along which the potential vanishes, known as flat directions\footnote{For example, the so-called $LH_u$ flat
direction \cite{Murayama:1993em}, where $L$ is a left-handed slepton $L_i=\frac{1}{\sqrt{2}} \begin{pmatrix}
0\\
\phi
\end{pmatrix}$ and $H_u$ is an up-type Higgs, $H_u=\frac{1}{\sqrt{2}} \begin{pmatrix}\phi\\ 0
\end{pmatrix}$}. 
The space of all the flat directions is known as the moduli space (see e.g. \cite{Gherghetta:1995dv} for the catalog of flat directions in the MSSM). By definition, these flat directions have no potential but in realistic situations, some induced terms appear which give a non-trivial dynamics to the field \cite{Dine:1995kz}. As we mentioned, these directions are never really flat, in fact, the potential is lifted by additional corrections that break supersymmetry  such as $m^2|\phi|^2$; by non-renormalizable terms like $|\phi|^{2n}$ with $n\geq 3$ or by thermal effects. These terms are important for setting the initial conditions of our field. In fact, let us assume that we have higher additional terms in the potential,
\begin{align}
    V=-a^2 |\phi|^2+b^2 |\phi|^{2n}
    \label{potentialLift}
\end{align}
The minimum $\phi_0$ of the potential lies at
\begin{align}
    \phi_0 \simeq \Bigl(\frac{a}{b}\Bigr)^{\frac{1}{n-1}},
\end{align}
from which, we can conclude that  larger the operator which lifts the flat direction,  larger would be the initial value of the field $\phi_0$. Therefore, we shall assume the existence of a field with large initial value.

As a toy model, let us consider a complex scalar field\cite{AD},
\begin{equation}
\mathcal{L}=-\partial_\mu \phi \partial^\mu \phi^*-m^2 |\phi |^2  
\label{ADF}
\end{equation}
which, as we know, is invariant under a global phase transformation; the corresponding conserved current is given by,
\begin{equation}
J^\mu_B=i(\phi^*\partial^\mu \phi-\phi  \partial^\mu \phi^*) 
\end{equation}
The field $\phi$  carries a baryon number by assumption and the Lagrangian (\ref{ADF}) is CP invariant ($\phi\to \phi^*$). It is easy to see that $\nabla_\mu J^\mu_B=0$ and  the baryon number density in the FLRW background is given by ,
\begin{equation}
n_B=J^0_B=2 \text{Im}(\phi^*\dot\phi) =2\rho^2(t) \dot{\theta};~~(\phi(t)=\rho(t) e^{i\theta(t)})   
\end{equation}
 Therefore, we can have a large baryon number if the field rotates in the complex plane with a large amplitude as seen in Fig.(\ref{spiral}).

Let us continue with this toy model, by adding higher order terms similar to what happens in supersymmetry with these additional contributions which lift the flat directions. 
\begin{align}
    \mathcal{L}=-\partial_\mu \phi \partial^\mu \phi^*-m^2 |\phi |^2 -\frac{i}{2}\lambda (\phi^4-\phi^{*4})
\end{align}
We can easily check that 
\begin{align}
    \nabla_\mu J^\mu_B=2\lambda (\phi^4+\phi^{*4})\neq 0
\end{align}
Therefore, this interaction clearly violates "B". Let us decompose the scalar field $\phi=(\phi_1+i\phi_2)/\sqrt{2}$ which gives for the equations of motion in FLRW spacetime
\begin{align}
    & \ddot\phi_1+3H\dot\phi_1+m^2\phi_1+\lambda\phi_2(\phi_2^2-3\phi_1^2)=0\\
    & \ddot\phi_2+3H\dot\phi_2+m^2\phi_2-\lambda\phi_1(\phi_1^2-3\phi_2^2)=0
\end{align}
As we said, let us suppose that initially, the field is very large\footnote{According to eq.(\ref{potentialLift}), the field acquires a large value if the mass term is negative, i.e. if the AD field has a tachyonic mass.}, $\phi_1\simeq \phi_0$, $\phi_2 \simeq 0$. At early times, the Hubble function is large, $H\gg m$ therefore we can neglect $\ddot\phi_i$ and $m^2\phi_i$ which gives
\begin{align}
    & \dot\phi_1 \simeq 0\\
    & \dot\phi_2 \simeq \frac{\lambda}{3H}\phi_0^3
\end{align}
Therefore, the additional term generates a motion in the imaginary direction. We obtain
\begin{align}
    n_B=i\phi\dot\phi^*-i\phi^*\dot\phi
    =\phi_1\dot\phi_2-\phi_2\dot\phi_1
    \simeq \frac{\lambda}{3H}\phi_0^4
\end{align}
By generating a motion in the imaginary direction, proportional to $\lambda$, we have generated a net baryon density. In the absence of coupling, the initial conditions on $\phi_1$ and $\phi_2$ imply that $\theta=const$. The introduction of coupling gives a kick to $\theta$ (angular motion in the complex plane) or equivalently to baryon number density.

As Universe evolves and much time has elapsed such that $H\ll m$, then field begins to oscillate around the minimum, with a decreasing amplitude. As a consequence, the effect of the $\lambda \phi^4$ term will become negligible. In this regime, baryon number violation becomes insignificant. The total baryon charge of the field $\phi$ will then be conserved and its density will fall as $a(t)^{-3}$. 

In a matter dominated Universe, for which $H=2/(3t)$, so for large time $t\gg m^{-1}$, the motion is damped, we get \cite{Affleck:1984fy,Dine:2003ax}
\begin{align}
    \phi_i = \frac{A_i}{m t} \sin(m t+\delta_i)\,,~~i=1,2
\end{align}
where $(A_i,\delta_i)$ depend on the parameters of our model, in our case {$A_i \propto \phi_0^3/m^2$.}

To conclude, the large baryon number will depend on the initial dynamics viz. on the initial value of the field. It is therefore important to restore this model in a more realistic context of inflation and reheating.
In these conditions, it would be impossible to give credit to all the different models and mechanisms during reheating. But as we have seen the field will oscillate and eventually decay which would transfer this baryon and lepton asymmetry to fermions\footnote{The instability of the AD field could directly produce  this excess of baryons or the process go through formation
of Q-ball/non-topological solitons, if the AD condensate carries a global charge. Eventually, this condensate could decay into baryons or if stable it could be a candidate for dark matter.}. The ratio of baryon density number to entropy density will depend on the reheating temperature $T_R$, the mass of the field $m$ and the coupling constant $\lambda$ \cite{Asaka:2000nb}
\begin{align}
    \eta = \frac{n_B}{s} =\frac{1}{4}\frac{T_R}{M_p^2 H(t_{\text{osc}})} n_B(t_{\text{osc}})
\end{align}
{where $n_B(t_{\text{osc}})$ and $H(t_{\text{osc}})$ refer to the values of these quantities when field oscillations commence.}
If we assume ${H(t_{\rm osc})\simeq m}$, $\lambda=\beta {m}/M$\footnote{$n_B$ depends on $\lambda$.} where $\beta=\mathcal{O}(1)$ and $M$ is a scale beyond which new physics should appear that we will assume to be $M_p$, we get \cite{Enqvist:2003gh}
\begin{align}
    \eta \simeq 10^{-10} \beta \Bigl(\frac{1~\text{TeV}}{m}\Bigr)\Bigl(\frac{T_R}{10^9~\text{GeV}}\Bigr)
\end{align}

As for the sphaleron washout, one can work with a grand unified theory that violates $B-L$ or suppress the decay of the field $\phi$ before the sphaleron becomes ineffective around electroweak phase transition.

\subsection{ Spontaneous Baryogenesis}
In this subsection, we shall briefly examine a scenario
dubbed "Spontaneous Baryogenesis" (SBG)
that evades all the three Sakharov's  conditions
for asymmetry generation\cite{SB1,SB2,Ar,sam}. Clearly, if such a framework is to work, there no option for it other than to effectively violate $CPT$. Since, this mechanism operates in thermal equilibrium, we need to recall that the Gibbs potential for multi-component system with variable number of particles is given by,
\begin{equation}
dG=-S dT+V dP+\sum_{i=n}^n{\mu_i dN_i}
\end{equation}
where $\mu$ dubbed chemical potential is the Gibbs potential per particle. For the system to be in thermal equilibrium, see section \ref{SGL},
\begin{equation}
\label{GM}
dG=0\, \Rightarrow\,  \sum_{i=1}^n{\mu_i dN_i} =0  \,\to \, \sum_{i=1}^n{\mu_{i}}=0
\end{equation}
 where the last relation holds as the total number of particles are supposed to be constant in a closed system. Early Universe is hot composed of the plasma of relativistic particles which are frequently interacting. Thermal equilibrium was established as the reaction rate of species took over the Hubble rate which happened when temperature fell to around $10^{16} GeV$, see \ref{SakharovE} for details. The equilibrium distributions of species depend upon two parameters only, namely, temperature $T$ and chemical potential $\mu$,
\begin{equation}
\label{DF}
 f(p)=\frac{1}{e^{\frac{E-\mu }{T} }  \pm 1 }
\end{equation}
where $E=\sqrt{p^2+m^2}$ is the energy of the particle with mass $m$ ; "+" refers to fermions and "-" to bosons. Let us note that  $\mu<E$ for a boson where as no such lower bound exists for fermions. Secondly, for particle species in thermal equilibrium,
\begin{equation}
 A_1+A_2+...+A_n \leftrightarrow{} B_1+B_2+....B_n  ,
 \label{nreaction}
\end{equation}
Eq.(\ref{GM}) implies that\footnote{Formally, in the context of (\ref{nreaction}), $N_{A_i}+N_{B_i}=const, i=1,n$ (additional constraint within each species) which implies that $dN_{A_i}=-dN_{B_i}\equiv dN_i$. Since equilibrium state is the minimum of Gibbs potential, we have (for P, T constant), $dG=0=\sum_{1}^n{(\mu_{A_i}}-\mu_{B_i})dN_i\Rightarrow  \sum_1^n{\mu_{A_i}}=   \sum_1^n{\mu_{B_i}} $.}, \begin{equation}
\label{muc}
 \sum_{i=1}^n{\mu_{A_i}}=   \sum_{i=1}^n{\mu_{B_i}}
\end{equation}
In case of photons, since they can be produced in a reaction in variable number, it suffice to assume that $\mu_\gamma=0$
to satisfy relation (\ref{muc}). Last but not least, if, 
\begin{equation}
A+\bar{A}\to 2\gamma \,\Rightarrow \, \mu_A+\mu_{\bar{A}}=0 \Rightarrow \, \mu_{\bar{A}}=-\mu_A
\end{equation}
At high temperatures, one is dealing with relativistic degrees of freedom for which mass can be ignored. Using the distribution (\ref{DF}), we have,
\begin{eqnarray}
 n_B=n_b-n_{\bar{b}}&=&
{g}\int{\frac{d^3\vec{p}}{(2\pi)^3}\left[f(p,\mu)-f(p,-\mu)\right]} \\
&=&\frac{g}{(2\pi)^3}\int_0^\infty{4\pi p^2dp\left( \frac{1}{e^{ (p-\mu)/T}}-\frac{1}{e^{ (p+\mu)/T} }\right)}\\
&=&\frac{2gT^3}{\pi^2}\sinh{\frac{\mu}{T}}
\end{eqnarray}
where ${g}$ denote the spin degree of freedom for baryon.
In the limit, $ \mu/T\ll 1$, we have,
\begin{equation}
\label{nbQ}
 n_B\simeq  g \frac{2{\mu}T^2}{\pi^2} \end{equation}
In what follows, we shall first describe the phenomenological implications of a non-conserved baryonic current in the paradigm of quintessential inflation. We would then show that such an effect can be generated through spontaneous symmetry breaking {\it {\`{a} la}}  spontaneous baryogenesis.
 \\
 
{\bf 1. Phenomenological Aspects:}

\begin{figure}[H]
\centering
\includegraphics[scale=.45]{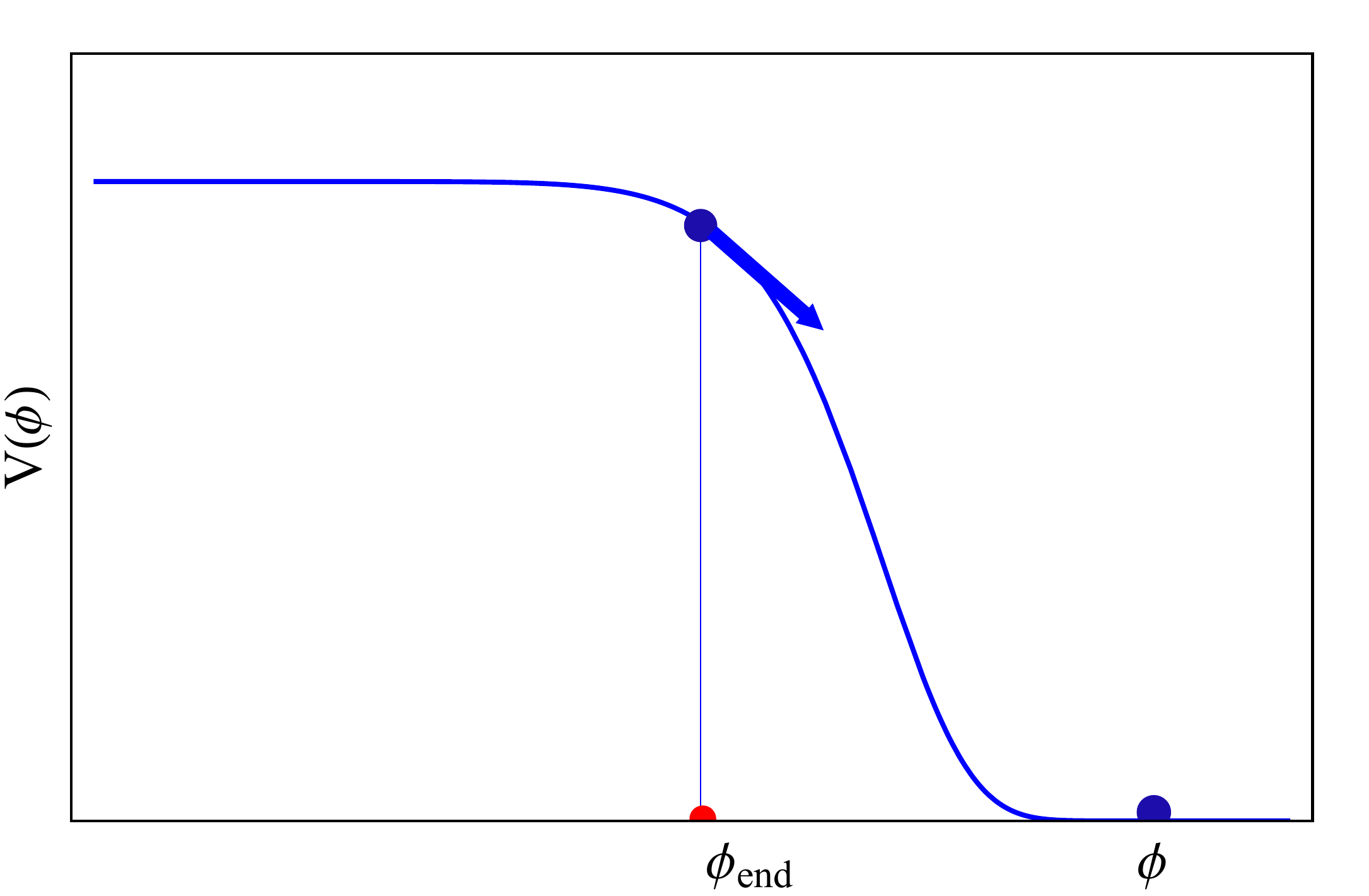}
\caption{A cartoon diagram for a field potential, suitable to quintessential inflation, which is shallow at early stages for $\phi\lesssim \phi_{end} $ followed by a steep behaviour thereafter and again shallow at late times. For $\phi>\phi_{end}$, the field enters into the  kinetic regime and only at late stages, slow roll commences
giving rise to late time acceleration. Potential  (\ref{QIPOT}) mimics the said behaviour after a late time feature is added to it .} 
\label{inf}
\end{figure}

In the paradigm of quintessential inflation, which we briefly describe below, inflaton $\phi$ survives
after inflation ends. In this framework, one can invoke an interaction of $\phi$ with non-conserved baryonic current which might be attributed to spontaneous  breaking of a hypothetical $U(1)$ baryon symmetry  {\it {\`{a} la}}  "spontaneous baryogenesis". Such an interaction is certainly absent in the standard model,
\begin{equation}
\label{JBNC}
\mathcal{L}_{eff}=\frac{\lambda'}{M}  \partial_\mu \phi J_B^\mu  
\end{equation}
where $M$ is the cut-off of the effective field theory construction and $\lambda'$ is coupling. Let us include a brief mention about quintessential inflation from where the field $\phi$, in particular, could originate. In the FRW background, the effective interaction takes the form\cite{sam},
\begin{equation}
\mathcal{L}_{eff}=\frac{\lambda'}{M}\dot{\phi}n_B\equiv\mu(t)n_B    
\end{equation}
where $\mu(t)$ is the effective chemical potential and $n_B=J^0$ is the baryon number density. Plugging the identified value of chemical potential in Eq.(\ref{nbQ}), we find,
\begin{equation}
    n_B=\frac{2 \lambda'{g}}{\pi^2 M}T^2 \dot{\phi}(t)
\end{equation}
Using the expression of entropy density,
\begin{equation}
s=\frac{2\pi^2}{45}g_{*s}T^3  ,  \end{equation}
we obtain the baryon to entropy ratio,
\begin{equation}
\label{eta}
\eta \equiv\frac{n_B}{s}\simeq 0.46 \lambda'\left(\frac{{g}}{g_{*s}}\right)\frac{\dot{\phi}(T) }{MT}   
\end{equation}
where $g_{*s}$ refers to effective entropy degrees of freedom.
Let us include here a brief mention on quintessential inflation from where the field $\phi$ originates and which would allow us to connect $\dot{\phi}$ to inflationary quantities.
Quintessential inflation refers to a framework which unifies inflation and late time acceleration using a single scalar field. At the onset, this idea can be realized by using an inflaton potential which is shallow at early times followed by a steep behaviour that continues to present when it again turns shallow necessary to capture  the late acceleration, see Fig.\ref{inf}. Steep behaviour of potential after inflation is required for non-interruption with thermal  history which is accurately known. Independence of late time physics from initial conditions asks for a particular type of steepness of the field potential. To this effect, let us consider the following form of the potential,
\begin{equation}
\label{QIPOT}
V(\phi)=V_0 e^{-\lambda \phi^n/\Mpl^n}    
\end{equation}
where $n$ is an integer to be fixed from observational constraints on inflation($n\gtrsim 6$ suffice \cite{sam,samad}). In order to meet the last cosmological requirement, one needs a late time feature in the potential (\ref{QIPOT}). One of the possibilities is provided by the introduction of coupling of massive neutrino matter to field $\phi$. 
We shall focus on model independent estimates from inflation to be used in context with  (\ref{eta}).
In this regard, let us note that the scale of inflation $H_{inf}$ is set by the COBE normalization,
\begin{equation}
H_{inf}\simeq 3\times 10^{14} r^{1/2} GeV    
\end{equation}
where $r$ is the tensor to scalar ratio. Since the present observational value of $r$ is quite small, field does not roll much to reach the destination, $H_{end}$ where inflation ends starting from a location where $H=H_{inf}$. Secondly,  potential (\ref{QIPOT}) is quite steep after inflation ends and this together implies that,
\begin{equation}
H_{inf}\simeq H_{end}\simeq H_{kin}    
\end{equation}
where $H_{kin}$($H_{kin}\sim a^{-3})$ refers to kinetic regime which immediately follows inflation. Numerical results with (\ref{QIPOT}) show that all the three quantities are within the same order of magnitude. Putting all this together, we have,
\begin{equation}
\dot{\phi} \simeq \Mpl H_{end}\left(  \frac{a_{end}}{a}\right)^3= \Mpl H_{end}\left(  \frac{T}{T_{end}}\right)^3
\label{phidot}
\end{equation}
where $T_{end}$ is temperature of radiation, due to alternative reheating mechanism\footnote{Since the potential (\ref{QIPOT}) under consideration is run away type, the standard reheating mechanism does not work in this case. One, in particular,  might employ instant preheating.},  at the end of inflation. Plugging $\dot{\phi}$ from Eq.(\ref{phidot}) into Eq.(\ref{eta}), we find,
\begin{equation}
\eta\simeq 4 \times 10^{-3}\left(\frac{\lambda'H_{end}\Mpl}{M} \right) \left(\frac{T^2}{T^3_{end}}\right) ;~~~~\left(\frac{g}{g_{*s}}\simeq 10^{-2} \right) 
\end{equation}
Since $\eta$ depends here upon temperature, we need to know the value of temperature at which asymmetry freezes. The latter requires the knowledge of the underlying process responsible for the generation of asymmetry. To achieve this, let us assume an effective four Fermi interaction that violates $B-L$,
\begin{equation}
\mathcal{L}_{B-L}=\frac{\bar{g}}{M_X} \psi_1\psi_2 \bar{\psi}_3\bar{\psi}_4    
\end{equation}
where $M_X$ is a cut off mass and $\bar{g}$ is a dimensionless coupling. The rate of baryon number violation in the process, $\psi_1+\psi_2\to \psi_3+\psi_4$, for $T<M_X$, is given by
\begin{equation}
\label{gammaBL}
\Gamma_{B-L}=\bar{g}^2\frac{T^5}{M^4_X}   
\end{equation}
This process goes out of thermal equilibrium when temperature drops below, $T=T_F$ such that,
\begin{equation}
 \Gamma(T_F)=H(T_F) \Rightarrow T_F=\frac{M^2_X}{\bar{g}}\left(\frac{H_{end}}{T^3_{end}} \right)^{1/2} 
 \label{TF}
 \end{equation}
 where we have assumed that freezing takes place in the kinetic regime where $H\sim T^3$. We should then replace $T$ in Eq.(\ref{eta}) by $T_F$ from Eq.(\ref{TF}) to yield the freezing value of the asymmetry $\eta_F$ that depends upon several parameters such as $\lambda' \Mpl/M$ and $M_X$\footnote{The coupling $\bar{g}$ might be taken to be of the order $10^{-1}$.}. Using then the Carroll bound, $\lambda' \Mpl/M\sim 8$, one can fix the cut off scale $M_X$ that would give rise to observed value of the asymmetry $\eta_F$.
  Let us note that the cut off mass $M_X$, in this case, turns out to much lower than the Planck mass, see Ref.\cite{sam} for details.

We next turn to spontaneous symmetry breaking for the derivation of (\ref{JBNC}) that we used for generation of asymmetry in the framework of quintessential inflation.
\\

{\bf 2. Non-conserved baryonic current from spontaneous symmetry breaking:}
\\
We shall now demonstrate that the effective interaction (\ref{JBNC}) can be generated by spontaneous breaking of a global $U(1)$ baryon symmetry.
To this effect, we consider a complex scalar field interacting with quarks and leptons\cite{AD,Ar},
\begin{eqnarray}
 \label{comphiA}
&& \mathcal{L}= -\frac{1}{2}  \partial_\mu \Phi \partial^\mu \Phi^* -V(\Phi \Phi^*)+\bar{Q}(i\gamma^\mu\partial_\mu-m_Q)Q+\bar{L}(i\gamma^\mu\partial_\mu-m_L)L\nonumber\\
 &&\qquad + \mathcal{L}_{int}\\
  \label{LintA}
  &&\mathcal{L}_{int}=\frac{1}{M^2_X}
 \frac{\Phi}{\langle \Phi\rangle_0}
 (\bar{L}\gamma_\mu Q)(\bar{Q}^c\gamma^\mu Q)+H C\\
 &&V(\Phi \Phi^*)=-\frac{\mu^2}{2}\Phi^*\Phi+\frac{\lambda}{4}(\Phi \Phi^*)^2
 \end{eqnarray}
 where $\Phi$ and quark fields carry non-zero baryon number; $M_X$ (to be fixed from observational constraints as in case of quintessential baryogenesis) is a  cut off mass scale and $Q^c$ refers to charge conjugated quark field. Baryon numbers $-1$ and $1/3$
 are assigned to $\Phi$ and quarks such that the interaction Lagrangian (\ref{LintA}) preserves the baryon number.
Let us note that the  Lagrangian (\ref{comphiA})  is invariant under a global $U(1)$ baryon symmetry, namely,
\begin{equation}
\label{U1A}
 \Phi\to e^{i\beta}\Phi; ~~Q\to e^{-i\alpha/3} Q;~~L\to L  
\end{equation}
thanks to which the total baryon number of $\Phi$ and quarks remains intact. We next assume that $\mu^2>0$; in that case, the vacuum manifold is non-trivial,
\begin{equation}
    \Phi_0\equiv \langle \Phi\rangle_0=ve^{i\beta}
\end{equation}
And by virtue of global $U(1)$ invariance (\ref{U1A}), we may choose for convenience, $\beta=0$ which is equivalent to choosing a particular ground state. In this context,
it would be convenient to write the complex field in Euler form,
 \begin{equation}
 \label{GSA}
 \Phi=\rho(x) e^{i\theta(x)} \Rightarrow \text{Ground state}:~\langle\rho\rangle_0=v,~~\langle\theta\rangle_0=0   \end{equation}
As a part of symmetry breaking ritual, we now carry out the field shifting, namely, $\Phi=(\rho(x)+v)Exp(i\theta)$ in the Lagrangian (\ref{comphiA}) and find,
\begin{eqnarray}
\mathcal{L}=& -&\frac{1}{2}  \partial_\mu \rho \partial^\mu \rho -\frac{1}{2}(v+\rho)^2    \partial_\mu \theta \partial^\mu \theta-V(\rho)+\bar{Q}(i\gamma^\mu\partial_\mu-m_Q)Q+\bar{L}(i\gamma^\mu\partial_\mu-m_L)L+\nonumber\\
& +&\frac{1}{M^2_X}
 \frac{(\rho+v)e^{i\theta(x)}}{v}
 (\bar{L}\gamma_\mu Q)(\bar{Q}^c\gamma^\mu Q)
\end{eqnarray}
This Lagrangian retains the original symmetry (\ref{U1A}) of the model; choosing a particular vacuum state, violates the vacuum symmetry but it can not fix or break the original symmetry ($U(1)$-global) of the Lagrangian, see section \ref{U1global} for details.
We can now read off the masses of fields: the radial field $\rho$ has standard mass, $m_\rho=\sqrt{2\lambda}v$; the angular field $\phi\equiv \theta/v$
is a massless Goldstone mode. The radial field is heavy and frozen and may be ignored from the dynamics below the energy scale set by the vacuum expectation value $v$,
\begin{align}
 \label{LvrhoA}
\mathcal{L} &=-\frac{1}{v^2}   \partial_\mu \theta \partial^\mu \theta+\bar{Q}(i\gamma^\mu\partial_\mu-m_Q)Q+\bar{L}(i\gamma^\mu\partial_\mu-m_L)L\nonumber\\
&+
 \frac{e^{i\theta(x)}}{ M^2_X}
 (\bar{L}\gamma_\mu Q)(\bar{Q}^c\gamma^\mu Q)
\end{align}
The last term in (\ref{LvrhoA}) contains  $\theta$ dependence  which can be removed by transforming the quark field, namely, $Q\to Exp(-i\theta/3)Q$.  However, a new interesting term, $\partial_\mu \theta J_B^\mu$, gets generated in this process in the Lagrangian that actually we were looking for,
\begin{align}
 \label{LvrhoA1}
\mathcal{L} =&-\frac{1}{v^2}   \partial_\mu \theta \partial^\mu \theta+\bar{Q}(i\gamma^\mu\partial_\mu-m_Q)Q+\bar{L}(i\gamma^\mu\partial_\mu-m_L)L+
 \frac{1}{ M^2_X}
 (\bar{L}\gamma_\mu Q)(\bar{Q}^c\gamma^\mu Q)\nonumber\\
 &+\partial_\mu \theta J_B^\mu\\
 J^\mu_B \equiv & 1/3(\bar{Q}\gamma_\mu Q),\nonumber
\end{align}

where $J^\mu_B$ is a baryonic current.
The last term in the Lagrangian (\ref{LvrhoA1}) essentially contains a non-conserved baryonic current otherwise it would have not been there. In fact, it is not difficult to check that the divergence of  $J^\mu_B$ is non-vanishing,
\begin{equation}
 \partial_\mu J_B^\mu=\frac{1}{M^2_X}  (\bar{Q}\gamma_\mu Q_1^c)(\bar{Q}\gamma^\mu L)+ H C 
\end{equation}
To summarise, we have briefly discussed the mechanism of spontaneous baryogenesis in the framework of quintessential inflation using the effective interaction given by (\ref{JBNC}) and also  demonstrated that such an interaction can be generated using spontaneous breaking of a $U(1))$ global baryon symmetry.

\section{Summary and conclusions}
\label{CON}
This review is devoted to the study of phase transitions in the early and late Universe. Its first part
describes  the efforts made by cosmologists to obtain late time acceleration from spontaneous symmetry breaking in the late Universe. In this paradigm, late time transition from deceleration to acceleration  is thought to be the result of a phase transition. Spontaneous symmetry breaking serves as the foundation of Unification of interactions and it is therefore plausible to think that it might  work in cosmology as well. This idea  is essentially inspired by thermodynamic theory of phase transitions. To this effect, we have briefly reviewed the Ginzburg$-$Landau theory of phase transitions. We clearly show that the phenomenon spontaneous symmetry breaking can not be realized in quantum mechanics$-$ in this case, the classical degeneracy of the ground state gets lifted by quantum tunnelling whose probability can be estimated using the instanton  solutions. We then mention that tunnelling probability exponentially vanishes in an infinite system such as field theory. The heuristic argument is supported by the fact that there exists no instanton solution in scalar field theory with multiple vacua that could facilitate transition between different vacua. 

As a first example from field theory, we consider the  $\lambda \phi^4$ theory with $Z_2$ symmetry but with wrong mass sign in the potential. In this case, we have two symmetric minima and choosing one of them breaks the symmetry of the Lagrangian. Shifting the field to new configuration with zero vacuum expectation value corrects the sign of mass term for the shifted field. This follows the discussion of  complex scalar field with $U(1)$ global continuous symmetry, again with the wrong mass sign. In this case, the vacuum state is infinitely degenerate; the vacuum manifold is a circle.  After symmetry breaking, one of the components of the scalar field has standard mass as in case of $Z_2$ symmetry breaking but the other component is a massless Goldstone boson. Miracle happens, if we consider the same problem in the presence of an electromagnetic field; the underlying symmetry , in this case, is $U(1)$ gauge symmetry. After "spontaneous symmetry breaking", the Goldstone boson gets gobbled up by the massless vector field turning it massive such that the degrees of freedom in the model are same before and after symmetry breaking. We should mention that this is a general feature of any gauge theory with symmetry breaking.

After the pedagogical introduction to "spontaneous symmetry breaking",  we get to the description of the first model of symmetry breaking in the Universe at large scales known as "symmetron". This model is based upon the $\lambda \phi^4$ theory with the wrong mass sign as before and with direct coupling of $\phi$ to cold matter. Such a coupling can be generated  using a conformal transformation from Jordan to Einstein frame. We describe  conformal transformation in detail to show that the   coupling  is proportional to the trace of  matter energy momentum tensor, $T_m$. In this  case,  coupling reduces to modifying the field potential which does facilitate symmetry breaking in the low density regime whereas in the high density regime, $Z_2$ symmetry is exact. Model has two parameters $\mu$ and $M$ to be fixed from observation. The first parameter gets gracefully fixed by demanding that symmetry breaking takes place when matter density drops below the critical density. As for the cut off mass $M$, it should be such that the field  rolls slowly around the true ground state, i.e., $m \lesssim H_0$ ($m$ being the field mass about the true minimum) that emerges after symmetry breaking. While studying direct coupling to cold matter, the most stringent restrictions on  model parameters are imposed by local gravity constraints which, in the case of "symmetron", imply   that, $m\gg H_0$. As a result, the field  never settles  in the ground state to mimic de-Sitter-like solution of interest to late time acceleration; field keeps overshooting the potential minimum. Thereby, the beautiful model gets killed by local gravity constraints which play a crucial role for testing models based upon direct coupling to matter. Keeping this in mind, we have included  a sub-section dealing with local gravity constraints which, in particular, give rise to the bound on the field mass,  in the symmetron model. 

We then review a model based upon $\phi$ coupling to massive neutrino matter proportional to the trace of its energy momentum tensor, $ T_\nu$. By virtue of their small masses, massive neutrinos are relativistic at early times and coupling vanishes. Coupling automatically builds up  at late stages as neutrinos turn non-relativistic. Let us mention that local gravity constraints do not apply to neutrino matter. In this case, mass of field, after symmetry breaking, gets naturally connected to neutrino masses such that $m\sim \alpha \Omega^{(0)}_\nu H_0$ ($\alpha$ fixes the cut off mass scale in our case) thereby, with little adjustment of $\alpha$, one can achieve the desired bound on $m$, namely, $m\lesssim H_0$ or slow roll around the true ground state,  evading  "NO GO"  to late acceleration faced by the symmetron model, see Fig.$\ref{weff}$.
The distinguished and generic feature of this proposal is related to the physical process of turning massive neutrinos to non-relativistic at late times which triggers  the breaking of $Z_2$ symmetry in the low density regime. 

Keeping in mind the electroweak phase transition, we included detailed discussion on first order phase transitions in the framework 
of Ginzburg-Landau theory, see section \ref{1order}. One of the distinguished feature of the first order phase transition is related to jump of the order parameter at the critical temperature $T_c$ which is linked to the jump of entropy. The first order phase transition proceeds through bubble nucleation followed by a non-equilibrium situation (entropy is generated during the phase transition) whose strength depends upon the size of the order parameter at $T_c$, see Eq.(\ref{entropy}).
We carried over the formalism  to  one loop finite temperature effective potential in electroweak theory and provided a pedagogical exposition of electroweak phase transition (EWPT) dynamics.
In electroweak theory, the sphaleron transitions give rise to baryon number violation. However, they are in equilibrium for $T > T_c$  such that the net asymmetry produced by them vanishes. The sphaleron transitions would continue to be in equilibrium as Universe cools further until and unless something extra ordinary happens there.  We mean the first order phase transition which could do the needful here.
Using the effective potential given by Eq.(\ref{VT}) or Eq.(\ref{VTF}) (which is same as Gibbs potential (\ref{GL1O}) provided the coefficients (a,b,c) are adapted from the standard model), we demonstrated that EWPT is a first order transition (see Fig.\ref{firstorder})) that takes place for $T=T_c\simeq 150$GeV . And this could provide us with an opportunity to push out the sphaleron transitions from equilibrium allowing to generate the required baryon asymmetry. In order to accomplish the said mission, the EWPT should be strong enough or satisfy the bound, namely, $\langle \Phi \rangle_{T_c}/T_c \gtrsim 1$ dubbed {\it bound for strongly first order phase transition}\footnote{We recall that $\langle \Phi \rangle_{T_c}/T_c $ is a measure of non-equilibrium which is vanishingly small for  a system close to  equilibrium }  but it misses this bound by a factor of seven or so. This led us to emphasize  the need to go beyond the standard model to satisfy the criteria of strongly first order phase transition, see Eq.(\ref{gammaext}).
 The failure of electroweak baryogenesis clearly suggests that the standard model of particle physics is incomplete and there is a need to look beyond. To this effect, we discussed
 two mechanisms for generation of baryon asymmetry known as
 Affleck-Dine mechanism and spontaneous baryogenesis which are inspired by grand unified theories.
 
Keeping in mind the pedagogical commitments, we have included a brief discussion on selected aspects of $SU(2)_L\times U_Y$ theory which is heavily based upon the understanding of Abelian Higgs model that we discussed in great detail in section \ref{Abelian}. 
We sincerely hope that the review would be helpful to young researchers working on the applications of spontaneous symmetry breaking to early and late Universe.

\section{Acknowledgements}
We are  indebted  to R. Kaul, Elena Arbuzova, R. Rangarajan, S. Nojiri and Rathin Adhikari for  their  comments on the draft. We  thank  S. Odintsov, A. Nizami, Shibesh Kumar,
Nur Jaman, Ron-gen Cai, Sang Pyo Kim, Anzhong Wang, Roy maartens, S. Capozziello, V. Soni, M. Azam and Marcus Högås for useful discussions. MS is supported by the Ministry of Education and Science of the Republic of Kazakhstan, Grant
No. 0118RK00935 and by NASI-Senior Scientist Platinum Jubilee Fellowship(2021).

\appendix
\section{Essential ingredients for baryogenesis {\it \`{a} la} Sakharov }
\label{SakharovE}

The baryon asymmetry\footnote{A non-exhaustive list of 44 different ways to create baryons in the Universe has been listed in \cite{Shaposhnikov:2009zzb}.} in the Universe can be defined as the baryon to photon ratio which remained constant during evolution at late stages (after neutrinos decoupled and positrons annihilated). The magic number, 
\begin{equation}
\eta\equiv \frac{n_b-n_{\bar{b}}}{n_\gamma}\equiv \frac{n_B}{n_\gamma}\simeq 6\times 10^{-10}
\end{equation}
whose origin yet remains to unveiled, has been a crucial input to the hot big bang nucleosynthesis and its estimate has been independently confirmed by  the
measurements of CMB temperature fluctuations. 
Since entropy density is a conserved quantity, it is certainly a better measure to be used for quantifying the baryon asymmetry\footnote{Since,
$s=(2\pi^2/45)g_{*s}(T)T^3 $ and $n_\gamma=(2\zeta(3)/\pi^2)T^3$; the conversion factor is given by, $(\pi^4g_{*s}(T)/45 \zeta(3))\simeq 7$ where we used $g_*(T)=3.909$ valid after neutrino decoupling and positron annihilation when $g_*\simeq 3.3$ ; before this epoch  $g_{*s}\simeq g_*$. Here, $g_{*s}$ and  $g_*$ denote the effective number of entropy degrees of freedom and effective number of relativistic degrees of freedom.},
\begin{equation}
Y_B\equiv \frac{n_b-n_{\bar{b}}}{s} \simeq \frac{\eta}{7}  
\end{equation}
where the conversion is valid at late times after neutrino decoupling and positrons annihilation.
For the sake of brevity, we shall use only 
one definition for asymmetry,
\begin{equation}
\label{etas}
\eta\equiv
\frac{n_B-n_{\bar{B}}}{s}\simeq  8\times 10^{-11}    
\end{equation}
Since, in the present Universe, antimatter is absent to great accuracy, $\eta\simeq n_b/s $. The fact that the Universe is baryon asymmetric now can be attributed to it being thus from the beginning or to early Universe baryon number violating phenomena. The first alternative appears to be unappealing, because any such mismatch would have been obliterated by inflation if it had existed. There are a plethora of ways to induce baryon asymmetry, but they all must satisfy the three necessary conditions set forth by Sakharov in 1967\cite{Sakharov}(see Ref.\cite{Jim} for details),\\
(1) Baryon (lepton) number violation.\\
(2) C and CP violation.\\
(3) Departure from thermal equilibrium.

 The third condition is perhaps most important as its absence can undo the outcome of the first two. A remark is in order about thermal equilibrium in the early Universe. Early Universe was hot composed of plasma of relativistic particles. Universe was expanding very fast with frequent interactions of its constituents. Clearly, here, we have two relevant rates to compare with for deciding  the onset of equilibrium, namely, the Hubble expansion rate $H(t)$ and the interaction or reaction rate of species, $\Gamma(t)$. The time when $\Gamma$ takes over the Hubble rate marks the commencement of equilibrium in the early Universe,
 \begin{equation}
 \label{eqc}
\Gamma=n\sigma \sim \alpha_w^2 T  >H\sim \frac{T^2  }{\Mpl^2}   \Rightarrow T\lesssim \alpha_w^2 \Mpl\simeq 10^{16} GeV~~(\alpha_w^2 =g^2/4\pi\simeq 1/30),
 \end{equation}
 where $n$ is the number density of relativistic species at temperature $T$; $\sigma$ is the   two body scattering cross section and $g$ refers to the gauge coupling. At lower temperatures, processes might fall out
 of equilibrium due to Boltzmann suppression. For instance, let us consider the decay of a massive particle $X$ with mass $M_X$:  $X\to a+b$. For $T\lesssim M_X$, the reverse reaction rate would be exponentially suppressed, ($\Gamma(a+b\to X)\sim e^{-M_X/T})$; thermal energy in this case is not sufficient to create back the massive particle $X$. At lower temperatures there will be other species and corresponding interaction to support the thermal  equilibrium and so on.

The first Sakharov condition is self evident. As for the second, let us consider the following process that does not conserve baryon number $B$,
 \begin{equation}
 \label{CV}
 X\, \to \, Y+B;~~ ~~C:    \bar{X}\, ~\to \, \bar{Y}+\bar{B}
 \end{equation}
 where by definition, $X$ and $Y$ carry no baryon number.
 Violation of $C$ in context with (\ref{CV}) implies that,
 \begin{equation}
\Gamma(  X\, \to \, Y+B)\neq \Gamma( \bar{X}\, ~\to \, \bar{Y}+\bar{B})  
 \end{equation}
 otherwise the produced excess of baryons would be balanced by the excess of antibaryons.
 Let us confirm that the violation of $C$ alone is not enough, one also needs $CP$ violation. As for $CP$, since it deals with parity along with charge conjugation, we consider the decays involving the left handed and right handed quarks,
 \begin{eqnarray}
&& C:~ q_{L}  \, \to \,  \bar{q_{L}} ; ~~  q_{R}  \, \to \,  \bar{q_{R}}  \nonumber\\
&& \Gamma(X\,\to\, q_L q_L)\neq \Gamma(\bar{X}\,\to\, \bar{q}_L \bar{q}_L)\\
 && \Gamma(X\,\to\, q_R q_R)\neq \Gamma(\bar{X}\,\to\, \bar{q}_R \bar{q}_R)
 \label{CPV})
 \end{eqnarray}
Under $CP$, left handed quark transforms into  right handed and vice versa. Conservation of $CP$  implies that\cite{Jim},
\begin{eqnarray}
 && CP:~q_L\,\to\, \bar{q}_R;~  q_R\,\to\, \bar{q}_L \nonumber \\
&& \Gamma(X\,\to q_Lq_L)= \Gamma(\bar{X}\,\to \bar{q}_R \bar{q}_R)  \nonumber\\
&&\Gamma(X\,\to q_Rq_R)= \Gamma(\bar{X}\,\to \bar{q}_L \bar{q}_L)  \nonumber\\
&& \Gamma(X\,\to q_Lq_L)+\Gamma(X\,\to q_Rq_R)=\Gamma(\bar{X}\,\to \bar{q}_R \bar{q}_R)+\Gamma(\bar{X}\,\to \bar{q}_L \bar{q}_L)
\label{CPC}
\end{eqnarray}
 The last equation means that the excess of left handed quarks over left handed antiquarks  would be balanced by excess of right handed antiquarks over right handed quarks. And no asymmetry can be generated in such a situation. Thereby one needs to violate $CP$ also. However, there is caveat here. As we understand, we need to violate both $C$ and $CP$ such that,
 \begin{equation}
  \Gamma(X\,\to qq)\neq \Gamma(\bar{X}\,\to \bar{q} {q})   
 \end{equation}
 Let us assume that initially, $n_X=n_{\bar{X}}$ and all the $X^{'s}$ and $\bar{X}^{'s}$ have decayed giving rise to same number of quarks ans antiquarks; same logic applies to (\ref{CPC}) also. Thus in this case, even after both the $C$ and $CP$ are violated, one does not generate  asymmetry. In order to avoid such a situation, one has to admit at least one partial decay channel of $X$ with baryon number violation  different from $X\to qq$, for instance, $X\,\to\, {q}\bar{l}$ where $l$ refers to lepton. Let us now assume that,
 \begin{equation}
 X\,\to\, A_1+B_1;~~X\,\to\, A_2+B_2~~(B_1\neq B_2);~~(\text{In particular} ~A_1\equiv qq~\text{and}~A_2\equiv q\bar{l})    
 \end{equation}
such that their partial widths are different. For the sake of convenience, the decay widths can be cast in the following form\cite{AD,Jim},
\begin{eqnarray}
&& \Gamma(X\,\to\, A_1+B_1)=\Gamma_1+\epsilon\Delta \Gamma;~ \Gamma(X\,\to\, A_2+B_2)=\Gamma_2-\epsilon\Delta \Gamma\\
&&\Gamma(\bar{X}\,\to\, \bar{A}_1-B_1)=\Gamma_1-\epsilon\Delta \Gamma;~ \Gamma(\bar{X}\,\to\, \bar{A}_2-B_2)=\Gamma_2+\epsilon\Delta \Gamma
\end{eqnarray}
 which respects $CPT$ as $\Gamma_X=\Gamma_{\bar{X}}=\Gamma_1+\Gamma_2$ and non-vanishing value of $\epsilon$ implies the violation of $C$ and $CP$. The decay of  all the $X^{'s}$ and $\bar{X}^{'s}$ will generate net baryon number,
 \begin{equation}
\Delta B=B_1\frac{\Gamma_1+\epsilon\Delta \Gamma}{\Gamma_X} +B_2\frac{\Gamma_2-\epsilon\Delta \Gamma}{\Gamma_X} 
-B_1\frac{\Gamma_1-\epsilon\Delta \Gamma}{\Gamma_X}-B_2\frac{\Gamma_1+\epsilon\Delta \Gamma}{\Gamma_X}=2\epsilon
\left(\frac{\Delta \Gamma}{\Gamma_X}\right)(B_1-B_2)
 \end{equation}
which apart from the requirement of $C$ and $CP$ violations tells us that we also need a competing channel which generates  baryon number different from the original channel, namely, $B_1\neq B_2$. However, in absence, of a non-equilibrium process, the outcome due to $C$ and $CP$ violations would be erased.
Indeed, if the processes: $X\to A_1+B_1$ and $X\to A_2+B_2$ along with their conjugate processes\footnote{Which take place with same probability as that of the original processes
thanks to $CPT$ invariance}
remain in equilibrium, the asymmetry generated by them would be erased by the reverse reactions which run with same probability as the forward processes. Thereby these processes need to be decoupled or get out from equilibrium to freeze the asymmetry. In a sense, the third Sakharov condition is most important as its absence would undo the outcome of the first two.

Let us formally demonstrate that the thermal average of baryon number operator vanishes identically in the equilibrium  state. A system in equilibrium is described by a density operator,
\begin{equation}
\hat{\rho} =e^{-\hat{H}/T}   
\end{equation}
 where quantities with overhead hats refer to operators in the notation of Landau.
 Let us note that the thermal average of an operator should be time independent in equilibrium state. Indeed,
 \begin{eqnarray}
\hat{B}(t) &&=e^{-i\hat{H}t}\hat{B}(0) e^{i\hat{H}t}\\
\langle\hat{B}(t)\rangle_T &&=Tr(e^{-\hat{H}/T} e^{i\hat{H}}\hat{B}(0)  e^{-i\hat{H}t}  )=Tr(e^{-i\hat{H}t} e^{-\hat{H}/T} e^{i\hat{H}}\hat{B}(0)) \nonumber\\
&&=Tr( e^{-\hat{H}/T} \hat{B}(0))=\langle\hat{B}(0)\rangle_T ,
 \end{eqnarray}
 which implies that we can ignore the time dependence of the operator $\hat{B}$ while computing the thermal average.
As $CPT$ is an exact symmetry of the theory\footnote{We do not put hats on $C,P$ and $T$. Rigorously speaking, when we deal with $C$, $CP$ or $CPT$ in relation to $\hat{H}$ or $\hat{B}$, we should promote the former  to their corresponding operators in the Hilbert space, for instance, $CP\to \hat{U}_{CP}$. nonetheless, for brevity, we did not  introduce  new notations.   }, it should commute with the Hamiltonian operator, $[\hat{H}, CPT]=0$. Further, let us also emphasize that the baryon number operator $\hat{B}$ is both, $C$ and $CP$ odd or it anti-commutes with them. With this information at hand, we can prove the assertion by few manipulations on $\langle\hat{B}\rangle_T$,
\begin{eqnarray}
 \langle \hat{B}\rangle_T &&=Tr(e^{-\hat{H}/T} \hat{B})=Tr\left((CPT)^{-1}(CPT)e^{-H/T} \hat{B} \right)  \nonumber \\
 &&=Tr \left(e^{-\hat{H}/T} (CPT)\hat{B}(CPT)^{-1}     \right)=-\langle \hat{B}\rangle_T\Rightarrow \langle\hat{B}\rangle_T\equiv 0
 \label{thermalav}
\end{eqnarray}
where we used the property of trace and the fact that $CPT$ also commutes with any function of the Hamiltonian operator, $\hat{H}$ and that the baryon number is $CP$ odd. Eq.(\ref{thermalav}) proves our assertion that was based upon intuition that asymmetry can not be generated in equilibrium as long as we adhere to $CPT$ invariance. 

\section{The one loop finite temperature effective potential}
\label{effpot}
The effective potential has 3 components. The original potential, $V_{tree}$, the 1-loop contribution at zero temperature known as the Coleman-Weinberg potential and finally a 1-loop contribution at finite temperature,
\begin{align}
    V_{\rm eff}=V_{\rm tree}+V_{\rm 1,T=0}+V_{\rm 1,T}
\end{align}
We have
\begin{align}
    V_{\rm tree}=-\frac{\mu^2}{2}\phi^2+\frac{\lambda}{4}\phi^4
\end{align}
The 1-loop contribution at zero temperature is
\begin{align}
    V_{\rm 1,T=0}=\frac{1}{64\pi^2}\sum_i \pm m_i^4(\phi)\Bigl[\ln \frac{m_i^2(\phi)}{\bar\mu^2}-\frac{3}{2}\Bigr]
\end{align}
where the sum runs over all degrees of freedom which couple to the Higgs field, while $\pm$ represents the bosonic and fermionic contributions respectively. Finally the 1-loop thermal contribution is (see Refs.\cite{Dolan:1973qd,Quiros:1999jp} for more details)
\begin{align}
    V_{\rm 1,T}&=\pm \frac{T}{(2\pi)^3}\sum_i\int {\rm d}^3p~\ln \Bigl(1\mp e^{-\sqrt{p^2+m_i^2(\phi)}/T}\Bigr)\\
    &=\pm \frac{T^4}{2\pi^2}\sum_i\int_0^\infty {\rm d}x ~x^2 \ln \Bigl(1\mp e^{-\sqrt{x^2+m_i^2(\phi)/T^2}}\Bigr)
\end{align}
At high temperatures, this last expression can be approximated
\begin{align}
    \int_0^\infty {\rm d}x ~x^2 \ln \Bigl(1- e^{-\sqrt{x^2+m_i^2(\phi)/T^2}}\Bigr)&\simeq -\frac{\pi^4}{45}+\frac{\pi^2}{12}\frac{m_i^2(\phi)}{T^2}-\frac{\pi}{6}\Bigl(\frac{m_i^2(\phi)}{T^2}\Bigr)^{3/2}\nonumber\\
    &-\frac{m_i^4(\phi)}{32T^4}\ln \frac{m_i^2(\phi)}{a_b T^2}
\end{align}

\begin{align}
    \int_0^\infty {\rm d}x ~x^2 \ln \Bigl(1+ e^{-\sqrt{x^2+m_i^2(\phi)/T^2}}\Bigr)\simeq \frac{7\pi^4}{360}-\frac{\pi^2}{24}\frac{m_i^2(\phi)}{T^2}-\frac{m_i^4(\phi)}{32T^4}\ln \frac{m_i^2(\phi)}{a_f T^2}
\end{align}
where we defined $a_b=16\pi^2e^{3/2-2\gamma_E}$, $a_f=\pi^2 e^{3/2-2\gamma_E}$ and $\gamma_E=0.5772$ which is the Euler-Mascheroni constant.

We see that only bosons will contribute to terms linear  in $T$  while fermions and bosons contribute to $T^2$ term respectively in the effective potential. At the leading orders, and at high temperature, the 1-loop Coleman-Weinberg contribution is negligible, we have
\begin{align}
    V_{\rm eff}\simeq -\frac{\mu^2}{2}\phi^2+\frac{\lambda}{4}\phi^4 +\frac{T^2}{48}\Bigl[\sum_{i=\text{bosons}}2m_i^2(\phi)+\sum_{i=\text{fermions}}m_i^2(\phi)\Bigr]-\frac{T}{12\pi}\sum_{i=\text{bosons}}m_i^3(\phi)
\end{align}
where we neglected an irrelevant constant, $-\frac{\pi^2}{48}T^4$.
In models based upon symmetry breaking, particles acquire masses due to scalar field,
\begin{equation}
 m_i(\phi)=g_i \phi ,
\end{equation}
where $g_i^{'s}$ are coupling constants. This relation acquires the following form in the standard model,
\begin{equation}
M_W= \frac{g}{\sqrt{2}}
\langle \phi \rangle_T  ;~~M_Z=\frac{\sqrt{g^2+g'^2}}{\sqrt{2}} \langle \phi \rangle_T ;~~ m_F={f}\langle \phi \rangle_T
\end{equation}
where $f$ is Yukawa coupling and ($g,g'$) are gauge couplings. It should be kept in mind that particle masses are temperature dependent as vacuum expectation value of the field  $\phi$ depends upon temperature.

\section{{Sphaleron}}
\label{Appsphaleron}

\begin{figure}[h]
\centering
\includegraphics[scale=.5]{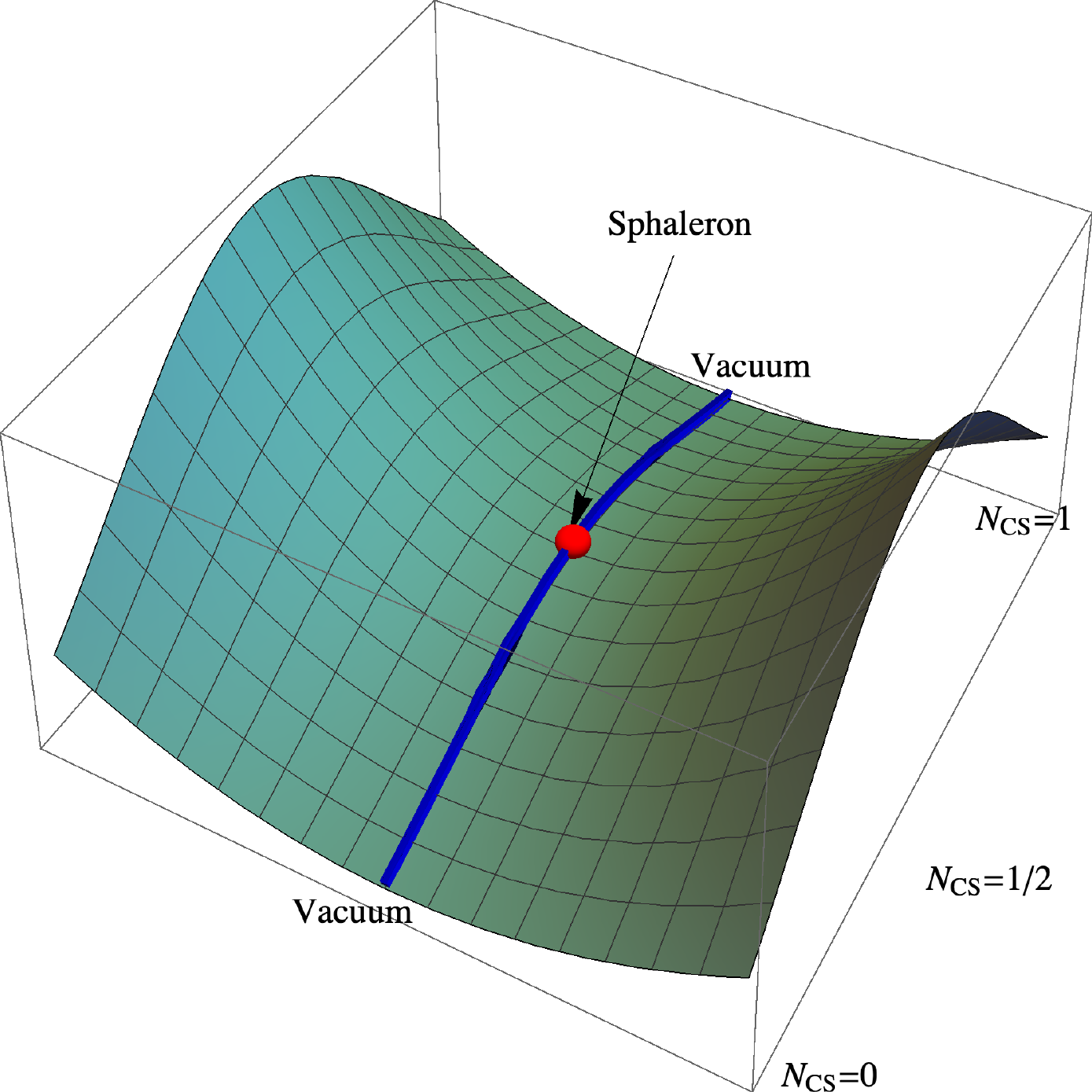}
\caption{The sphaleron corresponds in the configuration space to a saddle point along the least energy path connecting two vacua with $\Delta N_{CS}=1$}
\label{fig:sphaleron2}
\end{figure}

The sphaleron corresponds to a configuration of the gauge fields in such away that $N_{CS}=1/2$ and situated on the least energy path that connects the two different vacua with different Chern-Simons numbers. This configuration is located on a saddle point and at the top of the barrier (see Fig. \ref{fig:sphaleron2})

Classically, baryon number (B) and lepton number (L) currents are conserved in the electroweak theory but conservation is violated at the quantum level. The anomaly in baryon(lepton) number current is related to the fact that all quarks and leptons couple only to left handed $SU(2)$ gauge fields to reconcile with parity violation in weak interactions,
\begin{eqnarray}
&& \partial_\mu J^\mu_{B(L)}=N_F \frac{g^2}{32\pi^2} F^{\mu \nu}_a  \tilde{F}_{\mu \nu}^a =N_F\partial _\mu K^\mu \\
&& \tilde{F}^a_{\mu\nu}\equiv \frac{1}{2}F_a^{\rho\sigma}
\epsilon_{\mu\nu\rho\sigma}\\
&& K^\mu =\frac{g^2}{32 \pi^2} \epsilon^{\mu\nu\rho\sigma}\left(F^a_{\nu\rho} A^a_\sigma-\frac{g}{3} \epsilon_{abc}A^a_\nu A^b_\rho A^c_\sigma    \right)
\end{eqnarray}
where  $\epsilon$'s denote fully anti-symmetric Levi-Civita symbols (there is no difference in upper and lower indices with regard to the group indices, a,b and c) and $N_F$ denotes the number of fermion generation which is three in case of the standard model. It is interesting to note that $\partial_\mu (J^\mu_B- J^\mu_L)=0$ or $B-L$ is  a conserved quantity in the standard model.
Making an appropriate choice of gauge, one finds after some manipulations,
\begin{eqnarray}
   && \partial_t B(L)=N_F\partial_t N_{CS}\\
   && B(L)=\int{d^3xJ_0^{B(L)}};~~~N_{CS}=\int{d^3xK_0}
\end{eqnarray}
where $N_{CS}$ is a winding number dubbed Chern-Simons number which can take both positive and negative values. Thus the gauge fields, that evolve changing the winding number, would violate B(L) conservation,
\begin{equation}
   \Delta B \equiv B(t_f)-B(t_i)=\Delta L= N_F \Delta N_{CS}\equiv N_F(N_{CS}(t_f)-N_{CS}(t_i))
\end{equation}
 The B(L) violation is due to the vacuum structure of non-abelian gauge which is infinitely degenerate labelled by $N_{CS}$ such the adjacent vacua are separated by a finite barrier of energy $\mathcal{O}(M_W/\alpha_W)$ (see discussion below).
 The vacuum to vacuum transition changes $\Delta B$ and $\Delta L$ by multiples of 3 units;  't Hooft \cite{tHooft:1976snw} was first to interpret instantons as tunneling between vacuum states differing by a unit of topological charge ($N_{CS})$. Instantons, therefore are the excellent candidates to describe baryon number violation. The euclidean action evaluated at this trajectory gives the probability for this transition between vacua of different topological numbers
\begin{align}
    \Gamma \sim e^{-4\pi/\alpha_W}
\end{align}
where $\alpha_W=g^2/4\pi$, which gives $\Gamma \sim 10^{-162}$, see  Ref.\cite{tHooft:1976snw,tHooft:1976rip}. Therefore this process is exponentially suppressed. Interestingly, it was noticed \cite{Kuzmin:1985mm} that at finite temperature, instead of a tunneling process, the particle could go over the barrier by thermal fluctuations. If the temperature is large enough, the transition rate in the electroweak interaction is given by the sphaleron configuration \cite{Manton:1983nd,Klinkhamer:1984di}, which correspond to a static, unstable and finite energy solution. They are stationary point of the energy, and possess Chern-Simons number equal to $1/2$ (see Fig.(\ref{sphaleron})). In a simple interpretation,
the field configurations in the early universe at high temperature, can form a sphaleron which would subsequently decay into a vacuum state but with a different winding number.

For a simple understanding of the sphaleron configuration, let us consider the O$(3)$ sigma model in 2 dimensions defined by 
\begin{align}
    S=\frac{1}{2g^2}\int d^2x \partial_\mu \phi^a\partial^\mu \phi^a
\end{align}
with the restriction $\phi^a \phi^a=1$. Therefore, $\phi$ is a vector which takes values on a 2-sphere, $\mathbb{S}^2$. We can therefore see $\phi$ as an application  $\mathbb{R}^2 \rightarrow \mathbb{S}^2$. But to have a solution with finite energy, we need $\phi$ to be zero at infinity. The field should approach the same vacuum regardless of the direction. That implies, that all the points at infinity are similar and can be considered as the same point. All these points at infinity should be identified, transforming therefore $\mathbb{R}^2$ into $\mathbb{S}^2$. In conclusion, the field becomes a map $\phi: ~\mathbb{S}^2\rightarrow \mathbb{S}^2$ from a physical space to a target space. These mapings are classified by the homotopy group (in this case the second homotopy group) $\pi_2(\mathbb{S}^2)=\mathbb{Z}$ known as the winding number. Therefore, we can find nontrivial solutions classified by their winding number $Q$ known also as the topological charge which indicates the number of wrappings of the physical space $\mathbb{S}^2$ onto the target space $\mathbb{S}^2$. This topological number can be related to the field $\phi$ by defining a conserved current, $j^\mu$, which is always conserved independently of the solution such as a Bianchi identity, or we can define it directly by defining the unit volume element in the target space, which is for a sphere $\int_{\mathbb{S}^2} \frac{1}{4\pi}\sin{\theta} d\theta \wedge d\varphi$ where $(\theta,\varphi)$ are some coordinates on $\mathbb{S}^2$ and $\wedge$ is the wedge product. The vector field $\phi$ can always be parameterized\footnote{Because of the restriction $\phi^a\phi^a=1$} as $\phi^a=(\sin\theta\cos\varphi,\sin\theta\sin\varphi,\cos\theta)$. It is easy to show with this parametrization that
\begin{align}
    \sin\theta d\theta\wedge d\varphi = (\phi^1,\phi^2,\phi^3).(d\phi^2\wedge d\phi^3,d\phi^3\wedge d\phi^1,d\phi^1\wedge d\phi^2)\equiv \vec\phi.\vec{dS}
\end{align}
After pulling back this expression on the physical space, we obtain the topological charge
\begin{align}
    Q=\frac{1}{4\pi}\int d^2 x ~\vec \phi.\Bigl(\frac{\partial \vec \phi}{\partial x}\times \frac{\partial \vec \phi}{\partial y}\Bigr)
    \label{topo}
\end{align}
Notice that the topological charge is independent of the Lagrangian chosen. This model admits instanton solutions \cite{Polyakov:1975yp,Polyakov:1975rr} but as we have said they will play very negligible role.  

Because in $1+1$ dimensions, we can't break spontaneously a continuous symmetry, it is difficult to obtain a sphaleron configuration. For that we will break explicitly the O$(3)$ symmetry in the action into O$(2)$ by adding an additional term
\begin{align}
    S=\frac{1}{g^2}\int d^2x ~\Bigl[\frac{1}{2}\partial_\mu \phi^a\partial^\mu \phi^a +\omega^2 (1-\phi_2)\Bigr]
\end{align}
We will also consider the following mapping \cite{McLerran:1988ja}
\begin{align}
    \vec\phi=\Bigl(\sin\mu\sin\theta,\sin^2\mu\cos\theta+\cos^2\mu,\sin\mu\cos\mu(\cos\theta-1)\Bigr)
\end{align}
It is easy to check that $\phi^a\phi_a=1$. The variable $\mu$ ranges in $[0,\pi]$ and $\theta$ in $[0,2\pi]$. At infinity, we should have $\phi_2=0$ in order to have a finite energy (zero potential at infinity). Finally, we can calculate the topological charge using eq.(\ref{topo})
\begin{align}
    Q=\frac{1}{4\pi}\int \sin{\mu}(1-\cos{\theta})  d\mu d\theta
    \label{charge}
\end{align}
Considering the previous range, we would find $Q=1$, therefore this mapping has topological winding number $+1$, which can be used for instanton solutions. We should here mention that for instantons, we perform a Wick rotation so time becomes a euclidean coordinate. To describe sphaleron, we will trade that euclidean time by a number, the Chern-Simons number. So the configuration is described instead of time and space but by a topological number and space. From our eq.(\ref{charge}), we will consider our $\theta$ variable as the coordinate over a circle, so the space variable, $\theta(x)$, which runs from $0$ to $2\pi$ while $\mu$ will run in a range in such a way $Q=1/2$. For that we need $\mu$ to run until $\pi/2$. To find our configuration, we will restrict $\mu$ to a constant and let $\theta$ be a function of the space coordinate $\theta(x)$, so the action becomes
\begin{align}
    S=\frac{A\sin^2{\mu}}{g^2}\int dx ~\Bigl[\frac{1}{2}\Bigl(\frac{d\theta}{dx}\Bigr)^2 +\omega^2 (1-\cos{\theta})\Bigr]
\end{align}
where $A$ comes from the integral over time. Here, the action is similar to the energy because the configuration is static. We see again that the energy has a maximum for $\mu=\pi/2$. Extremizing the action, we find the equation $\theta''(x)-\omega^2\sin{\theta}=0$ with solution 
\begin{align}
    \theta_{\text{sphaleron}}(x)=2\sin^{-1}{(\text{sech} ~\omega x)}
\end{align}
and an energy $E_{sp}=8\omega/g^2$. In the case of the electroweak theory, it is difficult to find an exact solution. But the structure is similar, we need to find the constraints for which the energy is finite, to find a suitable ansatz for the fields $(A_\mu,\Phi)$ consistent with the restriction. It has been shown that the $U(1)$ gauge fields do not contribute much and can be introduced perturbatively. Therefore, it is easier to find the sphaleron for pure $SU(2)$ gauge field. \cite{Manton:1983nd,Klinkhamer:1984di} (which means for a zero mixing angle). Later, calculations were performed taking into account the mixing angle also \cite{Kunz:1992uh}; the sphaleron energy  was found to be,
\begin{align}
    E_{sp}(T)=B \frac{2 m_W}{\alpha_W}\frac{\langle \phi(T)\rangle}{\langle \phi(0)\rangle}
\end{align}
where $B$ is a constant which depends on the characteristics of the Higgs. Numerically it is found that $1.5<B<2.7$, the last part $\langle \phi(T)\rangle$ comes from the finite temperature effect calculated in \cite{Brihaye:1993ud,Braibant:1993is,Moreno:1996zm}. This energy is important because it appears in the calculation of baryon number violation rate, that contains the Boltzmann factor $e^{-E_{sp}(T)/T}$ along with a prefactor \cite{Carson:1990jm}. From which, we can find the bound\cite{Bochkarev:1990gb},
\begin{align}
    \frac{E_{sp}(T_c)}{T_c} \gtrsim 45
\end{align}
that can be translated into a bound on $\phi$
\begin{align}
    \frac{\langle\phi(T_c)\rangle}{T_c} \gtrsim 1
\end{align}
{as a necessary condition for sphaleron transitions to be out of equilibrium }
This bound implies that the phase transition under consideration should be a strongly first order phase transition (for a second order phase transition $\phi(T_c)=0$). 

\section{Sphaleron in $SU(2)$}
\label{sphaleronsu2}
In this section, we briefly summarize, the solution\footnote{As it is mentioned in \cite{Rubakov:1996vz} the solution was previously found in \cite{Dashen:1974ck,Soni:1980ps,Boguta:1983xs,Forgacs:1983yu} but its importance in topology was discovered only in \cite{Manton:1983nd}} found in \cite{Manton:1983nd,Klinkhamer:1984di}.

Considering the Lagrangian 
\begin{align}
  \mathcal{L} = -\frac{1}{4}\text{Tr}[F_{\mu\nu}F^{\mu\nu}]+(D_\mu\Phi)^\dag D^\mu\Phi-\lambda \left(\Phi^\dag\Phi-\frac{v^2}{2}\right)^2
\end{align}
with
\begin{align}
    F_{\mu\nu}^a &=\partial_\mu A_\nu^a-\partial_\nu A_\mu ^a+g \epsilon^{abc}A_\mu^b A_\nu^c\\
    D_\mu\Phi &= \partial_\mu\Phi+\frac{i}{2} g \tau^a A_\mu^a\Phi
\end{align}
where $A_\mu^a$ is the gauge field related to the $SU(2)_L$ symmetry and $\Phi$ the Higgs field. Working in the gauge $A_0=0$, we have for the energy
\begin{align}
    E=\int {\rm d}^3x \Bigl[\frac{1}{4}\text{Tr}[F_{ij}F^{ij}]+(D_i\Phi)^\dag D^i\Phi+\lambda (\Phi^\dag\Phi-\frac{v^2}{2})^2\Bigr]
\end{align}
In order to have a finite energy, we need at infinity to impose that $\Phi,A_i$ take the vacuum configuration. A first solution is to consider $\Phi^\infty=\frac{v}{\sqrt{2}} \begin{pmatrix} 0 \\ 1 \end{pmatrix} $, which is $\Phi_{\text{vac}}$. Therefore $\Phi^\infty$ can be seen as a map from $S^2$ (space at infinity defined by the angles $(\theta,\phi)$), to the vacuum manifold of the Higgs field which is topologically equivalent to $S^3$. Hence, we have $\Phi^\infty:~S^2\rightarrow S^3$. Since the 2-dimensional homotopy group is trivial $\pi_2(S^3)=I$, this map contracts to a single point. They are no non-trivial static, finite energy, configurations. To construct a non-trivial structure, we need to introduce an additional parameter $\mu\in [0,\pi]$ and build the following map
\begin{align}
    \Phi^\infty(\mu,\theta,\phi)=\frac{v}{\sqrt{2}}\begin{pmatrix} \sin\mu\sin\theta e^{i\phi} \\ e^{-i\mu}(\cos\mu+i\sin\mu\cos\theta) \end{pmatrix} 
\end{align}
in such away that $\Phi^\infty(0,\theta,\phi)=\Phi^\infty(\pi,\theta,\phi)=\Phi_{\text{vac}} $. $\Phi^\infty(\mu)$ starts and ends at the vacuum configuration and most interestingly, it is not contractible because it is a family of maps $\Phi^\infty(\mu):~S^2\times S^1\rightarrow S^3$ which is equivalent to $S^3\rightarrow S^3$ and therefore the 3-dimensional homotopy group is non-trivial $\pi_3(S^3)=\mathbb{Z}$. From, the expression of the energy, we see that we need to define some conditions for the energy not to diverge, $D_\theta\Phi^\infty =D_\phi\Phi^\infty=0$, which translates into $\partial_i\Phi+igA_i\Phi=0$ ($i=(\theta,\phi)$). Taking $A_\mu$ in the vacuum configuration, i.e. $A_\mu=0$ or any gauge transformation of it, i.e. $A_\mu=\frac{i}{g}(\partial_\mu U)U^{-1}$. We get
\begin{align}
    A^\infty_\theta =\frac{i}{g}\partial_\theta U^\infty (U^\infty)^{-1}\,,~~~~~~~
A^\infty_\phi =\frac{i}{g}\partial_\phi U^\infty (U^\infty)^{-1}
\end{align}
while $A_r=0$ (radial gauge). We can write $\Phi^\infty = \frac{v}{\sqrt{2}}U^\infty \begin{pmatrix} 0 \\ 1 \end{pmatrix}$ where
\begin{align}
    U^\infty=\frac{\sqrt{2}}{v}\begin{pmatrix} \Phi_2^\infty* & \Phi_1^\infty\\ -\Phi_1^\infty* & \Phi_2^\infty \end{pmatrix}=\begin{pmatrix} e^{i\mu}(\cos\mu-i\sin\mu\cos\theta) & \sin\mu\sin\theta e^{i\phi}\\ -\sin\mu\sin\theta e^{-i\phi} & e^{-i\mu}(\cos\mu+i\sin\mu\cos\theta) \end{pmatrix}
\end{align}
where we defined $\Phi^\infty(\mu,\theta,\phi)=\begin{pmatrix} \Phi_1^\infty\\  \Phi_2^\infty \end{pmatrix}$

From this behavior at infinity, we can define an ansatz for the whole space
\begin{align}
    \Phi(\mu,r,\theta,\phi) &=(1-h(r))\begin{pmatrix} 0\\  e^{-i\mu}\cos\mu \end{pmatrix}+h(r)\Phi^\infty(\mu,\theta,\phi)\\
    A_\theta(\mu,r,\theta,\phi) &=f(r)A_\theta^\infty (\mu,\theta,\phi)\\
     A_\phi(\mu,r,\theta,\phi) &=f(r)A_\phi^\infty (\mu,\theta,\phi)
\end{align}
In order to recover, the right behavior at infinity, we impose $f(\infty)=h(\infty)=1$ and in order not to have a singular behavior in zero, we impose $f(0)=h(0)=0$. Inserting this ansatz in the energy, we obtain a function $E[\mu,\theta,\phi]$. For most choices of $(f,h)$, this functional reaches the maximum for $\mu=\pi/2$. For this value, of $\mu$, we look for $(f,h)$ which minimizes the energy, giving us the sphaleron configuration. Variation of $E$ with respect to $(f,g)$ gives
\begin{align}
    &r^2\frac{d^2f}{dr^2}-2 f(1-f)(1-2f)+\frac{g^2 r^2 v^2}{4}(1-f)h^2=0\\
    &\frac{d}{dr}\Bigl(r^2 \frac{d h}{dr}\Bigr)-2 h (1-f)^2+\lambda v^2 r^2 h (1-h^2)=0
\end{align}
\begin{figure}[h]
\centering
\includegraphics[scale=.4]{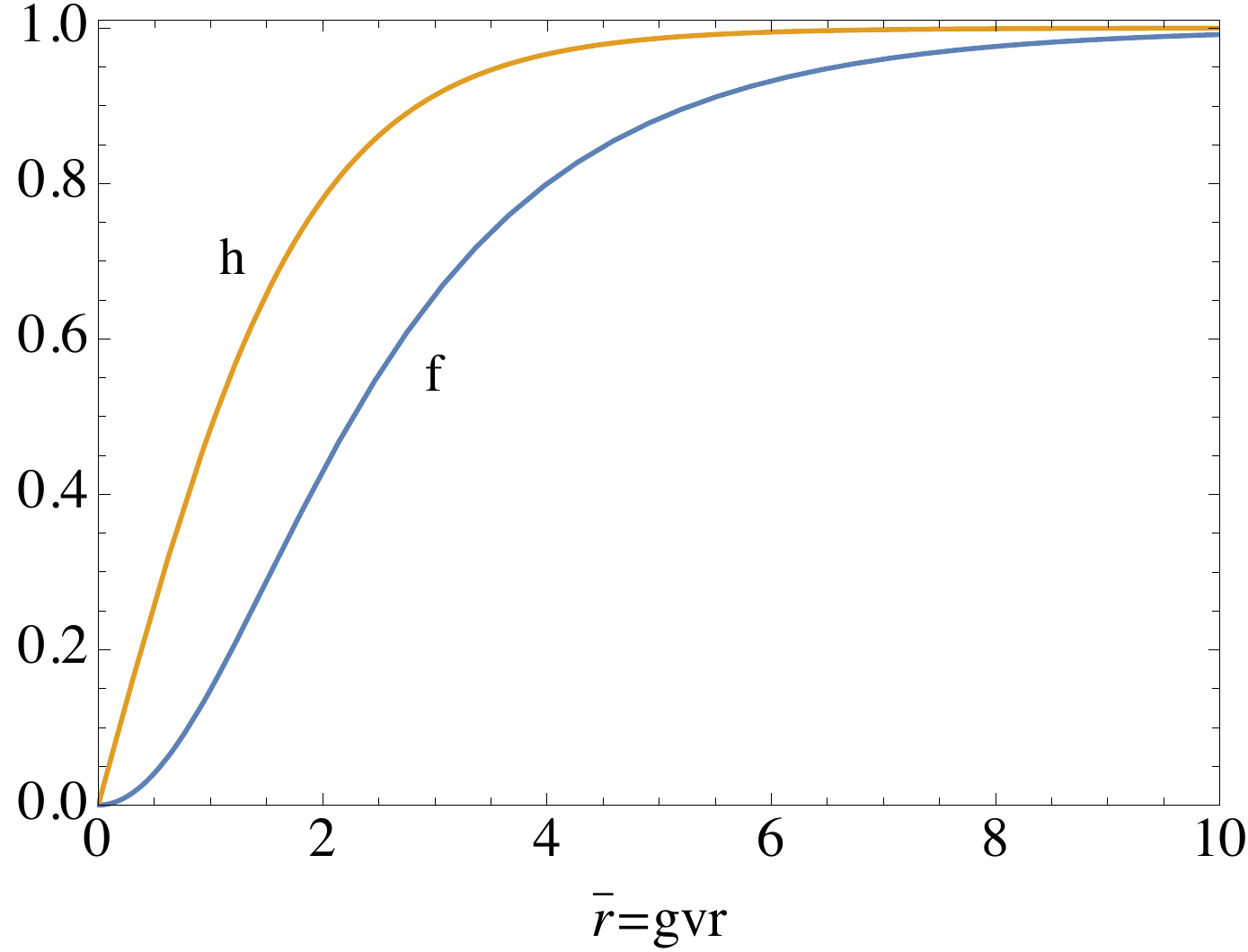}~~~~~
\includegraphics[scale=.42]{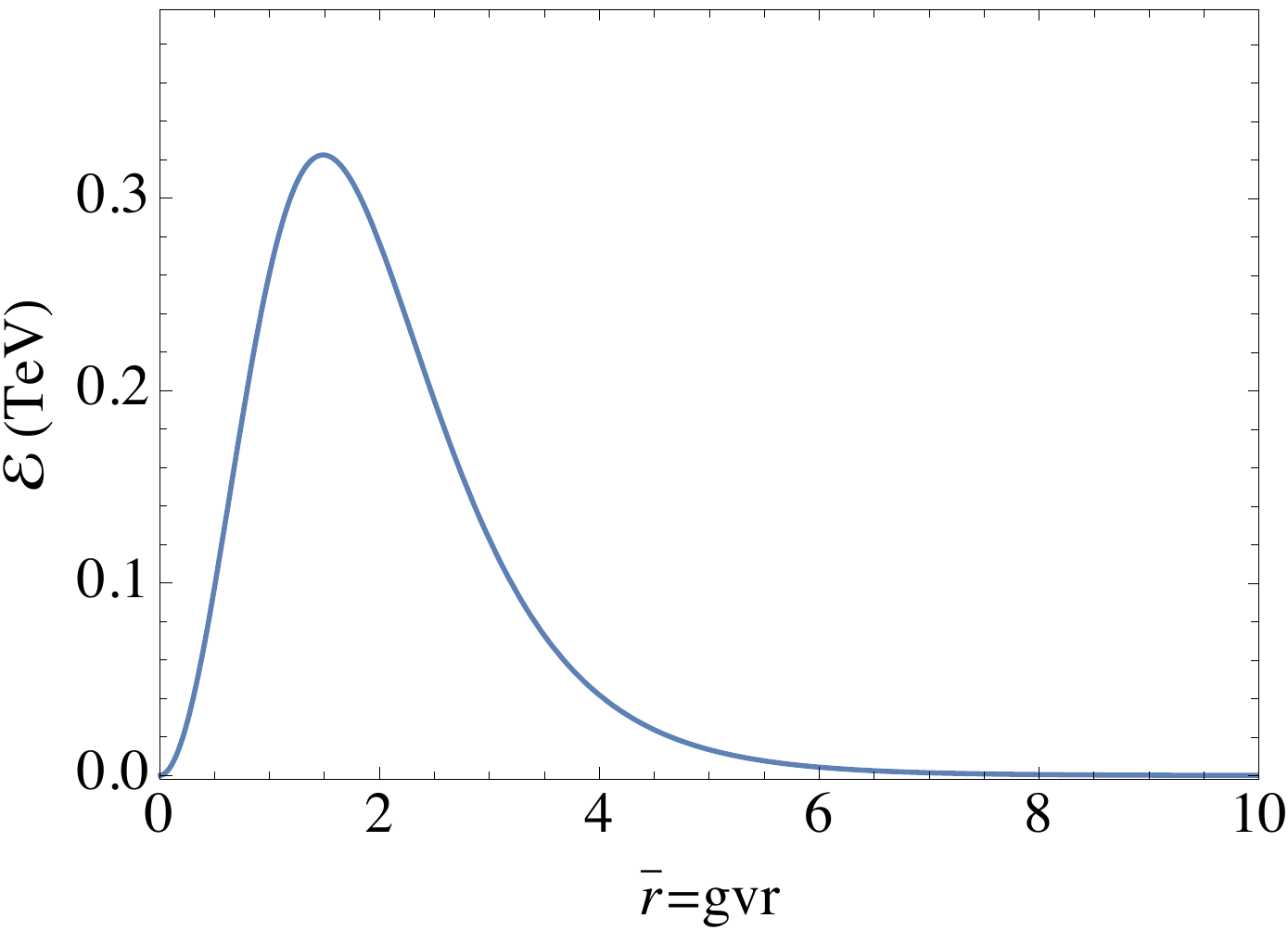}
\caption{Numerical profile of $(h(r),f(r))$ and the energy density of the sphaleron.} 
\end{figure}
These equations can be solved numerically. To match the asymptotic behavior, we found that initial conditions at $r=0$ should be of the order $f(r)\simeq 0.17 r^2$ and $h(r)\simeq 0.53 r$. From which we can obtain the energy of the sphaleron $E=4\pi\int_0^\infty \mathcal{E} {\rm d}r\simeq 9.09$ TeV where we used $v=246.22$ GeV, $m_W=80.379$ GeV and $m_H=125$ GeV from which we can obtain $g=2m_W/v$ and $\lambda=m_H^2/(2v^2)$.

\section{Dynamics of bubble nucleation}
\label{bubblenucleation}
{The first order phase transition proceeds through bubble nucleation. A system cooled below  critical temperature, will have small number of bubbles of the new phase or droplets, formed usually around some impurities. In the absence of these impurities, the system can be cooled to temperatures much below the critical temperature without any phase transition. The early Universe is supposed to be quite smooth, thereby we can not have this process there, except in the presence of primordial black holes. However, these bubbles could form by quantum or thermal fluctuations in some region of space. A bubble of the new phase, $\langle \phi \rangle \neq 0$ forms in the sea of the unbroken phase $\langle \phi\rangle =0$. As for the  quantum transitions due to instanton,  their probability is negligibly  small. These bubbles of the new phase appear  due to thermal fluctuations;  they grow if their  radius is larger than some critical value. Meanwhile,  different growing bubbles collide and merge and the process continues till the whole Universe is in the new phase}.

In this this appendix, we shall estimate the rate of bubble nucleation at a given temperature, in particular, at the critical temperature for growing bubbles.
It is well known  from quantum mechanics that the transition amplitude between 2 states described by a time independent Hamiltonian $H$ is given by,
\begin{align}
    \langle q_2,t| q_1,0\rangle=\langle q_2| e^{-iH t}|q_1\rangle = \sum_{n,m} \langle q_2|n\rangle \langle n|e^{-iH t}|m\rangle \langle m|q_1 \rangle=\sum_n \psi_n(q_2)\psi^*_n(q_1)e^{-i E_n t}
\end{align}
where we introduced the complete set of eigenvectors of the Hamiltonian, ${|n\rangle}$ such that, $H|n\rangle=E_n |n\rangle$. Also we defined, $\psi_n(q_i)\equiv \langle q_i|n\rangle$. If we  analytically continue this expression to Euclidean time, i.e. $t\rightarrow -i\beta$ also known as a Wick rotation, we obtain
\begin{align}
    \langle q_2,\beta| q_1,0\rangle=\sum_n \psi_n(q_2)\psi^*_n(q_1)e^{-\beta E_n }
    \label{partition}
\end{align}
We shall use the normalization condition, 
\begin{align}
   \int dq \psi_n(q)\psi^*_n(q)=1
\end{align}
We can take $q_2=q_1\equiv q$ and integrate Eq.(\ref{partition}) over $q$,
\begin{align}
    \int dq \langle q,\beta| q,0\rangle=\sum_n e^{-\beta E_n }\equiv Z(\beta),
\end{align}
where $Z$ is the partition function and $\beta=1/T$. This expression relates the statistical partition function to quantum mechanics, by considering closed paths of Euclidean time length $\beta$. But because in quantum field theory, the partition function is defined by the integral of $e^{iS}$ where $S$ is the action, we obtain
\begin{align}
    Z=\int \mathcal{D}q e^{-I}
\end{align}
where $I$ is the euclidean action and the integral is taken over all closed paths of Euclidean time  $\beta$. Therefore field theory at finite temperature is equivalent to Euclidean field theory, periodic in imaginary time.
In our case, we have
\begin{align}
    I=\int_0^\beta d\tau \int d^3x\Bigl[\frac{1}{2}\Bigl(\frac{\partial\phi}{\partial\tau}\Bigr)^2+\frac{1}{2}(\vec{\nabla}\phi)^2+V(\phi,T)\Bigr]
    \label{euclidean}
\end{align}
which at high temperature can be approximated as,
\begin{align}
    I\simeq \frac{1}{T}\int d^3x\Bigl[\frac{1}{2}(\vec{\nabla}\phi)^2+V(\phi,T)\Bigr]\equiv \frac{I_3}{T}
        \label{euclidean2}
\end{align}
The starting point here is to evaluate the partition function. For that, we use the saddle point approximation which means identifying the stationary point of the exponential which are dominant. That corresponds to $\delta I_3/\delta\phi=0$ which means $\phi$ should be solution of
\begin{align}
    \Box\phi-\frac{dV(\phi,T)}{d\phi}=0
\end{align}
Solutions of this equation which minimize $I_3$ (that corresponds to the energy of our system) turns out to be the radial configuration
\begin{align}
    \frac{d^2\phi}{dr^2}+\frac{2}{r}\frac{d\phi}{dr}-\frac{dV(\phi,T)}{d\phi}=0
\end{align}
We have neglected the expansion of the universe by assuming that the time of nucleation is much faster than the Hubble time scale $1/H$. Using potential (\ref{VTFC1}), we get
\begin{align}
    \frac{d^2\phi}{dr^2}+\frac{2}{r}\frac{d\phi}{dr}-\beta (T^2-T_0^2)\phi+\gamma T \phi^2-\lambda \phi^3=0
\end{align}
where the coefficient $\beta$ should not be confused with the temperature. We have natural units associated to the problem, we define $\phi_+$ as the value of the scalar field for the global minimum of $V(\phi,T)$ when $T<T_c$ and $r_0=1/\sqrt{\beta(T^2-T_0^2)}$ which is related to the mass scale. From which, we define the normalized parameters \cite{Cutting:2020nla}
$\bar r=r/r_0$ and $\bar \phi=\phi/\phi_+$. We get
\begin{align}
    \frac{d^2\bar\phi}{d\bar r^2}+\frac{2}{\bar r}\frac{d\bar\phi}{d\bar r}-\bar\phi+\frac{9(1+\sqrt{1-8\bar\lambda/9})}{4\bar\lambda} \bar\phi^2-\frac{9(1+\sqrt{1-8\bar\lambda/9})^2}{8\bar\lambda} \bar\phi^3=0
    \label{eq:kink}
\end{align}
where we see that the equation depends only on one parameter 
\begin{align}
    \bar \lambda = \frac{9\lambda \beta (T^2-T_0^2)}{2\gamma^2 T^2}
\end{align}
\begin{figure}[ht]
\centering
\includegraphics[scale=.4]{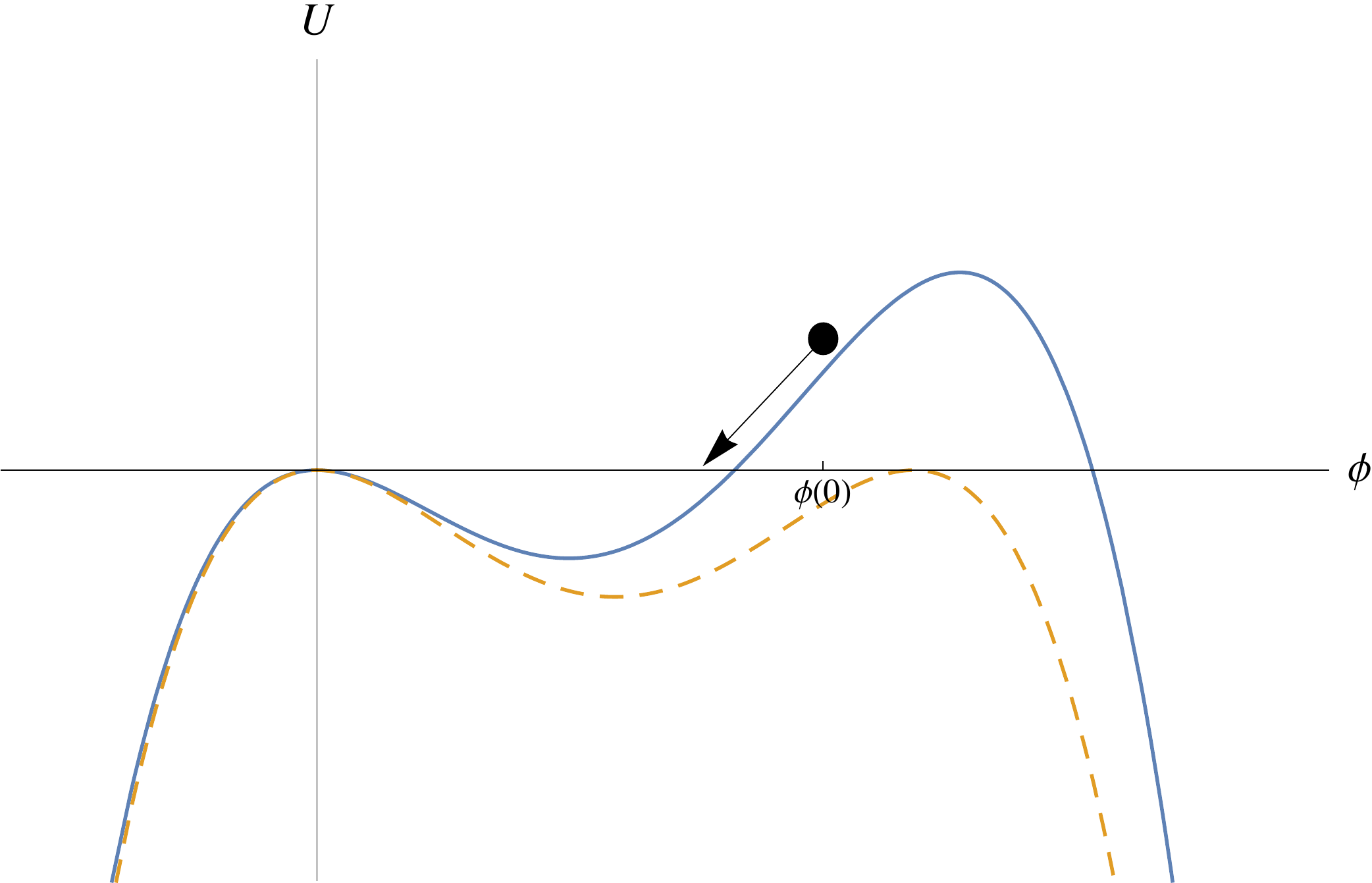}
\caption{Schematic representation of the potential $U=-V(\phi,T)$. The trajectory of the particle is shown to start at values larger than $U(\phi)=0$ because of the friction term. The dashed line represents the potential for the critical temperature.} 
\label{newton}
\end{figure}
This equation should be solved by imposing two boundary conditions. We assume $\phi'(r=0)=0$ to avoid singularity and $\phi(r=\infty)=0$ in order to have a finite energy of our system. In fact, if the scalar field is not asymptotically zero, the contribution of the potential to the total energy would be divergent $\int d^3x V(\phi,T)$. Of course, the trivial solution $\phi=0$ is solution to our problem and hence, contribute to the saddle point approximation but we have also another non-trivial solution. The eq.(\ref{eq:kink}) is similar to the Newton's equation describing the motion of a particle in a potential $U=-V(\phi,T)$ and under the effect of friction. The radial coordinate would correspond to the time variable, while the scalar field would correspond to the position of the particle. As seen from Fig.(\ref{newton}), the particle should finish its motion at the top of the potential $U$ (at $\phi=0$), but because of the presence of the friction term $(\phi'(r)/r)$, the initial condition should be taken at values where $U>0$ as shown in Fig.(\ref{newton}). For any condition, lower than this value $\phi(0)$, the particle would not be able to climb up the potential, while for initial values larger, we would overshoot, which means that we would reach the maximum of the potential $U$ with non-zero velocity. Therefore, using a shooting method, we can find the unique initial condition solving our problem. 
\begin{figure}[ht]
\includegraphics[scale=.28]{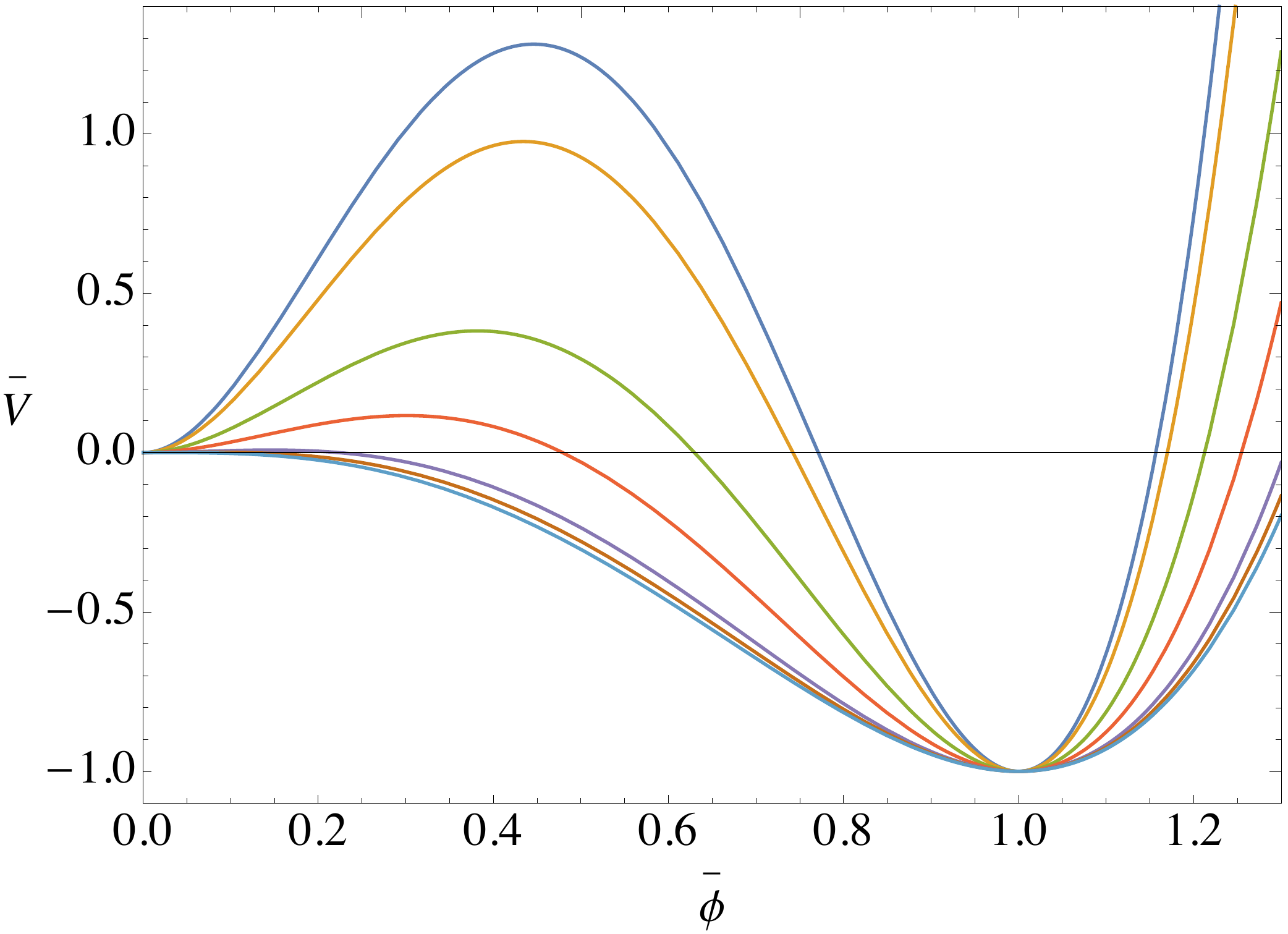}~~~~~
\includegraphics[scale=.28]{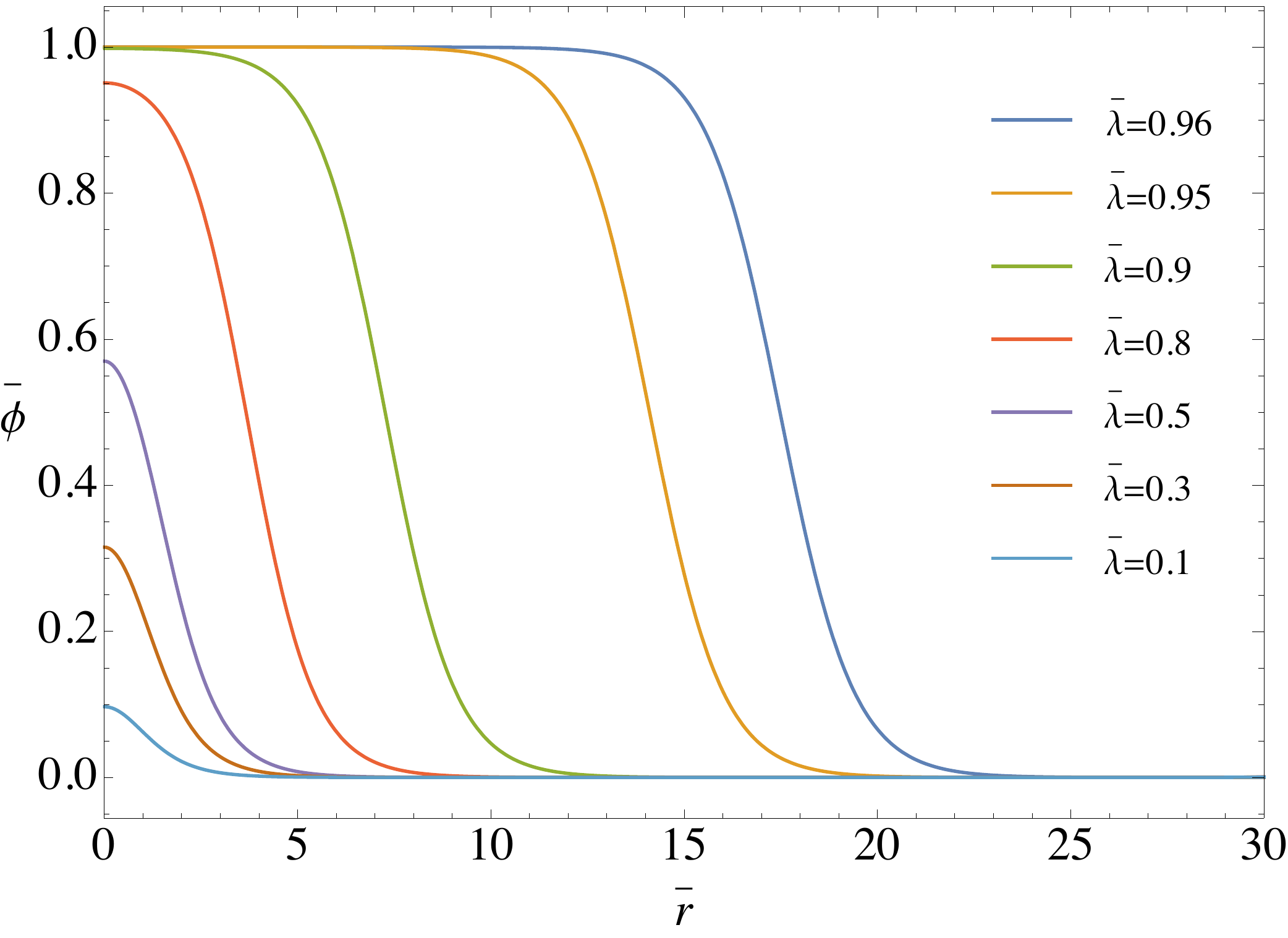}
\caption{On the left, the potential $V(\phi,T)$ normalized to the same minimum, for different values of $\bar\lambda$. On the right, the corresponding solution of the eq.(\ref{eq:kink}) with different values of the parameter $\bar\lambda$.} 
\label{fig:kink}
\end{figure}

As we can see from Fig.(\ref{fig:kink}), while $\bar \lambda\rightarrow 1$, the solution simplifies. In that limit, known as the thin wall limit, the scalar field is constant for $r<R$ for some radius $R$ until it changes abruptly to reach the asymptotic value $\phi=0$. In this limit, which corresponds to the critical temperature, the scalar field starts from $\phi=\phi_+(T_c)=\frac{2\gamma T_c}{3\lambda}$ which in the Fig.(\ref{newton}) corresponds to the condition where $U=0$. This solution is possible because in this regime, the "friction" term ($\phi'(r)/r$) is negligible. Therefore our equation reduces to $\phi''(r)-V(\phi,T)_{,\phi}=0$ (or equivalently $\phi'(r)^2=2V$) with
\begin{align}
    V(\phi,T)=\frac{\lambda}{4}\phi^2(\phi-\phi_+)^2
\end{align}
which is obtained from the potential (\ref{VTFC1}) using $\bar\lambda\rightarrow 1$. This equation admits an exact solution,
\begin{align}
    \phi=\frac{\phi_+}{2}\Bigl(1-\tanh{\frac{r-R}{2 r_0(T_c)}}\Bigr) 
    \label{tanh}
\end{align}
where $R$ is a constant of integration which represents the "radius" of the bubble. 

Using this solution, we can approximately obtain the euclidean action (\ref{euclidean2}). 

If we consider the situation where we are very close to the critical temperature, the potential changes very little between the initial value $\phi(0)=\phi_+$ and $\phi(\infty)=0$. Defining, $\Delta V_T=V_T(0)-V_T(\phi_+)=\frac{4 \gamma^4 }{81 \lambda^3}T_c^2 (T-T_c)^2$ which is zero for $T=T_c$, We have 
\begin{align}
    \int d^3x V_T=4\pi \int_0^\infty dr~ r^2 V_T\simeq 4\pi (V_T(\phi_+)-V_T(0)) \int_0^R dr~ r^2=-\frac{4}{3}\pi \Delta V_T R^3
\end{align}
In the last part of the calculation, we assumed that for $r>R$, we have $\phi=0$ and therefore $V=0$. The other contribution to the action depends on the kinetic energy which is negligible except around $r=R$ (because inside the bubble, $\phi=\phi_+$ and outside the bubble, $\phi=0$)
\begin{align}
    \int d^3x \frac{1}{2}\phi'(r)^2 &=2\pi \int_0^\infty dr~r^2 \phi'(r)^2\simeq 2\pi \int_{R-\epsilon}^{R+\epsilon} dr~r^2 \phi'(r)^2 \simeq 2\pi R^2 \int_{R-\epsilon}^{R+\epsilon} dr~\phi'(r)^2\nonumber\\
    &=-2\pi R^2 \int_{0}^{\phi_+} d\phi~\phi'(r)=2\pi R^2 \int_{0}^{\phi_+} d\phi~\sqrt{2V}=2\pi R^2 \int_{0}^{\phi_+} d\phi~\frac{\sqrt{\lambda}}{\sqrt{2}}\phi(\phi_+-\phi)\nonumber\\
    &=4\pi R^2\frac{\sqrt{\lambda}\phi_+^3}{12\sqrt{2}}
    \equiv 4\pi \sigma R^2
\end{align}
where we used the solution (\ref{tanh}) and we defined the surface tension
\begin{align}
    \sigma = \frac{\phi_+^3\sqrt{\lambda}}{12\sqrt{2}} = \frac{2\sqrt{2}\gamma^3}{81\sqrt{2}\lambda^{5/2}}T_c^3
\end{align}
Therefore, we have 
\begin{align}
    I_3=4 \pi \sigma R^2-\frac{4}{3}\pi \Delta V_T R^3
    \label{energy}
\end{align}
We can define a critical radius at which the bubble neither grows or shrinks. For a bubble larger, it will expand while for a smaller bubble, it will collapse. This critical value is defined as the stationary point of the energy $\delta I_3/\delta R=0$ which means 
\begin{align}
    R_c(T)=\frac{2\sigma}{\Delta V_T}
\end{align}
which gives for $I_3$ (that is also the energy)
\begin{align}
    E_c(T)\equiv E(T, R=R_c)=\frac{16\pi}{3}\frac{\sigma^3}{(\Delta V_T)^2},
    \label{EcE}
\end{align}
where T is close to $T_c$. Obviously energy of the bubble at $T=T_c$ is infinite and so is $R_c$ but the probability of  creation of such a bubble due to thermal fluctuations is zero.
From Eq.(\ref{EcE}), we can obtain the probability of nucleation per unit time and per unit volume, 
\begin{align}
    \Gamma (T) \propto T^4 \Bigl(\frac{E_c(T)}{2\pi T}\Bigr)^{3/2}e^{-E_c(T)/T}
\end{align}
For the derivation of this formula and approximation used to obtain it, reader is referred to Ref.\cite{Hindmarsh:2020hop}.

\end{document}